# Exact Lattice Supersymmetry at the Quantum Level for $N = 2$ Wess-Zumino Models in 1- and 2-Dimensions

Keisuke Asaka[b†], Alessandro D'Adda[a*], Noboru Kawamoto[b‡] and Yoshi Kondo[b§]

[a] *INFN Sezione di Torino, and*
*Dipartimento di Fisica Teorica, Universita di Torino*
*I-10125 Torino, Italy*

[b] *Department of Physics, Hokkaido University*
*Sapporo 060-0810, Japan*

**Abstract**

Supersymmetric lattice Ward-Takahashi identities are investigated perturbatively up to two-loop corrections for super doubler approach of $N = 2$ lattice Wess-Zumino models in 1- and 2-dimensions. In this approach notorious chiral fermion doublers are treated as physical particles and momentum conservation is modified in such a way that lattice Leibniz rule is satisfied. The two major difficulties to keep exact lattice supersymmetry are overcome. This formulation defines, however, nonlocal field theory. Nevertheless we confirm that exact supersymmetry on the lattice is realized for all supercharges at the quantum level. Delicate issues of associativity are also discussed.

†asaka@particle.sci.hokudai.ac.jp
*dadda@to.infn.it
‡kawamoto@particle.sci.hokudai.ac.jp
§proton@particle.sci.hokudai.ac.jp

## 1 Introduction

Supersymmetry is considered to be one of the most important guiding principles to find a formulation beyond the standard model. There is, however, only indirect evidence of realization of supersymmetry for a moment. Supersymmetry should be broken even if it is realized in nature. It is thus important to find a mechanism for the supersymmetry breaking. The origin of supersymmetry is mysterious[1]. There are some interesting examples that the origin of twisted supersymmetry is fundamentally related to a quantization of topological field theory[2, 3]

Supersymmetric field theories need a regularization for the constructive definition of the formulation. It has been a long standing challenge to realize exactly supersymmetric regularization on the lattice[4]. This problem is notoriously difficult to solve completely and can be compared with the possible practical solution of chiral fermion problem for lattice QCD[5].

We find it is important to reconsider the origin of difficulties of realizing exact lattice supersymmetry. We hope to find an origin of breaking mechanism of supersymmetry from regularization point of view. There are two major obstacles to realize exact supersymmetry on the lattice:

1. Breakdown of Leibniz rule for the difference operator.

2. Chiral fermion species doubler problem.

In the supersymmetry algebra bilinear product of supercharges is equal to a differential operator which could be replaced by difference operator on the lattice. The difference operator breaks distributive law of Leibniz rule as an operator while supercharges satisfy the rule, which causes immediate breakdown of the supersymmetry algebra. Secondly in the naive chiral fermion formulation on the lattice, the chiral fermion species doublers appear as real extra particles. The unbalance of the degrees of freedom between bosons and fermions would cause another source of supersymmetry breaking.

The difficulty 1. was already recognized by the first pioneering work by Dondi and Nikolai[4]. In order to overcome this difficulty they proposed a possible modification of lattice supersymmetry formulation in such a way that the conserved momentum should be the momentum representation of difference operator; $\frac{1}{a}\sin ap_\mu$ instead of lattice momentum $p_\mu$ itself. The coordinate representation of this formulation has been studied later[6]. To our knowledge this proposal has never been taken seriously and not been investigated further afterwards. In this paper we pursue along this line of lattice supersymmetry formulation by solving the second chiral fermion problem 2. at the same time.

Due to the above difficulties exact lattice supersymmetry was given up at the early stage of investigations. In two dimensional non-gauge theory, however, a partial restoration of exact supersymmetry was realized by Nicolai mapping[7, 8]. Later the connection between lattice chiral symmetry breaking and lattice supersymmetry breaking was claimed to recover simultaneously in the continuum limit by Curci and Veneziano[9]. Then perturbative recovery of the supersymmetry in the continuum limit was investigated intensively especially by the use of Ward-Takahashi identities (WT-id.) of supersymmetry[10, 11, 12].

The difficulties of the second problem 2. for lattice supersymmetry was already recognized by WT-id. analyses. In order to avoid chiral fermion species doublers of the naive lattice fermion formulation we may employ the Wilson lattice fermion formulation which gets rid of species doublers by sending their mass to the order of inverse lattice constant. In order to satisfy perturbative WT-id. it was shown by Bartels and Kramer that bosonic doublers were needed by introducing bosonic Wilson terms to compensate fermion doublers as counter terms[13]. This analyses suggested that we need bosonic species doublers to compensate fermionic counter parts. This point was rediscovered and stressed by Giedt et.al.[14] Even with a lattice gravity level with Wilson fermions the necessity of bosonic counter terms was recognized[15]. If we identify the fermionic species doublers as physical particles as super partner we need to introduce bosonic species doublers as well, which is precisely the formulation of our SUPER DOUBLER APPROACH[16, 17].

In solving the second problem 2. we may naively expect to use the lattice chiral fermion solution of QCD; modifying the definition of perturbative chiral transformation on the lattice and using Ginsparg-Wilson fermions. In this formulation fermions are treated differently on the lattice from bosons, which may cause another problem for exact lattice SUSY.

Obviously the practical solution of lattice QCD of chiral fermion problem triggered new investigations on lattice SUSY[18, 19, 20]. An exactness of WT-id. along this line has been investigated intensively and commented[21, 20]. Especially the renormalization group method of Ginsparg-Wilson relation for lattice SUSY gave a new insight into the criteria how to choose a derivative operator on the lattice[22, 23]. Unique solution of the SLAC derivative for Ginsparg-Wilson relation was stressed from exact lattice SUSY point of view[23].

Since it took long time for finding a practical solution of the chiral fermion problem of lattice QCD, it has been expected that the story may go similar or there may not be a good solution of exact lattice SUSY. A partial solution for exact supersymmetry was proposed by Kaplan et. al.[24, 25]. and intensively investigated[24, 26, 27, 28, 29]. The nilpotent part of extended twisted supersymmetry algebra can be kept exactly supersymmetric on the lattice. Link gauge version of nilpotent super charge formulation was proposed by Sugino[30]. In two dimensions this supersymmetric nature is strong enough to ensure the recovery of superymmety of other super charges in the continuum limit due to the super renormalizable nature of two dimensional field theory[24]. On the other hand this formulation cleverly avoids the fundamental problem since the nilpotent part of supersymmetry algebra does not include the lattice derivative operator which causes fundamental difficulties.

Although the incompatibility of Leibniz rule and lattice difference operator of 1. has been widely recognized, it was shown explicitly by Kato, Sakamoto and So that it is in fact impossible within the framework of locality, associativity and lattice translational invariance[31]. In order to pursue to finding exact lattice SUSY formulation we need to give up some of the fundamental principles of lattice field theory. In solving lattice chiral fermion problem the definition of perturbative lattice chiral transformation was modified and the notion of exponentially damping locality was introduced. We may need to widen the meaning of exact lattice SUSY as well. There was a trial in this direction as a proposal of exact lattice SUSY with infinite flavors interpretation [32].

Super charge exact formulation of twisted supersymmetric Wess-Zumino model and super Yang-Mills were intensively studied in continuum formulation. The close connection between Dirac-Kaehler twisting procedure of supersymmetry algebra and quantization was cleared up and formulated successfully[3]. There appeared twisted version of lattice SUSY formulation[33]. Based on the above mentioned continuum formulations the link approach formulation of lattice SUSY were proposed, where exact lattice SUSY was claimed[34]. It was, however, pointed out that there is an ordering ambiguity of products of fields in the link approach formulation, which made the exact lattice SUSY statement doubtful[35, 36]. It turned out that this ordering ambiguity can be treated correctly by introducing link size noncommutative nature of fields with respect to the space time[16]. We thus have to give up the commutative nature of fields in this formulation contrary to the standard field theory. Algebraic consistency of this formulation with noncommutativity was confirmed in the framework of Hopf Algebra[37]. There remained a possible problem of gauge invariance for the link approach of lattice super Yang-Mills. This link approach with particular choice of parameter coincides with the nilpotent charge treatment of Kaplan et. al. The relation between these two formulations was cleared up[38].

Considering the experiences of link approach formulation, we reached to a notion of introducing intermediate lattice sites which are related to the species doublers of chiral fermion, to be compatible with twisted supersymmetry algebra. We then formulated it in the momentum space, which we call the SUPER DOUBLER APPROACH. It turned out that we modified the momentum conservation similar as Dondi and Nicolai proposed. Then formulation turns out to be nonlocal in the coordinate space. We claim that this formulation of Wess-Zumino models in 1- and 2-dimensions has exact N=2 twisted lattice supersymmetry for all super charges and does not have chiral fermion species doubler problem[16, 17].

If we formulate massless fermion on the lattice naively the appearance of the chiral fermion species doublers is unavoidable[39, 40]. Then we let them appear and identify them as super partners or truncate them consistently by chiral conditions so that chiral fermion problems disappear[16, 17]. The modified momentum conservation leads to define a new nonlocal product in the coordinate space. It was shown that the Lebniz rule of difference operator is satisfied exactly on this product. Although naive lattice translational invariance is lost, it does not necessarily mean that Poincare invariance of the action is lost[41].

After it has been recognized that breakdown of Leibniz rule is unavoidable and a possible solution of Ginsparg-Wilson relation for lattice SUSY is the SLAC derivative[23], it was stressed by Bergner that a nonlocal derivative operator and a nonlocal interaction term are unavoidable[42]. The questions were numerically investigated[43, 44]. It was then natural to ask a question if the nonlocal nature of the definition of a derivative operator and a interaction term or an introduction of momentum cut off generate any problems for the definition of local supersymmetric field theory in the continuum limit. Careful analyses by Kadoh and Suzuki showed that at least for non-gauge formulation of lattice SUSY in lower dimensions they do not generate any problems for the constructive definition of lattice SUSY[48]. A numerical criteria how supersymmetry can be recovered in the continuum limit was proposed[49].

In this paper we investigate if the supersymmetric WT-id's. are exactly satisfied at the quantum level by the the super doubler models of N=2 Wess-Zumino modles in 1- and 2-dimensions. The super doubler models are claimed to have exact supersymmetry for all super charges of N=2 algebra on the lattice. The models have nonlocal derivative and interaction terms and thus interesting to ask if supersymmetry is kept at the quantum level with the perspective of the above statements.

The paper is organized as follows: We first explain a simple derivation of WT-id. in section 2. We then briefly present the super doubler formulation of 1-dimensional Wess-Zumino model and the corresponding graphical representation of propagators and vertices in section 3. The corresponding tree, one-loop and two-loop analyses of WT-id's are given in section 4. The two dimensional model and the graphic representation of propagators and verices are given in section 5 and the analyses of WT-id's are given in section 6. In section 7 we discuss the breakdown of associativity of the super doubler approach of this paper. In the final section 8 we briefly summarize the results.

## 2 Ward-Takahashi identity for quantum level check of exact lattice supersymmetry

When a model possesses a symmetry, there exists a corresponding conservation of Noether current classically. It is not obvious if this symmetry is exactly kept at the quantum level. In order to see if the symmetry is kept at the quantum level we usually examine Ward-Takahashi identity perturbatively or even nonperturbatively. To derive general form of Ward-Takahashi identity, we start with functional formalism. The partition function and the expectation value of an operator $\mathcal{O}$ can be defined as,

$$\mathcal{Z} = \int \mathcal{D}[\Phi] e^{i\mathcal{S}[\Phi]}, \tag{2.1}$$

$$\langle \mathcal{O} \rangle = \frac{1}{\mathcal{Z}} \int \mathcal{D}[\Phi] \mathcal{O}[\Phi] e^{i\mathcal{S}[\Phi]}, \tag{2.2}$$

where $\Phi$ denotes any fields in the model. We then consider a variation of fields $\Phi \to \Phi' = \Phi + \delta\Phi$, and the operator changes as $\mathcal{O}[\Phi] \to \mathcal{O}[\Phi'] = \mathcal{O}[\Phi] + \delta\mathcal{O}[\Phi]$. We also assumes that the action receives the variation as $\mathcal{S} \to \mathcal{S}' = \mathcal{S} + \delta\mathcal{S}$. The redefinition of the fields does not change the value of the path integral and thus give the following Ward-Takahashi identity:

$$\begin{aligned}\langle \mathcal{O} \rangle = \frac{1}{\mathcal{Z}} \int \mathcal{D}[\Phi] \mathcal{O}[\Phi] e^{i\mathcal{S}[\Phi]} &= \frac{1}{\mathcal{Z}} \int \mathcal{D}[\Phi'] \mathcal{O}[\Phi'] e^{i\mathcal{S}[\Phi']} \\ &= \frac{1}{\mathcal{Z}} \int \mathcal{D}[\Phi] \left(\mathcal{O}[\Phi] + \delta\mathcal{O}[\Phi]\right) e^{i\mathcal{S}[\Phi] + i\delta\mathcal{S}} \\ &= \langle \mathcal{O} \rangle + \langle \delta\mathcal{O}[\Phi] \rangle + \langle \mathcal{O}[\Phi] \delta\mathcal{S}[\Phi] \rangle + \cdots , \end{aligned} \tag{2.3}$$

where we assume functional measure is not anomalous under this symmetry. If the action has a symmetry under this variation; $\delta\mathcal{S}[\Phi] = 0$, we obtain the following identity:

$$\langle \delta\mathcal{O}[\Phi] \rangle = 0. \tag{2.4}$$

In this paper we investigate the validity of this identity at the quantum level by perturbation. If the regularization of the field theoretical model is systematically defined the identity should hold even if quantum corrections are included at the loop level. It plays a role of fundamental check if the symmetry is exactly preserved with the lattice regularization. In this paper we examine the exactness of the supersymmetry at the quantum level with a lattice regularization.

Since we consider supersymmetry transformation for $\delta$ we need to choose $\mathcal{O}$ as fermionic operator to obtain bosonic composite fields. In this paper we investigate a case where $\mathcal{O}$ has fermion and boson bi-linear form as $\mathcal{O} = \phi_A \psi_B$ where $\phi_A$ and $\psi_B$ are bosonic and fermionic fields, respectively. Then the Ward-Takahashi identity (2.4) for two point operator leads the following form:

$$\langle \delta\phi_A \psi_B \rangle + \langle \phi_A \delta\psi_B \rangle = 0. \tag{2.5}$$

## 3 Super doubler approach for $N = 2$ Wess-Zumino model in 1-dimension

We introduce recently proposed lattice N=2 Wess-Zumino action of 1-dimension in the momentum space. In this formulation we postulate the lattice momentum $\hat{p} = \sin\frac{ap}{2}$ (which we often call sine momentum) as a conserved quantity instead of periodic lattice momentum $p$ itself. We then found that the following action has exact N=2 supersymmetry on the lattice for all super charges[16].

$$
\begin{aligned}
S^{(n)} &= \frac{4g_n^{(0)}a^n}{n!}\int_{-\frac{\pi}{a}}^{\frac{3\pi}{a}}\prod_{j=1}^{n}\left(\frac{dp_j}{2\pi}\right)G(p_1,\cdots,p_n)(2\pi)\delta\left(\sum_{j=1}^{n}\sin\frac{ap_j}{2}\right) \\
&\quad\times\frac{i}{2n}Q_1Q_2\left(\Phi(p_1)\Phi(p_2)\cdots\Phi(p_n)\right) \\
&= \frac{4g_n^{(0)}a^n}{n!}\int\prod_{j=1}^{n}\left(\frac{dp_j}{2\pi}\cos\frac{ap_j}{2}\right)(2\pi)\delta\left(\sum_{j=1}^{n}\sin\frac{ap_j}{2}\right) \\
&\quad\times\left[2\sin^2\frac{ap_1}{4}\Phi(p_1)\prod_{j=2}^{n}\Phi(p_j)+\frac{n-1}{4}\sin\frac{a(p_1-p_2)}{4}\Psi(p_1)\Psi(p_2)\prod_{j=3}^{n}\Phi(p_j)\right]. \qquad (3.1)
\end{aligned}
$$

where $G(p_1,\cdots,p_n)$ is a function satisfying the following conditions to keep exact lattice supersymmetry: it is symmetric under the permutation of $p_i$, and periodic in $\frac{4\pi}{a}$ for all momenta. In particular for interaction terms ($n \geq 3$) we choose

$$
G(p_1,\cdots,p_n) = \prod_{i=1}^{n}\cos\frac{ap_i}{2}. \qquad (3.2)
$$

This factor cancels the singularity at $p_i = \pm\frac{\pi}{a}$ caused from the "sine" momentum conservation of the delta function. Here $\Phi(p)$ and $\Psi(p)$ are bosonic and fermionic dimensionless composite fields in the momentum space including species doublers as super partners of $N = 2$ supersymmetry.

This action is invariant under the following N=2 supersymmetry transformation:

$$
\begin{aligned}
\delta_1\Phi(p) &= i\alpha_1\cos\frac{ap}{4}\Psi(p), & \delta_1\Psi(p) &= -4i\alpha_1\sin\frac{ap}{4}\Phi(p), \\
\delta_2\Phi(p) &= \alpha_2\cos\frac{ap}{4}\Psi\left(\frac{2\pi}{a}-p\right), & \delta_2\Psi\left(\frac{2\pi}{a}-p\right) &= 4\alpha_2\sin\frac{ap}{4}\Phi(p), \qquad (3.3)
\end{aligned}
$$

where $\alpha_1$ and $\alpha_2$ are super parameters.

Considering the lattice constant dependence with dimensionless nature of the action, we define the component fields of N=2 super multiplets by

$$
\begin{aligned}
\Phi(p) &= a^{-\frac{3}{2}}\phi(p), & \Phi\left(\frac{2\pi}{a}-p\right) &= -\frac{a^{-\frac{1}{2}}}{4}D(p), \\
\Psi(p) &= a^{-1}\psi_1(p), & \Psi\left(\frac{2\pi}{a}-p\right) &= ia^{-1}\psi_2(p), \qquad (3.4)
\end{aligned}
$$

where the species doubler of fermion $\psi_1$ is identified as a second fermion $\psi_2$ and the species doubler of boson $\phi$ is identified as an auxiliary field $D$ in the same supermultiplet of N=2 supersymmetry. We thus introduce the species doubler counterparts as super multiplets equally for boson and fermion. It is a characteristic of our formulation that we identify the species doublers of the original boson and fermion as super partners. Thus we naturally call this formulation as SUPER DOUBLER APPROACH[16].

The supersymmetry transformation by the component fields lead:

$$\begin{aligned} \delta_1\phi(p) &= i\eta_1 \cos\frac{ap}{4}\psi_1(p), & \delta_1\psi_1(p) &= -\frac{4i}{a}\eta_1 \sin\frac{ap}{4}\phi(p), \\ \delta_1 D(p) &= \frac{4}{a}\eta_1 \sin\frac{ap}{4}\psi_2(p), & \delta_1\psi_2(p) &= \eta_1 \cos\frac{ap}{4}D(p), \end{aligned} \tag{3.5}$$

and

$$\begin{aligned} \delta_2\phi(p) &= i\eta_2 \cos\frac{ap}{4}\psi_2(p), & \delta_2\psi_1(p) &= -\eta_2 \cos\frac{ap}{4}D(p), \\ \delta_2 D(p) &= -\eta_2\frac{4}{a} \sin\frac{ap}{4}\psi_1(p), & \delta_2\psi_2(p) &= -\frac{4i}{a}\eta_2 \sin\frac{ap}{4}\phi(p), \end{aligned} \tag{3.6}$$

where super parameters are parametrized as $\alpha_i = a^{-\frac{1}{2}}\eta_i$ $(i = 1, 2)$.

One can check that the supersymmetry transformation of supercharges $\delta_1 = \eta_1 Q_1$ and $\delta_2 = \eta_2 Q_2$ satisfy the momentum representation of the following algebra on the lattice:

$$Q_1^2 = Q_2^2 = 2\sin\frac{ap}{2}, \qquad \{Q_1, Q_2\} = 0. \tag{3.7}$$

This is the lattice momentum version of $D = 1, N = 2$ supersymmetry algebra.

For $n = 2$ the action (3.1) admits a generalization containing a free parameter and generates both kinetic and mass term. From (3.1) we have:

$$\begin{aligned} S^{(2)} = 2a^2 \int_{-\frac{\pi}{a}}^{\frac{3\pi}{a}} \frac{dp_1}{2\pi}\frac{dp_2}{2\pi} 2\pi\delta\left(\sin\frac{ap_1}{2} + \sin\frac{ap_2}{2}\right) G(p_1, p_2) \\ \times \left[2\sin^2\frac{ap_1}{4}\Phi(p_1)\Phi(p_2) + \frac{1}{4}\sin\frac{a(p_1 - p_2)}{4}\Psi(p_1)\Psi(p_2)\right], \end{aligned} \tag{3.8}$$

where we have set $g_2^{(0)} = 1$ and chosen $G(p_1, p_2)$ of the form given in (3.2). For $n = 2$, and only in that case, the argument of the delta function vanishes if one of the two linear conditions $p_1 + p_2 = 0$ and $p_1 - p_2 - \frac{2\pi}{a} = 0$ are satisfied modulo $\frac{4\pi}{a}$. So in eq. (3.8) we can replace the delta function according to:

$$\frac{a}{2}\delta\left(\sin\frac{ap_1}{2} + \sin\frac{ap_2}{2}\right) \to \frac{1}{\left|\cos\frac{ap_1}{2}\right|}\delta(p_1 + p_2) + \frac{m_0}{\left|\cos\frac{ap_1}{2}\right|}\delta\left(p_1 - p_2 - \frac{2\pi}{a}\right), \tag{3.9}$$

where we have introduced a dimensionless arbitrary mass parameter $m_0$*. With this replace-

*Notice that for $m_0 = 1$ the r.h.s. and the l.h.s. of (3.9) coincide, for $m_0 \neq 1$ this replacement amounts to a redefinition of the arbitrary function $G(p_1, p_2)$ ment $S^{(2)}$ splits into the sum of a kinetic and a mass term given respectively by:

$$
\begin{aligned}
S_k =&\int_{-\frac{\pi}{a}}^{\frac{\pi}{a}} \frac{dp}{2\pi} \cos\frac{ap}{2} \left[ \frac{4}{a^2}\left(1-\cos\frac{ap}{2}\right)\phi(-p)\phi(p) + \frac{1}{4}\left(1+\cos\frac{ap}{2}\right) D(-p)D(p) \right] \\
&+ \int_{-\frac{\pi}{a}}^{\frac{\pi}{a}} \frac{dp}{2\pi} \cos\frac{ap}{2} \left[ -\frac{1}{a}\sin\frac{ap}{2}\psi_1(-p)\psi_1(p) - \frac{1}{a}\sin\frac{ap}{2}\psi_2(-p)\psi_2(p) \right], \qquad (3.10) \\
S_m =& 2m \int_{-\frac{\pi}{a}}^{\frac{\pi}{a}} \frac{dp}{2\pi} \cos\frac{ap}{2} \left[ -\phi(-p)D(p) - i\psi_1(-p)\psi_2(p) \right], \qquad (3.11)
\end{aligned}
$$

where a dimensional mass parameter $m = \frac{m_0}{a}$ has been introduced. The integration domain is divided into $p \in \left[-\frac{\pi}{a}, \frac{\pi}{a}\right]$ and $p \in \left[\frac{\pi}{a}, \frac{3\pi}{a}\right]$ in (3.1) and the identifications of (3.4) are used.

Since we identify the lattice momentum $\hat{p}$ as conserved quantity it is convenient to reformulate by

$$
\hat{p}_i \equiv \frac{2}{a}\sin\frac{ap_i}{2}. \qquad (3.12)
$$

Hereafter we often use the following notation for

$$
C(\hat{p}_i) \equiv \sqrt{1-\frac{a^2}{4}\hat{p}_i^2} = \cos\frac{ap_i}{2}, \qquad (3.13)
$$

in the momentum region $p_i \in \left[-\frac{\pi}{a}, \frac{\pi}{a}\right]$. Then the kinetic and mass terms of the action for boson and fermion are given by

$$
S_B = \frac{1}{2}\int_{-\frac{2}{a}}^{\frac{2}{a}} \frac{d\hat{p}}{2\pi} \begin{pmatrix} \phi(-p) & D(-p) \end{pmatrix} K_B \begin{pmatrix} \phi(p) \\ D(p) \end{pmatrix}, \qquad (3.14)
$$

$$
S_F = \frac{1}{2}\int_{-\frac{2}{a}}^{\frac{2}{a}} \frac{d\hat{p}}{2\pi} \begin{pmatrix} \psi_1(-p) & \psi_2(-p) \end{pmatrix} K_F \begin{pmatrix} \psi_1(p) \\ \psi_2(p) \end{pmatrix}, \qquad (3.15)
$$

where

$$
K_B = \begin{pmatrix} \frac{8}{a^2}(1-C(\hat{p})) & -2m \\ -2m & \frac{1}{2}(1+C(\hat{p})) \end{pmatrix}, \qquad (3.16)
$$

$$
K_F = \begin{pmatrix} -\hat{p} & -2im \\ 2im & -\hat{p} \end{pmatrix}. \qquad (3.17)
$$

Then the propagators for the component fields are given as:

$$
K_B^{-1} = \frac{1}{\hat{p}^2-4m^2} \begin{pmatrix} \frac{1}{2}(1+C(\hat{p})) & 2m \\ 2m & \frac{8}{a^2}(1-C(\hat{p})) \end{pmatrix}, \qquad (3.18)
$$

$$
K_F^{-1} = \frac{1}{\hat{p}^2-4m^2} \begin{pmatrix} -\hat{p} & 2im \\ -2im & -\hat{p} \end{pmatrix}. \qquad (3.19)
$$

The interaction terms of action can be derived from $S^{(n)}$ in (3.1) for $n = 3$ and $n = 4$.

Here we consider cubic $\Phi^3$ and quartic $\Phi^4$ interactions. The cubic interaction term is given

$$
\begin{aligned}
S^{(3)} &= \frac{4g_0^{(3)}a^3}{3!}\int\prod_{j=1}^{3}\left(\frac{dp_j}{2\pi}\cos\frac{ap_j}{2}\right)(2\pi)\delta\left(\sum_{j=1}^{3}\sin\frac{ap_j}{2}\right) \\
&\quad\times\left[2\sin^2\frac{ap_1}{4}\Phi(p_1)\Phi(p_2)\Phi(p_3)+\frac{1}{2}\sin\frac{a(p_1-p_2)}{4}\Psi(p_1)\Psi(p_2)\Phi(p_3)\right] \qquad (3.20)\\
&= S_B^{(3)}+S_F^{(3)}, \qquad (3.21)
\end{aligned}
$$

where

$$
\begin{aligned}
S_B^{(3)} &= \frac{4g_3a^2}{3}\int\frac{d\hat{p}_1}{2\pi}\frac{d\hat{p}_2}{2\pi}\frac{d\hat{p}_3}{2\pi}2\pi\delta(\hat{p}_1+\hat{p}_2+\hat{p}_3) \\
&\quad\times\left[\phi_+(p_1)\phi_+(p_2)\phi_+(p_3)-C(\hat{p}_1)\phi_-(p_1)\phi_+(p_2)\phi_+(p_3)\right], \qquad (3.22)\\
S_F^{(3)} &= \frac{2g_3a}{3}\int\frac{d\hat{p}_1}{2\pi}\frac{d\hat{p}_2}{2\pi}\frac{d\hat{p}_3}{2\pi}2\pi\delta(\hat{p}_1+\hat{p}_2+\hat{p}_3) \\
&\quad\times\Big[\sin\frac{a}{4}(p_1-p_2)\psi_1(p_1)\psi_1(p_2)\phi_+(p_3) \\
&\quad+\sin\frac{a}{4}(p_1-p_2)\psi_2(p_1)\psi_2(p_2)\phi_+(p_3) \\
&\quad+2i\cos\frac{a}{4}(p_1+p_2)\psi_1(p_1)\psi_2(p_2)\phi_+(p_3)\Big], \qquad (3.23)
\end{aligned}
$$

where we use the notation for $C(\hat{p}_1)$ as given in (3.13) and

$$\phi_\pm(p)=\frac{1}{a}\phi(p)\pm\frac{1}{4}D(p). \qquad (3.24)$$

This combination of fields appears from the following decomposition:

$$
\begin{aligned}
\int_{-\frac{\pi}{a}}^{\frac{3\pi}{a}}dp\cos\frac{ap}{2}\Phi(p) &= \int_{-\frac{\pi}{a}}^{\frac{\pi}{a}}dp\cos\frac{ap}{2}\left(\Phi(p)-\Phi(\frac{2\pi}{a}-p)\right) \\
&= \int_{-\frac{\pi}{a}}^{\frac{\pi}{a}}dp\cos\frac{ap}{2}\frac{1}{\sqrt{a}}\left(\frac{1}{a}\phi(p)+\frac{1}{4}D(p)\right). \qquad (3.25)
\end{aligned}
$$

As we can see, the interaction terms have more compact form with respect to the component fields $\phi_\pm(p)$ than those of $\phi(p)$ and $D(p)$. The corresponding propagators for these fields can be given as

$$\phi_+(p) \text{ - - - - - - - - - - } \phi_+(-p) \quad = \frac{1+am}{D(\hat{p})}, \qquad (3.26)$$

$$\phi_-(p) \cdots\cdots\cdots\cdots \phi_-(-p) \quad = \frac{1-am}{D(\hat{p})}, \qquad (3.27)$$

$$\phi_+(p) \text{ - - - - - }\cdots\cdots \phi_-(-p) \quad = \frac{C(\hat{p})}{D(\hat{p})}, \qquad (3.28)$$

where we introduce the following notation:

$$D(\hat{p})=a^2\hat{p}^2-4a^2m^2. \qquad (3.29)$$

The fermionic propagators are given as

$\psi_{1,2}(p)$  $\psi_{1,2}(-p)$ $= \dfrac{-a^2\hat{p}}{D(\hat{p})},$ (3.30)

$\psi_1(p)$  $\psi_2(-p)$ $= \dfrac{2ia^2m}{D(\hat{p})}.$ (3.31)

Feynman rules for three point vertices are assigned as,

$\phi_+(p_1)$, $\phi_+(p_2)$ 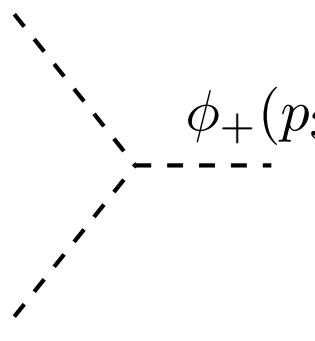 $\phi_+(p_3)$ $= \dfrac{4ig_3a^2}{3},$

$\phi_-(p_1)$, $\phi_+(p_2)$ 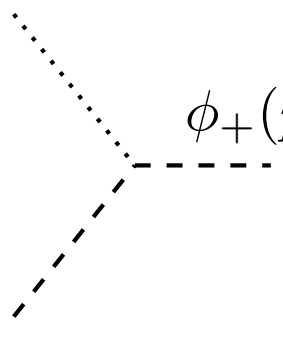 $\phi_+(p_3)$ $= -\dfrac{4ig_3a^3}{3}C(\hat{p}_1),$

$\psi_{1,2}(p_1)$, $\psi_{1,2}(p_2)$ 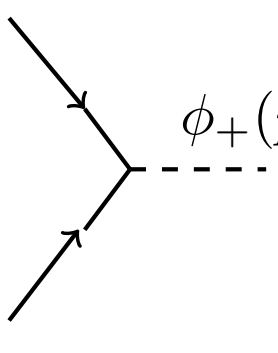 $\phi_+(p_3)$ $= \dfrac{2ig_3a}{3}\sin\dfrac{a}{4}(p_1-p_2),$

$\psi_1(p_1)$, $\psi_2(p_2)$ 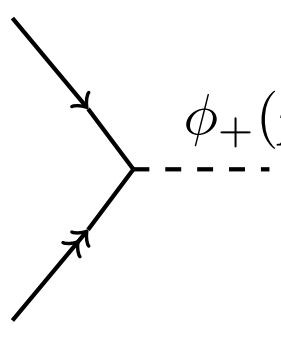 $\phi_+(p_3)$ $= -\dfrac{4g_3a}{3}\cos\dfrac{a}{4}(p_1+p_2).$

Similarly we obtain quartic interaction terms as:

$$S^{(4)} = \frac{4g_0^{(4)}a^4}{4!}\int\prod_{j=1}^{4}\left(\frac{dp_j}{2\pi}\cos\frac{ap_j}{2}\right)(2\pi)\delta\left(\sum_{j=1}^{4}\sin\frac{ap_j}{2}\right)$$
$$\times\left[2\sin^2\frac{ap_1}{4}\Phi(p_1)\Phi(p_2)\Phi(p_3)\Phi(p_4)+\frac{3}{4}\sin\frac{a(p_1-p_2)}{4}\Psi(p_1)\Psi(p_2)\Phi(p_3)\Phi(p_4)\right] \tag{3.32}$$
$$= S_B^{(4)} + S_F^{(4)}, \tag{3.33}$$

where

$$S_B^{(4)} = \frac{4g_0^{(4)}a^4}{4!}\int\prod_{j=1}^{4}\left(\frac{dp_j}{2\pi}\cos\frac{ap_j}{2}\right)(2\pi)\delta\left(\sum_{j=1}^{4}\sin\frac{ap_j}{2}\right)$$
$$= a^{-2}\left[\phi_+(p_1) - C(\hat{p}_1)\phi_-(p_1)\right]\phi_+(p_2)\phi_+(p_3)\phi_+(p_4), \tag{3.34}$$
$$S_F^{(4)} = \frac{4g_0^{(4)}a^4}{4!}\int\prod_{j=1}^{4}\left(\frac{dp_j}{2\pi}\cos\frac{ap_j}{2}\right)(2\pi)\delta\left(\sum_{j=1}^{4}\sin\frac{ap_j}{2}\right)$$
$$= a^{-3}\left(\sin\frac{a}{4}(p_1-p_2)\psi_1(p_1)\psi_1(p_2) + \sin\frac{a}{4}(p_1-p_2)\psi_2(p_1)\psi_2(p_2)\right.$$
$$\left. +2i\cos\frac{a}{4}(p_1+p_2)\psi_1(p_1)\psi_2(p_2)\right)\phi_+(p_3)\phi_+(p_4). \tag{3.35}$$

Feynman rules for four point vertices are assigned as,

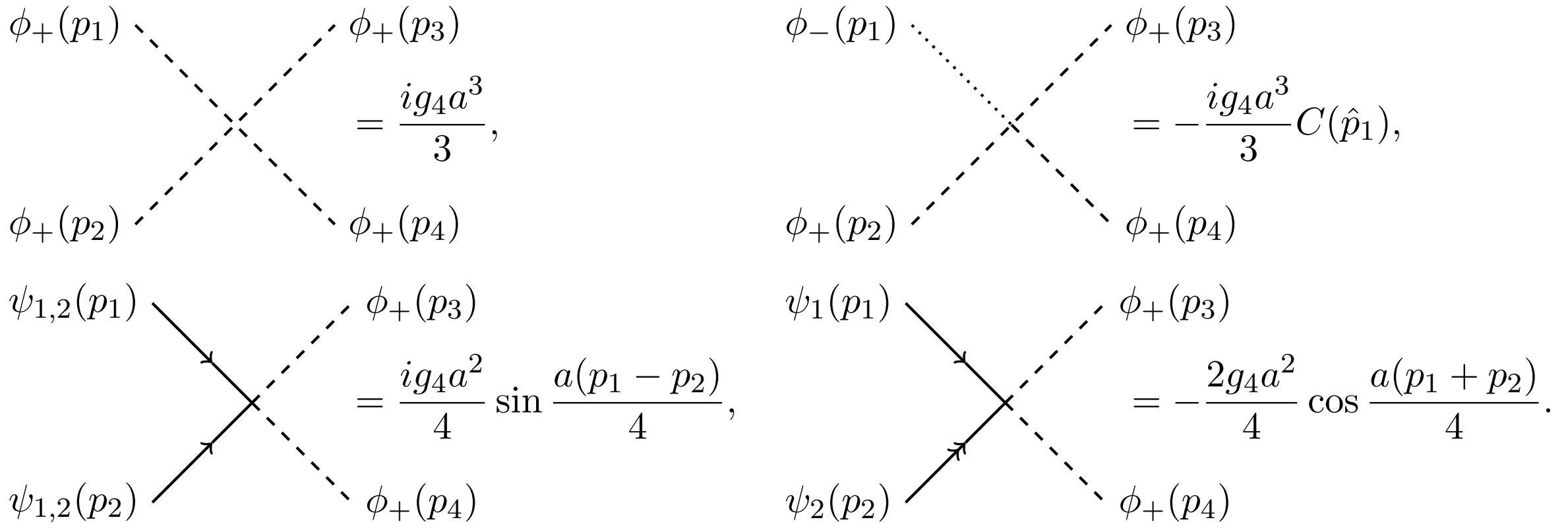

$$= \frac{ig_4a^3}{3}, \qquad = -\frac{ig_4a^3}{3}C(\hat{p}_1),$$

$$= \frac{ig_4a^2}{4}\sin\frac{a(p_1-p_2)}{4}, \qquad = -\frac{2g_4a^2}{4}\cos\frac{a(p_1+p_2)}{4}.$$

We have now prepared all the necessary tools for perturbative calculation of quantum corrections for $N=2$ supersymmetric Wess-Zumino model in 1-dimention.

# 4 Ward-Takahashi identity in 1-dimention

## 4.1 Tree level Ward-Takahashi identity

We first show that Ward-Takahashi identities for supersymmetry transformation for $\delta$ are satisfied at the tree level. We consider all fermionic bilinear fields $\mathcal{O} = \phi_A\psi_B$ of 1-dimensional $N=2$ Wess-Zumino model of previous section,

$$\phi_A\psi_B = \phi\psi_1,\ \phi\psi_2,\ D\psi_1,\ D\psi_2. \tag{4.1}$$

We can then obtain Ward-Takahashi identities of supersymmetry of two point operators for $\delta = \delta_1, \delta_2$ leading (2.4). For example considering $\phi_A\psi_B = \phi\psi_1$ for $\delta = \delta_1$, we can show that tree level Ward-Takahashi identity is satisfied as follows:

$$\begin{aligned}\langle\delta_1(\phi(p)\psi_1(-p))\rangle_{\text{tree}} &= i\eta_1\left[\cos\frac{ap}{4}\langle\psi_1(p)\psi_1(-p)\rangle_{\text{tree}} + \frac{4}{a}\sin\frac{ap}{4}\langle\phi(p)\phi(-p)\rangle_{\text{tree}}\right] \\ &= \frac{i\eta_1}{D(\hat{p})}\left[\cos\frac{ap}{4}\left(-4a\sin\frac{ap}{4}\cos\frac{ap}{4}\right) + \frac{4}{a}\sin\frac{ap}{4}\left(a^2(\cos\frac{ap}{4})^2\right)\right] = 0,\end{aligned} \tag{4.2}$$

where we have used the fact that the tree level two point functions of $\phi$ and $\psi_1$ are propagators of these fields given in (3.18) and (3.19). Here we have used the notation for inverse propagator $D(\hat{p})$ in (3.29). We have thus shown that the following Ward-Takahashi identity is satisfied at the tree level:

$$\cos\frac{ap}{4}\langle\psi_1(p)\psi_1(-p)\rangle_{\text{tree}} + \frac{4}{a}\sin\frac{ap}{4}\langle\phi(p)\phi(-p)\rangle_{\text{tree}} = 0. \tag{4.3}$$

Similarly we can show that the following tree level Ward-Takahashi identities for super-symmetry variation of $\delta_1$ are satisfied:

$$\begin{aligned}
\langle\delta_1(\phi(p)\psi_2(-p))\rangle_{\rm tree} &= \eta_1 \cos\frac{ap}{4}\left[i\left\langle\psi_1(p)\psi_2(-p)\right\rangle_{\rm tree} + \left\langle\phi(p)D(-\hat{p})\right\rangle_{\rm tree}\right] = 0, \\
\langle\delta_1(D(p)\psi_1(-p))\rangle_{\rm tree} &= \frac{4}{a}\eta_1 \sin\frac{ap}{4}\left[\left\langle\psi_2(p)\psi_1(-p)\right\rangle_{\rm tree} + i\left\langle D(p)\phi(-p)\right\rangle_{\rm tree}\right] = 0, \\
\langle\delta_1(D(p)\psi_2(-p))\rangle_{\rm tree} &= \eta_1\left[\frac{4}{a}\sin\frac{ap}{4}\left\langle\psi_2(p)\psi_2(-p)\right\rangle_{\rm tree} + \cos\frac{ap}{4}\left\langle D(p)D(-p)\right\rangle_{\rm tree}\right] = 0.
\end{aligned} \tag{4.4}$$

Similarly for supersymmetry variation $\delta_2$ we can show:

$$\begin{aligned}
\langle\delta_2(\phi(p)\psi_1(-p))\rangle_{\rm tree} &= \eta_2 \cos\frac{ap}{4}\left[i\left\langle\psi_2(p)\psi_1(-p)\right\rangle_{\rm tree} - \left\langle\phi(p)D(-p)\right\rangle_{\rm tree}\right] = 0, \\
\langle\delta_2(\phi(p)\psi_2(-p))\rangle_{\rm tree} &= i\eta_2\left[\cos\frac{ap}{4}\left\langle\psi_2(p)\psi_2(-p)\right\rangle_{\rm tree} + \frac{4}{a}\sin\frac{ap}{4}\left\langle\phi(p)\phi(-p)\right\rangle_{\rm tree}\right] = 0, \\
\langle\delta_2(D(\hat{p})\psi_1(-\hat{p}))\rangle_{\rm tree} &= -\eta_2\left[\frac{4}{a}\sin\frac{ap}{4}\left\langle\psi_1(p)\psi_1(-p)\right\rangle_{\rm tree} + \cos\frac{ap}{4}\left\langle D(p)D(-p)\right\rangle_{\rm tree}\right] = 0, \\
\langle\delta_2(D(p)\psi_2(-p))\rangle_{\rm tree} &= -\eta_2\frac{4}{a}\sin\frac{ap}{4}\left[\left\langle\psi_1(p)\psi_2(-p)\right\rangle_{\rm tree} - i\left\langle D(p)\phi(-p)\right\rangle_{\rm tree}\right] = 0.
\end{aligned} \tag{4.5}$$

Among the above eight Ward-Takahashi identities we can find the following four independent identity relations at the tree level:

$$\cos\frac{ap}{4}\left\langle\psi_1(p)\psi_1(-p)\right\rangle_{\rm tree} + \frac{4}{a}\sin\frac{ap}{4}\left\langle\phi(p)\phi(-p)\right\rangle_{\rm tree} = 0, \tag{4.6}$$

$$i\left\langle\psi_1(p)\psi_2(-p)\right\rangle_{\rm tree} + \left\langle\phi(p)D(-p)\right\rangle_{\rm tree} = 0, \tag{4.7}$$

$$\frac{4}{a}\sin\frac{ap}{4}\left\langle\psi_2(p)\psi_2(-p)\right\rangle_{\rm tree} + \cos\frac{ap}{4}\left\langle D(p)D(-p)\right\rangle_{\rm tree} = 0, \tag{4.8}$$

$$\left\langle\phi(p)\phi(-p)\right\rangle_{\rm tree} - \left(\frac{a}{4}\cot\frac{ap}{4}\right)^2\left\langle D(p)D(-p)\right\rangle_{\rm tree} = 0. \tag{4.9}$$

In the following we investigate if these Ward-Takahashi identities are satisfied even at the quantum level with loop corrections.

## 4.2 1-loop corrections

### 4.2.1 1-loop corrections: cubic interaction $\Phi^3$

We consider the following loop diagrams for two point operators:

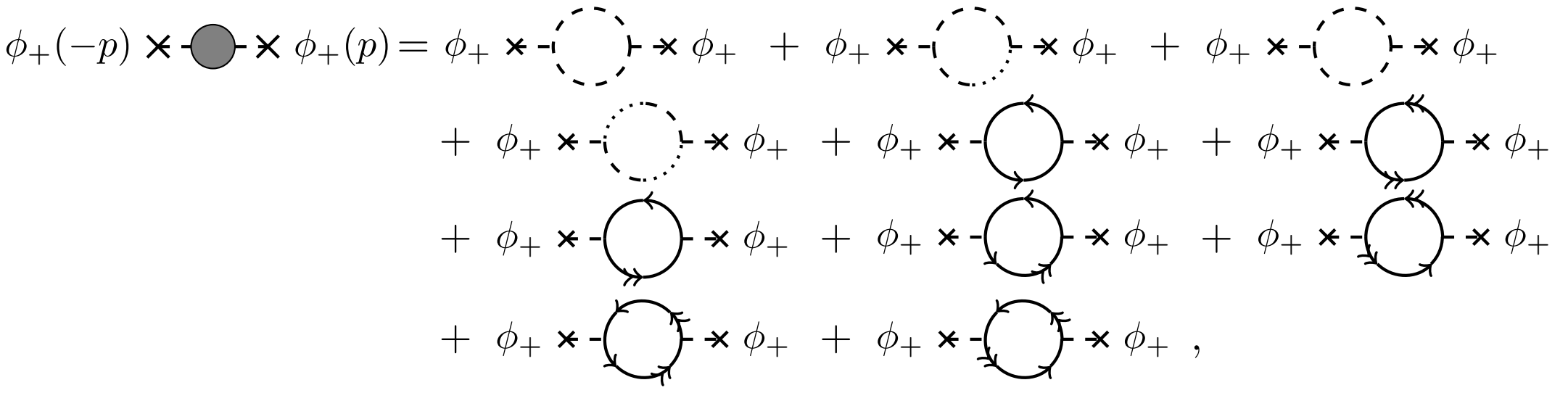

$$
= \phi_+(-p)\phi_+(p) \int \frac{d\hat{k}}{2\pi} \frac{1}{D(\hat{k})} \frac{1}{D(\hat{p}-\hat{k})} (1+M) \times \left( \frac{-16 g_3^2 a^4}{9} \right) \left( 8 + 6M - (3+M) C(\hat{p})^2 - 4C(\hat{k})^2 - KP \right), \tag{4.10a}
$$

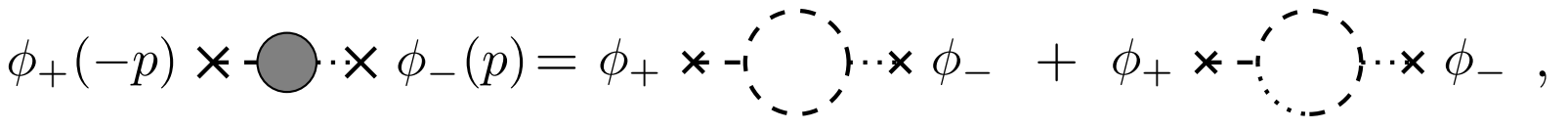

$$
= \phi_+(-p)\phi_-(p) C(\hat{p}) \int \frac{d\hat{k}}{2\pi} \frac{1}{D(\hat{k})} \frac{1}{D(\hat{p}-\hat{k})} (1+M) \times \left( \frac{16 g_3^2 a^4}{9} \right) \left( 8 + 6M - 2C(\hat{p})^2 - 4C(\hat{k})^2 - KP \right), \tag{4.10b}
$$

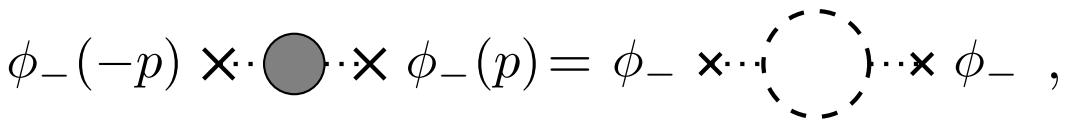

$$
= \phi_-(-p)\phi_-(p) C(\hat{p})^2 \int \frac{d\hat{k}}{2\pi} \frac{1}{D(\hat{k})} \frac{1}{D(\hat{p}-\hat{k})} \left( -\frac{16 g_3^2 a^4}{9} \right) (1+M)^2, \tag{4.10c}
$$

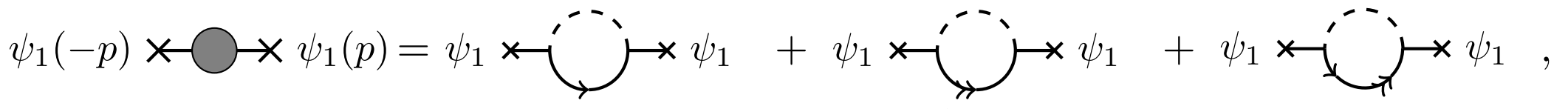

$$
= \psi_1(-p)\psi_1(p) \left( \frac{4 g_3^2 a^3}{9} \right) \int \frac{d\hat{k}}{2\pi} \frac{1}{D(\hat{k})} \frac{1}{D(\hat{p}-\hat{k})} (1+M) \times \left[ (4+3M)P - PC(\hat{p})^2 - 2PC(\hat{k})^2 - 2(1 - C(\hat{p})^2)K \right], \tag{4.10d}
$$

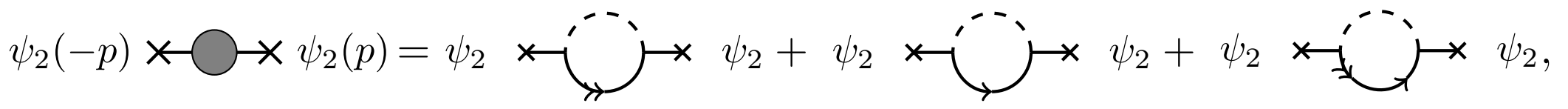

$$
= \psi_2(-p)\psi_2(p) \left( \frac{4 g_3^2 a^3}{9} \right) \int \frac{d\hat{k}}{2\pi} \frac{1}{D(\hat{k})} \frac{1}{D(\hat{p}-\hat{k})} (1+M) \times \left[ (4+3M)P - PC(\hat{p})^2 - 2PC(\hat{k})^2 - 2(1 - C(\hat{p})^2)K \right], \tag{4.10e}
$$

$\psi_1(-p)$ 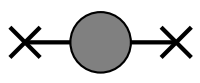 $\psi_2(p) = \psi_1$ 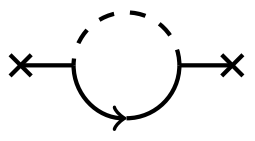 $\psi_2 + \psi_1$ 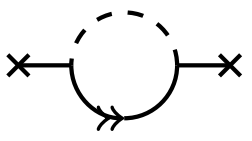 $\psi_2 + \psi_1$ 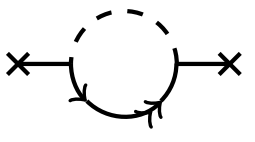 $\psi_2$

$+ \ \psi_1$ 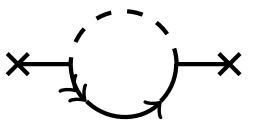 $\psi_2$ ,

$$= \psi_1(-p)\psi_2(p) \int \frac{d\hat{k}}{2\pi} \frac{1}{D(\hat{k})} \frac{1}{D(\hat{p}-\hat{k})} \left(-\frac{8ig_3^2 a^3}{9}\right)(1+M) \times \Big[8 + 6M - 4C(\hat{k})^2 - 2(2+M)C(\hat{p})^2 - KP\Big]. \tag{4.10f}$$

In these derivations of two point operators with 1-loop corrections, we have used the following non-trivial equalities:

$$\begin{aligned}
&\int_{-\frac{2}{a}}^{\frac{2}{a}} \frac{d\hat{k}_1}{2\pi} \frac{d\hat{k}_2}{2\pi} \frac{C^2(\hat{k}_2)}{D(\hat{k}_1)D(\hat{k}_2)} (2\pi)\delta(\hat{k}_1 + \hat{k}_2 - \hat{p}) \\
&= \int_{-\frac{2}{a}}^{\frac{2}{a}} \frac{d\hat{k}_1}{2\pi} \frac{d\hat{k}_2}{2\pi} \frac{C^2(\hat{k}_1)}{D(\hat{k}_1)D(\hat{k}_2)} (2\pi)\delta(\hat{k}_1 + \hat{k}_2 - \hat{p}) \\
&= \left[\Theta(\hat{p}) \int_{\hat{p}-\frac{2}{a}}^{\frac{2}{a}} \frac{d\hat{k}}{2\pi} + \Theta(-\hat{p}) \int_{-\frac{2}{a}}^{\hat{p}+\frac{2}{a}} \frac{d\hat{k}}{2\pi}\right] \frac{C^2(\hat{k})}{D(\hat{k})D(\hat{p}-\hat{k})} \\
&= \left[\Theta(\hat{p}) \int_{\hat{p}-\frac{2}{a}}^{\frac{2}{a}} \frac{d\hat{k}}{2\pi} + \Theta(-\hat{p}) \int_{-\frac{2}{a}}^{\hat{p}+\frac{2}{a}} \frac{d\hat{k}}{2\pi}\right] \frac{C^2(\hat{p}-\hat{k})}{D(\hat{k})D(\hat{p}-\hat{k})},
\end{aligned} \tag{4.11}$$

where $\Theta(\hat{p})$ is a step function. The first equality in the above relations can be understood by just exchanging the labels of integrated momenta; $\hat{k}_1 \leftrightarrow \hat{k}_2$. Then we can carry out the $\hat{k}_2$ integration where the integration range is limited:

$$-\frac{2}{a} \leq \hat{k}_2 = \hat{p} - \hat{k}_1 \leq \frac{2}{a}, \tag{4.12}$$

which leads:

$$\hat{p} - \frac{2}{a} \leq \hat{k}_1 \leq \hat{p} + \frac{2}{a}. \tag{4.13}$$

Since $\hat{k}_1$ has also the same integration range $-\frac{2}{a} \leq \hat{k}_1 \leq \frac{2}{a}$ the necessary changes of integration range are needed in the second and the third equalities. Thus due to the conservation of the lattice "sine" momentum in this new approach we need to take care of the integration range of the lattice momenta carefully. Explicit derivation of the equivalence of these relations are given in Appendix A.1.

We can now explicitly evaluate the 1-loop contribution for two point functions. For

example

$$
\begin{aligned}
&\langle\phi_+(p)\phi_+(-p)\rangle_{\text{1-loop}} = \\
&\underbracket{\phi_+(p)\phi_+}(-p) \times\!\!-\!\bigcirc\!-\!\!\times \underbracket{\phi_+(p)\phi_+}(-p) + \underbracket{\phi_+(p)\phi_+}(-p) \times\!\!-\!\bigcirc\!-\!\!\times \underbracket{\phi_-(p)\phi_+}(-p) \\
&+ \underbracket{\phi_+(p)\phi_-}(-p) \times\!\!-\!\bigcirc\!-\!\!\times \underbracket{\phi_-(p)\phi_+}(-p) \\
&= \left(-\frac{32g_3^2a^4}{9}\right)\frac{1}{D(\hat{p})^2}\int\frac{d\hat{k}}{2\pi}\frac{1}{D(\hat{k})}\frac{1}{D(\hat{p}-\hat{k})}(1+M)^2 \\
&\qquad \times\Big[(1+M)(8+6M)-(11+10M+M^2)C(\hat{p})^2+3C(\hat{p})^4 \\
&\qquad\qquad -4C(\hat{k})^2(1+M-C(\hat{p})^2)-KP(1+M-C(\hat{p})^2))\Big]\,,
\end{aligned}
\tag{4.14}
$$

where $\underbracket{\phi_+(p)\phi_+}(-p)$ represents Wick contraction and thus we need to take into account all the possible combinations for 1-loop two point function of $\phi_+$.

Similarly we can evaluate one-loop contribution of other two point functions as,

$$
\begin{aligned}
&\langle\phi_-(p)\phi_-(-p)\rangle_{\text{1-loop}} \\
&= \left(-\frac{32g_3^2a^4}{9}\right)\frac{1}{D(\hat{p})^2}\int\frac{d\hat{k}}{2\pi}\frac{1}{D(\hat{k})}\frac{1}{D(\hat{p}-\hat{k})}C(\hat{p})^2(1+M) \\
&\qquad \times\Big[(M^3+5M^2+7M+1)-(1+3M)C(\hat{p})^2-4MC(\hat{k})^2-MKP\Big]\,,
\end{aligned}
\tag{4.15a}
$$

$$
\begin{aligned}
&\langle\phi_+(p)\phi_-(-p)\rangle_{\text{1-loop}} \\
&= \left(-\frac{16g_3^2a^4}{9}\right)\frac{1}{D(\hat{p})^2}\int\frac{d\hat{k}}{2\pi}\frac{1}{D(\hat{k})}\frac{1}{D(\hat{p}-\hat{k})}C(\hat{p})(1+M) \\
&\qquad \times\big[(1+M)^2(8+6M)-2C(\hat{p})^2(3M^2+7M+5)+2C(\hat{p})^4 \\
&\qquad\qquad -4C(\hat{k})^2\left((1+M)^2-C(\hat{p})^2\right)-KP\left((1+M)^2-C(\hat{p})^2\right)\Big]\,,
\end{aligned}
\tag{4.15b}
$$

$$
\begin{aligned}
&\langle\psi_1(p)\psi_1(-p)\rangle_{\text{1-loop}} = \langle\psi_2(p)\psi_2(-p)\rangle_{\text{1-loop}} \\
&= \left(\frac{16g_3^2a^5}{9}\right)\frac{1}{D(\hat{p})^2}\int\frac{d\hat{k}}{2\pi}\frac{1}{D(\hat{k})}\frac{1}{D(\hat{p}-\hat{k})}(1+M)P \\
&\quad \times\Big[-4C(\hat{k})^2\left((1+M)^2-C(\hat{p})^2\right)-KP\left((1+M)^2-C(\hat{p})^2\right) \\
&\qquad\qquad +2C(\hat{p})^4-2C(\hat{p})^2(3M^2+7M+5)+2(1+M)^2(4+3M)\big]\,,
\end{aligned}
\tag{4.15c}
$$

$$
\begin{aligned}
&\langle\psi_1(p)\psi_2(-p)\rangle_{\text{1-loop}} \\
&= \left(-\frac{32ig_3^2a^5}{9}\right)\frac{1}{D(\hat{p})^2}\int\frac{d\hat{k}}{2\pi}\frac{1}{D(\hat{k})}\frac{1}{D(\hat{p}-\hat{k})}(1+M) \\
&\quad \times\Big[-4C(\hat{k})^2\left((1+M)-(1+2M)C(\hat{p})^2\right)-KP\left((1+M)^2-(1+2M)C(\hat{p})^2\right) \\
&\qquad +2(2+3M)C(\hat{p})^4-2C(\hat{p})^2\left(M^3+8M^2+14M+6\right)+2(1+M)^2(4+3M)\big]\,.
\end{aligned}
\tag{4.15d}
$$

To consider the 1-loop corrections for original two point functions, we need to consider linear combination of the above two point functions as follows:

$$\frac{4}{a^2}\langle\phi\phi\rangle = \langle\phi_+\phi_+\rangle + \langle\phi_-\phi_-\rangle + 2\langle\phi_+\phi_-\rangle\,, \tag{4.16a}$$
$$\frac{1}{4}\langle DD\rangle = \langle\phi_+\phi_+\rangle + \langle\phi_-\phi_-\rangle - 2\langle\phi_+\phi_-\rangle\,, \tag{4.16b}$$
$$\frac{1}{a}\langle D\phi\rangle = \frac{1}{a}\langle\phi D\rangle = \langle\phi_+\phi_+\rangle - \langle\phi_-\phi_-\rangle\,. \tag{4.16c}$$

We can then find the following non-trivial proportionality between 1-loop and tree level two point functions, where only two overall multiplicative factors appear for the same field and the different field two point functions:

$$\langle\phi(p)\phi(-p)\rangle_{\text{1-loop}} = \langle\phi(p)\phi(-p)\rangle_{\text{tree}}\,F(\hat{p}), \tag{4.17a}$$
$$\langle D(p)D(-p)\rangle_{\text{1-loop}} = \langle D(p)D(-p)\rangle_{\text{tree}}\,F(\hat{p}), \tag{4.17b}$$
$$\langle\psi_1(p)\psi_1(-p)\rangle_{\text{1-loop}} = \langle\psi_1(p)\psi_1(-p)\rangle_{\text{tree}}\,F(\hat{p}), \tag{4.17c}$$
$$\langle\psi_2(p)\psi_2(-p)\rangle_{\text{1-loop}} = \langle\psi_2(p)\psi_2(-p)\rangle_{\text{tree}}\,F(\hat{p}), \tag{4.17d}$$
$$\langle\phi(p)D(-p)\rangle_{\text{1-loop}} = \langle D(p)\phi(-p)\rangle_{\text{1-loop}} = (\phi(p)D(-p))_{\text{tree}}\,G(\hat{p}), \tag{4.17e}$$
$$\langle\psi_1(p)\psi_2(-p)\rangle_{\text{1-loop}} = -\langle\psi_2(p)\psi_1(-p)\rangle_{\text{1-loop}} = \langle\psi_1(p)\psi_2(-p)\rangle_{\text{tree}}\,G(\hat{p}), \tag{4.17f}$$

where

$$\begin{aligned}F(\hat{p}) = {} & \frac{1}{D(\hat{p})}\int\frac{d\hat{k}}{2\pi}\frac{1}{D(\hat{k})}\frac{1}{D(\hat{p}-\hat{k})}\left(-\frac{16g_3^2a^4}{9}\right)(1+M)\\ & \times\Big(-4C(\hat{k})^2\left((1+M)^2-C(\hat{p})^2\right)-KP\left((1+M)^2-C(\hat{p})^2\right)\\ & \qquad +2C(\hat{p})^4-2C(\hat{p})^2(3M^2+7M+5)+2(1+M)^2(4+3M)\Big)\,,\end{aligned} \tag{4.18}$$

$$\begin{aligned}G(\hat{p}) = {} & \frac{1}{D(\hat{p})}\int\frac{d\hat{k}}{2\pi}\frac{1}{D(\hat{k})}\frac{1}{D(\hat{p}-\hat{k})}\left(-\frac{16g_3^2a^4}{9}\right)\frac{1+M}{M}\\ & \times\Big(-4C(\hat{k})^2\left((1+M)^2-(1+2M)C(\hat{p})^2\right)-KP\left((1+M)^2-(1+2M)C(\hat{p})^2\right)\\ & \quad +2C(\hat{p})^4(2+3M)-2C(\hat{p})^2(M^3+8M^2+14M+6)+2(1+M)^2(4+3M)\Big)\,.\end{aligned} \tag{4.19}$$

We can now ask a question if the Ward-Takahashi identities of two point functions in (4.6-4.9) are satisfied at one loop level or not. We can show that identities of (4.6-4.9) at the 1-loop level are proportional to the tree level Ward-Takahashi identity with the multiplicative factor $F(\hat{p})$ or $G(\hat{p})$ given in (4.18) and (4.19):

$$\begin{aligned}
&\cos\frac{ap}{4}\left\langle\psi_1(p)\psi_1(-p)\right\rangle_{\text{1-loop}}+\frac{4}{a}\sin\frac{ap}{4}\left\langle\phi(p)\phi(-p)\right\rangle_{\text{1-loop}}\\
&=\left[\cos\frac{ap}{4}\left\langle\psi_1(p)\psi_1(-p)\right\rangle_{\text{tree}}+\frac{4}{a}\sin\frac{ap}{4}\left\langle\phi(p)\phi(-p)\right\rangle_{\text{tree}}\right]F(\hat{p})=0, \qquad (4.20)\\
&i\left\langle\psi_1(p)\psi_2(-p)\right\rangle_{\text{1-loop}}+\left\langle\phi(p)D(-p)\right\rangle_{\text{1-loop}}\\
&=\left[i\left\langle\psi_1(p)\psi_2(-p)\right\rangle_{\text{tree}}+\left\langle\phi(p)D(-p)\right\rangle_{\text{tree}}\right]G(\hat{p})=0, \qquad (4.21)\\
&\frac{4}{a}\sin\frac{ap}{4}\left\langle\psi_2(p)\psi_2(-p)\right\rangle_{\text{1-loop}}+\cos\frac{ap}{4}\left\langle D(p)D(-p)\right\rangle_{\text{1-loop}}\\
&=\left[\frac{4}{a}\sin\frac{ap}{4}\left\langle\psi_2(p)\psi_2(-p)\right\rangle_{\text{tree}}+\cos\frac{ap}{4}\left\langle D(p)D(-p)\right\rangle_{\text{tree}}\right]F(\hat{p})=0, \qquad (4.22)\\
&\left\langle\phi(p)\phi(-p)\right\rangle_{\text{1-loop}}-\left(\frac{a}{4}\cot\frac{ap}{4}\right)^2\left\langle D(p)D(-p)\right\rangle_{\text{1-loop}}\\
&=\left[\left\langle\phi(p)\phi(-p)\right\rangle_{\text{tree}}-\left(\frac{a}{4}\cot\frac{ap}{4}\right)^2\left\langle D(p)D(-p)\right\rangle_{\text{tree}}\right]F(\hat{p})=0. \qquad (4.23)
\end{aligned}$$

The proportionality of 1-loop level and tree level Ward-Takahashi identities ensures that 1-loop level Ward-Takahashi identities are satisfied since the tree level counterparts are satisfied as we have seen in (4.6-4.9).

### 4.2.2 1-loop corrections: quartic interaction $\Phi^4$

We next consider 1-loop contribution with quartic interactions for two point functions. Contracting two fields in a interaction term, we obtain the following two point operators:

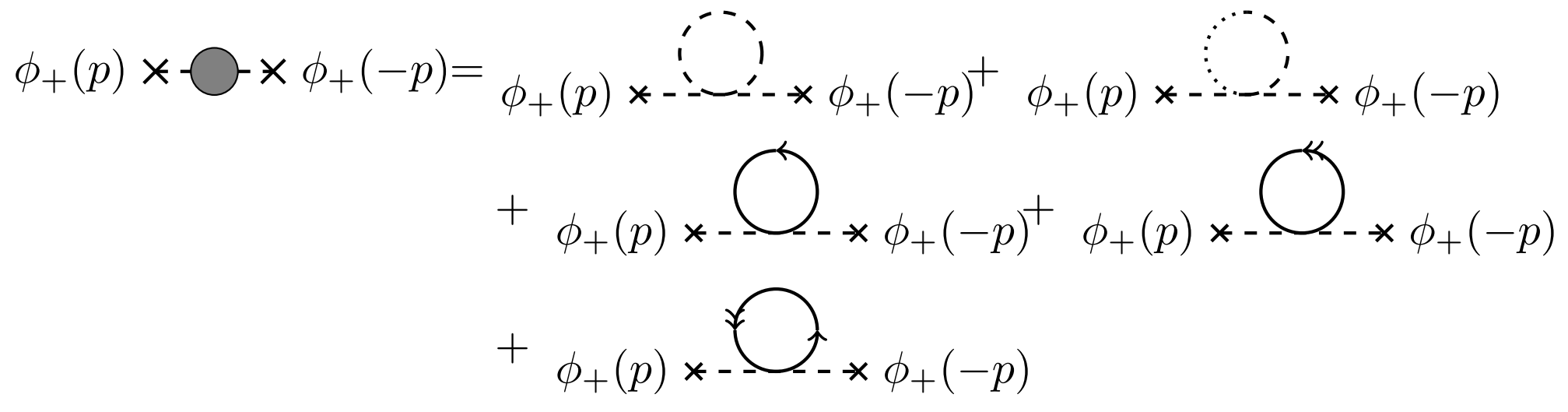

$$=\phi_+(p)\phi_+(-p)\left[-ig_4a^3(1+am)\int\frac{d\hat{k}}{2\pi}\frac{1}{D(\hat{k})}\right], \qquad (4.24a)$$

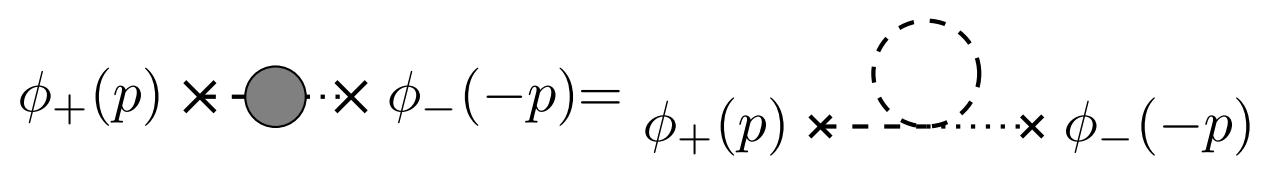

$$=\phi_+(p)\phi_-(-p)C(\hat{p})\left[ig_4a^3(1+am)\int\frac{d\hat{k}}{2\pi}\frac{1}{D(\hat{k})}\right], \qquad (4.24b)$$

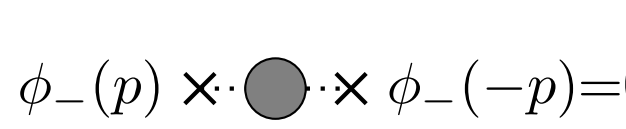

$=0,$ (4.24c)

$\psi_1(p)$ 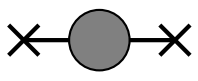 $\psi_1(-p)=$ $\psi_1(p)$ 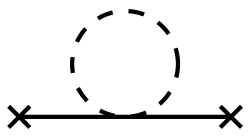 $\psi_1(-p)$

$$=\sin\frac{ap}{2}\psi_1(p)\psi_1(-p)\left[-\frac{ig_4}{4}a^2(1+am)\int\frac{d\hat{k}}{2\pi}\frac{1}{D(\hat{k})}\right], \qquad (4.24d)$$

$\psi_2(p)$ 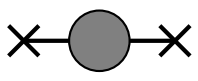 $\psi_2(-p)=\sin\frac{ap}{2}\psi_2(p)\psi_2(-p)\left[-\frac{ig_4}{4}a^2(1+am)\int\frac{d\hat{k}}{2\pi}\frac{1}{D(\hat{k})}\right], \qquad (4.24e)$

$\psi_1(p)$ 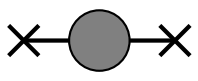 $\psi_2(-p)=$ $\psi_1(p)$ 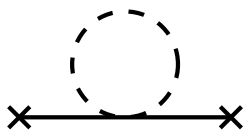 $\psi_2(-p)$

$$=\psi_1(p)\psi_2(-p)\left[\frac{g_4}{2}a^2(1+am)\int\frac{d\hat{k}}{2\pi}\frac{1}{D(\hat{k})}\right]. \qquad (4.24f)$$

We then contract these two point operators with other two external fields to get 1-loop corrections of two point functions as in (4.14). We then obtain two point functions with one-loop quantum corrections,

$$\langle\phi_+(p)\phi_+(-p)\rangle_{\text{1-loop}} = -2ig_4a^3\frac{1}{D(\hat{p})^2}\left[(1+am)^2-(1+am)C(\hat{p})^2\right]K, \qquad (4.25a)$$

$$\langle\phi_-(p)\phi_-(-p)\rangle_{\text{1-loop}} = -2ig_4a^3\frac{1}{D(\hat{p})^2}\left[C(\hat{p})^2(am)\right]K, \qquad (4.25b)$$

$$\langle\phi_+(p)\phi_-(-p)\rangle_{\text{1-loop}} = -ig_4a^3\frac{1}{D(\hat{p})^2}C(\hat{p})\left[(1+am)^2-C(\hat{p})^2\right]K, \qquad (4.25c)$$

$$\langle\psi_1(p)\psi_1(-p)\rangle_{\text{1-loop}} = ig_4a^4\frac{1}{D(\hat{p})^2}a\hat{p}\left[(1+am)^2-C(\hat{p})^2\right]K, \qquad (4.25d)$$

$$\langle\psi_2(p)\psi_2(-p)\rangle_{\text{1-loop}} = ig_4a^4\frac{1}{D(\hat{p})^2}a\hat{p}\left[(1+am)^2-C(\hat{p})^2\right]K, \qquad (4.25e)$$

$$\langle\psi_1(p)\psi_2(-p)\rangle_{\text{1-loop}} = 2g_4a^4\frac{1}{D(\hat{p})^2}\left[(1+am)^2-(1+2am)C(\hat{p})^2\right]K, \qquad (4.25f)$$

where $K$ is the following integral and can be evaluated explicitly as:

$$K=\int_{-\frac{2}{a}}^{\frac{2}{a}}\frac{d\hat{k}}{2\pi}\frac{1}{D(\hat{k})}(1+am)=\frac{1+am}{4a^2m\pi}\log\left|\frac{1-am}{1+am}\right|. \qquad (4.26)$$

Similar to the 1-loop contribution of two point functions for $\Phi^3$ in (4.17a-4.17f) we obtain the similar proportionality relations between the loop level and tree level two point functions for $\Phi^4$ theory:

$$\langle\phi(p)\phi(-p)\rangle_{\text{1-loop}} = \langle\phi(p)\phi(-p)\rangle_{\text{tree}}F_1(\hat{p}), \qquad (4.27a)$$

$$\langle D(p)D(-p)\rangle_{\text{1-loop}} = \langle D(p)D(-p)\rangle_{\text{tree}}F_1(\hat{p}), \qquad (4.27b)$$

$$\langle\psi_1(p)\psi_1(-p)\rangle_{\text{1-loop}} = \langle\psi_1(p)\psi_1(-p)\rangle_{\text{tree}}F_1(\hat{p}), \qquad (4.27c)$$

$$\langle\psi_2(p)\psi_2(-p)\rangle_{\text{1-loop}} = \langle\psi_2(p)\psi_2(-p)\rangle_{\text{tree}}F_1(\hat{p}), \qquad (4.27d)$$

$$\langle\phi(p)D(-p)\rangle_{\text{1-loop}} = \langle D(p)\phi(-p)\rangle_{\text{1-loop}} = \langle\phi(p)D(-p)\rangle_{\text{tree}}G_1(\hat{p}), \qquad (4.27e)$$

$$\langle\psi_1(p)\psi_2(-p)\rangle_{\text{1-loop}} = -\langle\psi_2(p)\psi_1(-p)\rangle_{\text{1-loop}} = \langle\psi_1(p)\psi_2(-p)\rangle_{\text{tree}}G_1(\hat{p}), \qquad (4.27f)$$

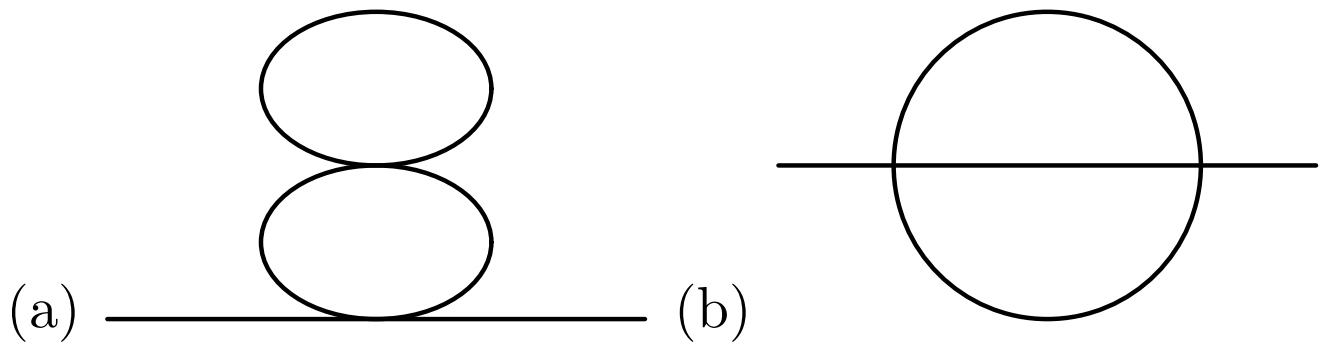

Figure 1: (a) Snowman-like diagram. (b) Sunset diagram

where

$$F_1(\hat{p}) = -ig_4a^3\frac{1}{D(\hat{p})}\left[(1+am)^2 - C(\hat{p})^2\right]K, \tag{4.28}$$

$$G_1(\hat{p}) = -\frac{ig_4a^4}{am}\frac{1}{D(\hat{p})}\left[(1+am)^2 - (1+2am)C(\hat{p})^2\right]K. \tag{4.29}$$

Since the two point functions of the same fields at 1-loop level and those of different fields are proportional to the tree level counterparts with the same corresponding multipicative factor, respectively, the Ward-Takahashi identities of (4.6 -4.9) for $\Phi^4$ theory are exactly satisfied at the 1-loop level as in the case of $\Phi^3$. In other words the relation between the 1-loop level and tree level Ward-Takahashi identities of $\Phi^4$ case are exactly the same as (4.20-4.23) of $\Phi^3$ case except for the replacement of the multiplicative factors $F_1(\hat{p})$ and $G_1(\hat{p})$ instead of $F(\hat{p})$ and $G(\hat{p})$. In the previous paper [16] the Ward-Takahashi identities for the two point function of $\Phi^4$ theory of (4.27a-4.27f) have already been investigated and shown to have the structure that 1-loop propagators are proportional to tree level propagators. Here we have confirmed that the Ward-Takahashi identities of two point functions for $\Phi^3$ theory have the same structure in the previous subsection.

It is interesting to recognize that it is a universal nature that Ward-Takahashi identities of two point functions with loop corrections are proportional to the tree level two point functions with the same multiplicative factor. In the following we investigate if this feature of the Ward-Takahashi identities is universal for the cases of higher loop corrections and 2-dimensional Wess-Zumino model.

### 4.3 2-loop corrections

We expect that the proportionality nature of the 2-loop contributions and tree level contributions for two point functions is universal like in the case of 1-loop corrections. Here we examine the simplest case of 2-loop contributions for quartic interaction shown in Fig. 1. To consider 2-loop corrections of the type of the diagram (a) in Fig.1, we can utilize the results of 1-loop correction given in the previous subsection 4.2.2. In the evaluation of 2-loop contributions for the type of diagram (b), we need to be careful for the integration range of lattice "sine" momenta similar to the case of 1-loop correction with cubic interaction which is discussed in Appendices A.1 and A.2. Here we do not treat 2-loop corrections with cubic interactions in this 1-dimensional example while we calculate them in 2-dimensional Wess-Zumiono model in the later section. We notice that 2-loop contributions of cubic interactions are higher order in the coupling constant compared with the corresponding counter parts of quartic interaction case.

### 4.3.1 Snowman diagrams: quartic interaction $\Phi^4$

Since the diagram (a) in Figure 1 includes 1-loop correction as a subdiagram, we can utilize the results in subsection 4.2.2. Contracting the fields of two point operators of 1-loop corrections with two fields from quartic interaction with another vertex yields the following two point operators with 2-loop corrections:

$\phi_+(q_1)$ 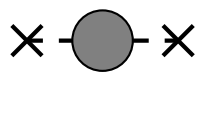 $\phi_+(q_2) = \phi_+(q_1)\phi_+(q_2)\left[-2g_4^2a^6\Pi_1\Pi_2\right],$ (4.30a)

$\phi_+(q_1)$ 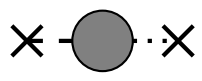 $\phi_-(q_2) = \phi_-(q_1)\phi_+(q_2)C(\hat{q}_1)\left[2g_4^2a^6\Pi_1\Pi_2\right],$ (4.30b)

$\psi_1(q_1)$ 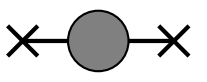 $\psi_1(q_2) = \psi_1(q_1)\psi_1(q_2)Sin(\hat{q}_1,\hat{q}_2)\left[-\frac{g_4^2a^5}{2}\Pi_1\Pi_2\right],$ (4.30c)

$\psi_2(q_1)$ 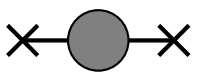 $\psi_2(q_2) = \psi_2(q_1)\psi_2(q_2)Sin(\hat{q}_1,\hat{q}_2)\left[-\frac{g_4^2a^5}{2}\Pi_1\Pi_2\right],$ (4.30d)

$\psi_1(q_1)$ 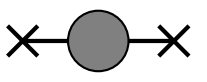 $\psi_2(q_2) = \psi_1(q_1)\psi_2(q_2)Cos(\hat{q}_1,\hat{q}_2)\left[-ig_4^2a^5\Pi_1\Pi_2\right].$ (4.30e)

where

$$\Pi_1 = \int_{-\frac{2}{a}}^{\frac{2}{a}} \frac{d\hat{k}}{2\pi}\frac{1+am}{D(\hat{k})} = \frac{1+am}{4ma^2\pi}\log\left|\frac{1-am}{1+am}\right|, \tag{4.31}$$

$$\begin{aligned}\Pi_2 &= \int_{-\frac{2}{a}}^{\frac{2}{a}} \frac{d\hat{k}'}{2\pi}\frac{1}{D(\hat{k}')^2}\left[(1+am)^2-(1+am)C(\hat{k}')^2\right] \\ &= \frac{(3am-1)(1+am)}{64\pi m^2a^3}\log\left|\frac{1-am}{1+am}\right| - \frac{1}{16\pi ma^2}\frac{1+am}{1-am},\end{aligned} \tag{4.32}$$

and

$$\begin{aligned}Sin(\hat{k}_1,\hat{k}_2) &= \sin\frac{a}{4}(k_1-k_2) \\ &= \sin\frac{ak_1}{4}\cos\frac{ak_2}{4} - \cos\frac{ak_1}{4}\sin\frac{ak_2}{4} \\ &= \sqrt{\frac{1}{4}(1-C(\hat{k}_1))(1+C(\hat{k}_2))}(\Theta(\hat{k}_1)-\Theta(-\hat{k}_1)) \\ &\qquad - \sqrt{\frac{1}{4}(1+C(\hat{k}_1))(1-C(\hat{k}_2))}(\Theta(\hat{k}_2)-\Theta(-\hat{k}_2)),\end{aligned} \tag{4.33}$$

$$\begin{aligned}Cos(\hat{k}_1,\hat{k}_2) &= \cos\frac{a}{4}(k_1+k_2) \\ &= \cos\frac{ak_1}{4}\cos\frac{ak_2}{4} - \sin\frac{ak_1}{4}\sin\frac{ak_2}{4} \\ &= \sqrt{\frac{1}{4}(1+C(\hat{k}_1))(1+C(\hat{k}_2))} \\ &\qquad - \sqrt{\frac{1}{4}(1-C(\hat{k}_1))(1-C(\hat{k}_2))}(\Theta(\hat{k}_1)-\Theta(-\hat{k}_1))(\Theta(\hat{k}_2)-\Theta(-\hat{k}_2)).\end{aligned} \tag{4.34}$$

Here $\Theta(\hat{k})$ is a step function. We then eventually obtain the following two point functions which are again proportional to the tree propagators:

$$\langle\phi(p)\phi(-p)\rangle_{\text{2-loop}} = \langle\phi(p)\phi(-p)\rangle_{\text{tree}}\, A(\hat{p}), \tag{4.35a}$$
$$\langle D(p)D(-p)\rangle_{\text{2-loop}} = \langle D(p)D(-p)\rangle_{\text{tree}}\, A(\hat{p}), \tag{4.35b}$$
$$\langle\psi_1(p)\psi_1(-p)\rangle_{\text{2-loop}} = \langle\psi_1(p)\psi_1(-p)\rangle_{\text{tree}}\, A(\hat{p}), \tag{4.35c}$$
$$\langle\psi_2(p)\psi_2(-p)\rangle_{\text{2-loop}} = \langle\psi_2(p)\psi_2(-p)\rangle_{\text{tree}}\, A(\hat{p}), \tag{4.35d}$$
$$\langle\phi(p)D(-p)\rangle_{\text{2-loop}} = \langle\phi(p)D(-p)\rangle_{\text{tree}}\, B(\hat{p}), \tag{4.35e}$$
$$\langle\psi_1(p)\psi_2(-p)\rangle_{\text{2-loop}} = \langle\psi_1(p)\psi_2(-p)\rangle_{\text{tree}}\, B(\hat{p}), \tag{4.35f}$$

where

$$A(\hat{p}) = 2g_4^2 a^6 \frac{1}{D(\hat{p})}\left((1+am)^2 - C(\hat{p})^2\right)\Pi_1\Pi_2, \tag{4.36}$$

$$B(\hat{p}) = \frac{2g_4^2 a^6}{am}\frac{1}{D(\hat{p})}\left((1+am)^2 - (1+2am)C(\hat{p})^2\right)\Pi_1\Pi_2. \tag{4.37}$$

As we expected that the 2-loop two point functions are proportional to the corresponding tree level 2-point functions with the common multiplicative factors for the same fields and the different fields of two point functions. They have exactly the same structure as the 1-loop contributions in (4.27a-4.27f) except for the replacements of the multiplicataive factors: $F_1(\hat{p}) \to A(\hat{p})$ and $G_1(\hat{p}) \to B(\hat{p})$. Thus we can prove Ward-Takahashi identity with 2-loop tadpoles just like the previous examples.

### 4.3.2 Sunset diagrams: quartic interaction $\Phi^4$

We evaluate the type of 2-loop diagram (b) of Figure 1 in this subsection. The following diagrams contribute to two point operators with 2-loop corrections for quartic interaction:

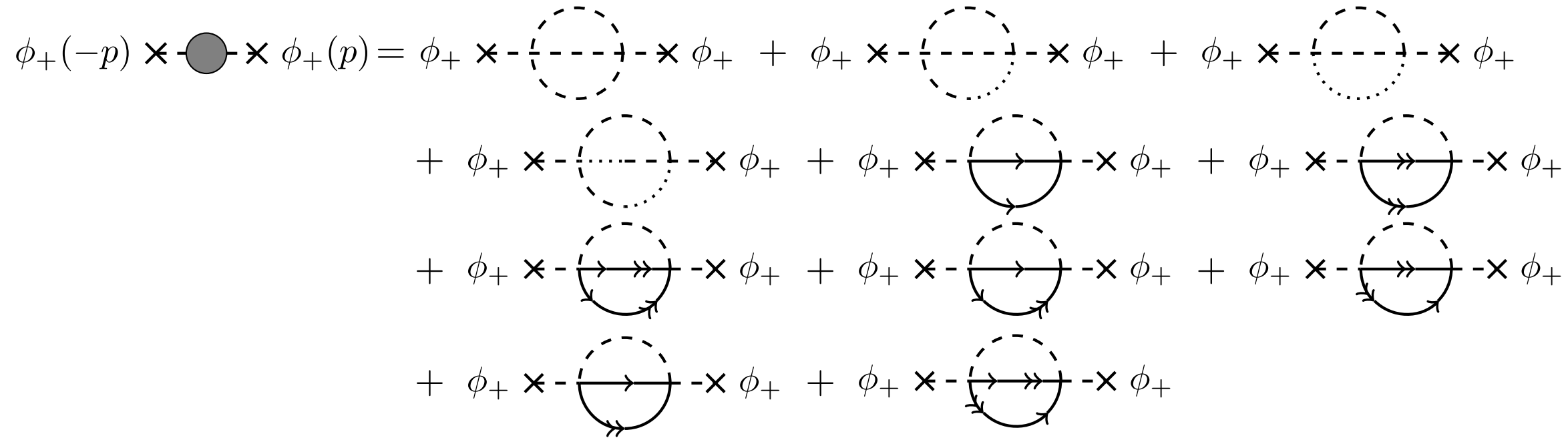

$$= \phi_+(-p)\phi_+(p)\left(\frac{g_4^2 a^6}{3}\right)\int \frac{d\hat{k}_1}{2\pi}\frac{d\hat{k}_2}{2\pi}\frac{(1+M)^2}{D(\hat{k}_1)D(\hat{k}_2)D(\hat{p}-\hat{k}_1-\hat{k}_2)}$$
$$\times\left[(3+M)C_p^2+4(C_1^2+C_2^2)-4(2M+3)+(P(K_1+K_2)-K_1K_2)\right], \tag{4.38a}$$

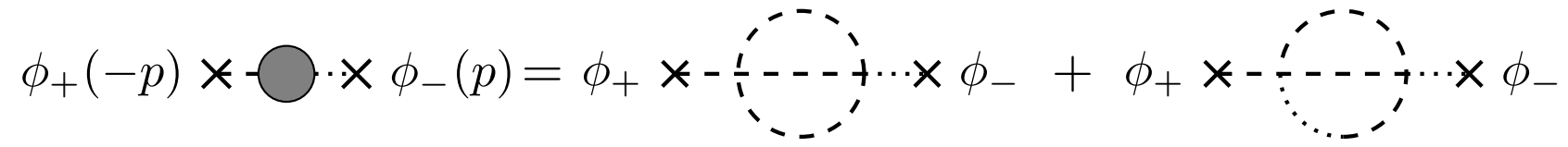

$$= \phi_+(-p)\phi_-(p)C(\hat{p})\left(-\frac{g_4^2 a^6}{3}\right)\int \frac{d\hat{k}_1}{2\pi}\frac{d\hat{k}_2}{2\pi}\frac{(1+M)^2}{D(\hat{k}_1)D(\hat{k}_2)D(\hat{p}-\hat{k}_1-\hat{k}_2)}$$
$$\times\left[2C_p^2+4(C_1^2+C_2^2)-4(2M+3)+(P(K_1+K_2)-K_1K_2)\right], \tag{4.38b}$$

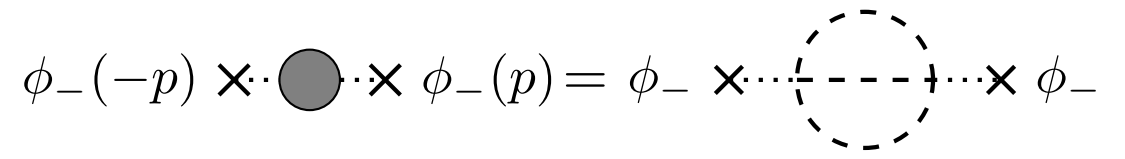

$$= \phi_-(-p)\phi_-(p)C(\hat{p})^2\left(-\frac{g_4^2 a^6}{3}\right)\int \frac{d\hat{k}_1}{2\pi}\frac{d\hat{k}_2}{2\pi}\frac{(1+M)^3}{D(\hat{k}_1)D(\hat{k}_2)D(\hat{p}-\hat{k}_1-\hat{k}_2)}, \tag{4.38c}$$

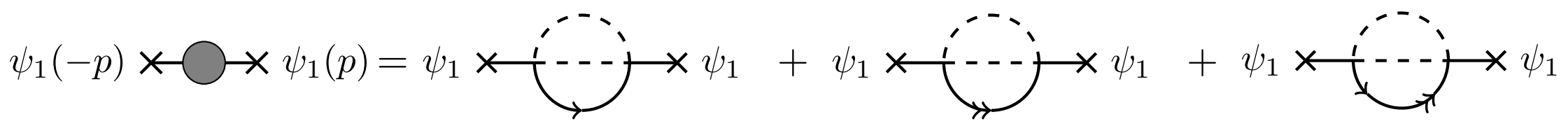

$$= \psi_1(-p)\psi_1(p)\int \frac{d\hat{k}_1}{2\pi}\frac{d\hat{k}_2}{2\pi}\frac{(1+M)^2}{D(\hat{k}_1)D(\hat{k}_2)D(\hat{p}-\hat{k}_1-\hat{k}_2)}\left(\frac{g_4^2a^5}{12}\right)P \times\left[2(3+2M)-C_p^2-2(C_1^2+C_2^2)-\frac{1}{2}(P(K_1+K_2)-K_1K_2)\right], \tag{4.38d}$$

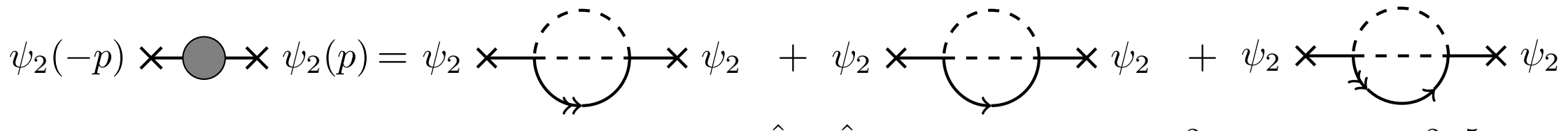

$$= \psi_2(-p)\psi_2(p)\int \frac{d\hat{k}_1}{2\pi}\frac{d\hat{k}_2}{2\pi}\frac{(1+M)^2}{D(\hat{k}_1)D(\hat{k}_2)D(\hat{p}-\hat{k}_1-\hat{k}_2)}\left(\frac{g_4^2a^5}{12}\right)P \times\left[2(3+2M)-C_p^2-2(C_1^2+C_2^2)-\frac{1}{2}(P(K_1+K_2)-K_1K_2)\right], \tag{4.38e}$$

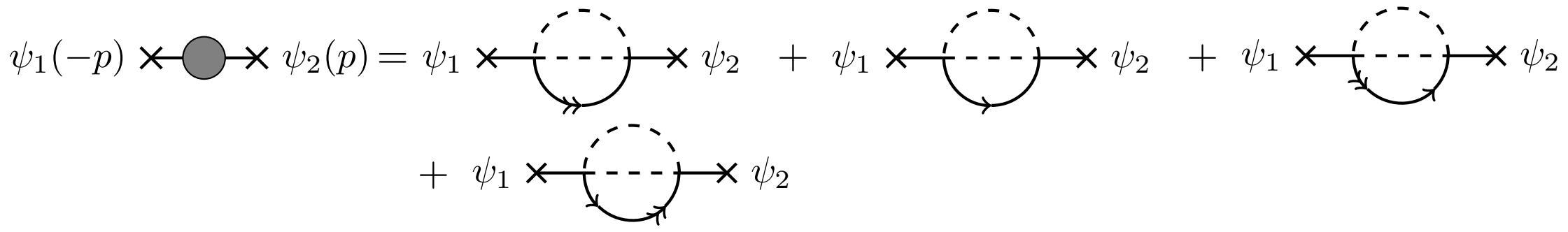

$$= \psi_1(-p)\psi_2(p)\int \frac{d\hat{k}_1}{2\pi}\frac{d\hat{k}_2}{2\pi}\frac{(1+M)^2}{D(\hat{k}_1)D(\hat{k}_2)D(\hat{p}-\hat{k}_1-\hat{k}_2)}\left(-\frac{ig_4^2a^5}{6}\right) \times\left[4(3+2M)-2(2+M)C_p^2-4(C_1^2+C_2^2)-(P(K_1+K_2)-K_1K_2)\right], \tag{4.38f}$$

where $C_p$ and $C_i$ denote $C(\hat{p})$ and $C(\hat{k}_i)$ respectively. In the integrand of fermionic two point functions we have utilized the following relations:

$$
\begin{aligned}
&\int \prod_{i=1}^{3} \frac{d\hat{k}_i}{2\pi} \frac{C^2(\hat{k}_1)}{D(\hat{k}_1)D(\hat{k}_2)D(\hat{k}_3)} (2\pi)\delta(\hat{k}_1+\hat{k}_2+\hat{k}_3-\hat{p}) \\
&= \int \prod_{i=1}^{3} \frac{d\hat{k}_i}{2\pi} \frac{C^2(\hat{k}_2)}{D(\hat{k}_1)D(\hat{k}_2)D(\hat{k}_3)} (2\pi)\delta(\hat{k}_1+\hat{k}_2+\hat{k}_3-\hat{p}) \\
&= \int \prod_{i=1}^{3} \frac{d\hat{k}_i}{2\pi} \frac{C^2(\hat{k}_3)}{D(\hat{k}_1)D(\hat{k}_2)D(\hat{k}_3)} (2\pi)\delta(\hat{k}_1+\hat{k}_2+\hat{k}_3-\hat{p}) \\
&= \left[\int_{-\frac{2}{a}}^{\hat{p}} \frac{d\hat{k}_1}{2\pi} \int_{\hat{p}-\hat{k}_1-\frac{2}{a}}^{\frac{2}{a}} \frac{d\hat{k}_2}{2\pi} + \int_{\hat{p}}^{\frac{2}{a}} \frac{d\hat{k}_1}{2\pi} \int_{-\frac{2}{a}}^{\hat{p}-\hat{k}_1+\frac{2}{a}} \frac{d\hat{k}_2}{2\pi}\right] \frac{C^2(\hat{k}_1)}{D(\hat{k}_1)D(\hat{k}_2)D(\hat{p}-\hat{k}_1-\hat{k}_2)}, \\
&= \left[\int_{-\frac{2}{a}}^{\hat{p}} \frac{d\hat{k}_1}{2\pi} \int_{\hat{p}-\hat{k}_1-\frac{2}{a}}^{\frac{2}{a}} \frac{d\hat{k}_2}{2\pi} + \int_{\hat{p}}^{\frac{2}{a}} \frac{d\hat{k}_1}{2\pi} \int_{-\frac{2}{a}}^{\hat{p}-\hat{k}_1+\frac{2}{a}} \frac{d\hat{k}_2}{2\pi}\right] \frac{C^2(\hat{k}_2)}{D(\hat{k}_1)D(\hat{k}_2)D(\hat{p}-\hat{k}_1-\hat{k}_2)}, \\
&= \left[\int_{-\frac{2}{a}}^{\hat{p}} \frac{d\hat{k}_1}{2\pi} \int_{\hat{p}-\hat{k}_1-\frac{2}{a}}^{\frac{2}{a}} \frac{d\hat{k}_2}{2\pi} + \int_{\hat{p}}^{\frac{2}{a}} \frac{d\hat{k}_1}{2\pi} \int_{-\frac{2}{a}}^{\hat{p}-\hat{k}_1+\frac{2}{a}} \frac{d\hat{k}_2}{2\pi}\right] \frac{C^2(\hat{p}-\hat{k}_1-\hat{k}_2)}{D(\hat{k}_1)D(\hat{k}_2)D(\hat{p}-\hat{k}_1-\hat{k}_2)}, \quad (4.39)
\end{aligned}
$$

where the equivalence of the first three relations can be understood by the relabeling of the integration variables $\hat{k}_i$. The next three relations can be obtained by integrating $\hat{k}_3$ which has the following limited integration range:

$$
-\frac{2}{a} \leq \hat{k}_3 = \hat{p} - \hat{k}_1 - \hat{k}_2 \leq \frac{2}{a}, \tag{4.40}
$$

and thus

$$
\hat{p} - \frac{2}{a} - \hat{k}_1 \leq \hat{k}_2 \leq \hat{p} + \frac{2}{a} - \hat{k}_1. \tag{4.41}
$$

The details of the equivalence of these relations are explained in Appendix A.2. Due to the lattice "sine" momentum conservation, we must treat the multiple integration range carefully.

Carrying out the similar Wick contraction as (4.14), we can obtain 2-loop contributions which are again proportional to the tree level propagators with common multiplicative factors for the same fields and different fields of two point functions:

$$
\begin{aligned}
\langle \phi(-p)\phi(p)\rangle_{\text{2-loop}} &= \langle \phi(-p)\phi(p)\rangle_{\text{tree}}\, I(\hat{p}), && (4.42a)\\
\langle D(-p)D(p)\rangle_{\text{2-loop}} &= \langle D(-p)D(p)\rangle_{\text{tree}}\, I(\hat{p}), && (4.42b)\\
\langle \psi_1(-p)\psi_1(p)\rangle_{\text{2-loop}} &= \langle \psi_1(-p)\psi_1(p)\rangle_{\text{tree}}\, I(\hat{p}), && (4.42c)\\
\langle \psi_2(-p)\psi_2(p)\rangle_{\text{2-loop}} &= \langle \psi_2(-p)\psi_2(p)\rangle_{\text{tree}}\, I(\hat{p}), && (4.42d)\\
\langle \phi(-p)D(p)\rangle_{\text{2-loop}} &= \langle \phi(-p)D(p)\rangle_{\text{tree}}\, J(\hat{p}), && (4.42e)\\
\langle \psi_1(-p)\psi_2(p)\rangle_{\text{2-loop}} &= \langle \psi_1(-p)\psi_2(p)\rangle_{\text{tree}}\, J(\hat{p}), && (4.42f)
\end{aligned}
$$

where,

$$I(\hat{p}) = \frac{1}{D(\hat{p})} \int \frac{d\hat{k}_1}{2\pi}\frac{d\hat{k}_2}{2\pi}\frac{1}{D(\hat{k}_1)D(\hat{k}_2)D(\hat{p}-\hat{k}_1-\hat{k}_2)}\left(\frac{g_4^2 a^6}{3}\right)(1+M)^2$$
$$\times\left[-2C_p^4 + 2C_p^2(3M^2+8M+7) - 4(1+M)^2(2M+3)\right.$$
$$\left.+\left(4(C_1^2+C_2^2)+P(K_1+K_2)-K_1K_2\right)\left((1+M)^2-C_p^2\right)\right], \qquad (4.43)$$

$$J(\hat{p}) = \frac{1}{D(\hat{p})} \int \frac{d\hat{k}_1}{2\pi}\frac{d\hat{k}_2}{2\pi}\frac{1}{D(\hat{k}_1)D(\hat{k}_2)D(\hat{p}-\hat{k}_1-\hat{k}_2)}\left(\frac{g_4^2 a^6}{3}\right)\frac{(1+M)^2}{M}$$
$$\times\left[-2(2+3M)C_p^4 + 2C_p^2(M^3+10M^2+19M+8) - 4(1+M)^2(2M+3)\right.$$
$$\left.+\left(4(C_1^2+C_2^2)+P(K_1+K_2)-K_1K_2\right)\left((1+M)^2-(1+2M)C_p^2\right)\right]. \qquad (4.44)$$

Here we again find the same structure of the proportionality of 2-loop level and the tree level 2-point functions with common multiplicative factors similar as 1-loop contributions (4.17a-4.17f ). Since one can find a diagrammatic similarity of loop diagrams between the 1-loop contributions of $\Phi^3$ theory and 2-loop contributions of $\Phi^4$ theory, this fact can be expected. It is also natural to expect that 2-loop contributions of loop diagrams in Fig.1(a) have the similar structure with the same reason. Indeed we have found the same structures in (4.35a-4.35f) with different multiplicative factors. The 2-loop Ward-Takahashi identities are thus satisfied just like 1-loop case. In these derivation of the common multiplicative factors we have to take into account the integration range of the multiple integrations carefully so that non trivial equalities appear due to the limited range of the "sine" momentum conservation. In addition, to properly extract out free propagators, it is also necessary to take into account all the possible equivalent relabeling of integrated loop momenta.

In this section we have focused on $N = 2$ Wess-Zumino model in 1-dimension. We have shown that two point functions with quantum loop corrections are proportional to the corresponding tree level two point functions, and thus Ward-Takahashi identities of two point functions with loop corrections are also proportional to the tree level two point functions. We have thus shown explicitly that Ward-Takahashi identities of lattice supersymmetry with quantum loop corrections are satisfied exactly up to the 2-loop levels for lattice 1-dimentional Wess-Zumino model.

# 5 Super doubler approach for $N = 2$ Wess-Zumino model in 2-dimensions

## 5.1 $N = 2, D = 2$ supertransformation

2-dimensional $N = 2$ Euclidean algebra is given by,

$$\{Q_{\alpha i}, Q_{\beta j}\} = 2\delta_{ij}\gamma^{\mu}_{\alpha\beta} i\partial_{\mu}, \qquad (5.1)$$

where $\gamma^{\mu} = (\sigma_3, \sigma_1)$ and $(i, j) = (1, 2)$. This algebra can be rearranged as

$$\{Q^{(i)}_{\pm}, Q^{(j)}_{\pm}\} = 2\delta_{ij} i\partial_{\pm}, \qquad \{Q^{(i)}_{\pm}, Q^{(j)}_{\mp}\} = 0, \qquad (5.2)$$

by taking linear combinations of supercharges and derivatives,

$$Q^{(j)}_{\pm} = \frac{1}{2}(Q_{1j} \pm iQ_{2j}), \qquad \partial_{\pm} = \frac{1}{2}(\partial_1 \pm i\partial_2). \qquad (5.3)$$

| | $Q_+^{(+)}$ | $Q_+^{(-)}$ | $Q_-^{(+)}$ | $Q_-^{(-)}$ |
|---|---|---|---|---|
| $\mathbf{\Phi}(p)$ | $i\mathbf{\Psi}_1(p)$ | 0 | $i\mathbf{\Psi}_2(p)$ | 0 |
| $\mathbf{\Psi}_1(p)$ | 0 | $-2i\sin\frac{ap_+}{2}\mathbf{\Phi}(p)$ | $-\mathbf{F}(p)$ | 0 |
| $\mathbf{\Psi}_2(p)$ | $\mathbf{F}(p)$ | 0 | 0 | $-2i\sin\frac{ap_-}{2}\mathbf{\Phi}(p)$ |
| $\mathbf{F}(p)$ | 0 | $2\sin\frac{ap_+}{2}\mathbf{\Psi}_2(p)$ | 0 | $-2\sin\frac{ap_-}{2}\mathbf{\Psi}_1(p)$ |

Table 1: Chiral $D = N = 2$ supersymmetry transformation

| | $Q_+^{(+)}$ | $Q_+^{(-)}$ | $Q_-^{(+)}$ | $Q_-^{(-)}$ |
|---|---|---|---|---|
| $\bar{\mathbf{\Phi}}(p)$ | 0 | $i\bar{\mathbf{\Psi}}_1(p)$ | 0 | $i\bar{\mathbf{\Psi}}_2(p)$ |
| $\bar{\mathbf{\Psi}}_1(p)$ | $-2i\sin\frac{ap_+}{2}\bar{\mathbf{\Phi}}(p)$ | 0 | 0 | $-\bar{\mathbf{F}}(p)$ |
| $\bar{\mathbf{\Psi}}_2(p)$ | 0 | $\bar{\mathbf{F}}(p)$ | $-2i\sin\frac{ap_-}{2}\bar{\mathbf{\Phi}}(p)$ | 0 |
| $\bar{\mathbf{F}}(p)$ | $2\sin\frac{ap_+}{2}\bar{\mathbf{\Psi}}_2(p)$ | 0 | $-2\sin\frac{ap_-}{2}\bar{\mathbf{\Psi}}_1(p)$ | 0 |

Table 2: Anti-chiral $D = N = 2$ supersymmetry transformation

As we can see from the algebra, 2-dimensional $N = 2$ superalgebra is decomposed into the direct sum of two one-dimensional $N = 2$ algebra. Thus we can use the one-dimensional formulation to each light-cone direction to construct a two-dimensional model. We now consider the half-lattice structure in each light-cone direction. To make each coordinate to be real ($x_\pm = x_1 \pm x_2$), we have to go from Euclidean space to Minkowski space.

We can equivalently express the above algebra in the following chiral form:

$$\{Q_+^{(+)}, Q_+^{(-)}\} = i\partial_+, \quad \{Q_-^{(+)}, Q_-^{(-)}\} = i\partial_-, \quad \{\text{others}\} = 0, \tag{5.4}$$

where

$$Q_\pm^{(+)} = \frac{Q_\pm^{(1)} + iQ_\pm^{(2)}}{2}, \quad Q_\pm^{(-)} = \frac{Q_\pm^{(1)} - iQ_\pm^{(2)}}{2}. \tag{5.5}$$

Replacing the derivative operators into symmetric difference operator of single lattice distance, we can rewrite momentum representation of the algebra as:

$$\left\{Q_\pm^{(+)}, Q_\pm^{(-)}\right\} = 2\sin\frac{ap_\pm}{2}, \qquad \{\text{others}\} = 0, \tag{5.6}$$

with all other anticommutators vanishing. The supersymmetry transformations for chiral and anti-chiral superfields are then given respectively in Table 1 and 2. In Table 2 the components of the anti-chiral superfield are denoted with over line to distinguish them from the chiral counterparts.

We then construct 2-dimensional $N = 2$ Wess-Zumino action. Kinetic term is obtained by $Q$-exact form of all supercharges on bi-linear product of anti-chiral and chiral field:

$$
\begin{aligned}
S_k &= \frac{a^4}{4^2}\int_{-\frac{\pi}{a}}^{\frac{3\pi}{a}} \frac{d^2\hat{p}}{(2\pi)^2}\frac{d^2\hat{q}}{(2\pi)^2}(2\pi)^2\delta^2\left(\sin\frac{aq}{2}+\sin\frac{ap}{2}\right) Q_+^{(-)}Q_-^{(-)}Q_+^{(+)}Q_-^{(+)}\left[\overline{\mathbf{\Phi}}(q)\mathbf{\Phi}(p)\right] \\
&= \frac{a^4}{4^2}\int_{-\frac{\pi}{a}}^{\frac{3\pi}{a}} \frac{d^2\hat{p}}{(2\pi)^2}\frac{d^2\hat{q}}{(2\pi)^2}(2\pi)^2\delta^2\left(\sin\frac{aq}{2}+\sin\frac{ap}{2}\right) \\
&\qquad \times\Big[-4\overline{\mathbf{\Phi}}(q)\sin\frac{ap_+}{2}\sin\frac{ap_-}{2}\mathbf{\Phi}(p)-\overline{\mathbf{F}}(q)\mathbf{F}(p) \\
&\qquad\qquad +2\overline{\mathbf{\Psi}}_1(q)\sin\frac{aq_-}{2}\mathbf{\Psi}_1(p)+2\overline{\mathbf{\Psi}}_2(q)\sin\frac{aq_+}{2}\mathbf{\Psi}_2(p)\Big], \qquad (5.7)
\end{aligned}
$$

where $d^2\hat{p}=d\hat{p}_+d\hat{p}_-$ and $\delta^2\left(\sin\frac{aq}{2}+\sin\frac{ap}{2}\right)=\prod_{i=\pm}\delta\left(\sin\frac{aq_i}{2}+\sin\frac{ap_i}{2}\right)$. The supersymmetry invariance is manifest due to $Q$-exact form and "sine" momentum conservation. Here we impose that anti-chiral fields are hermitian conjugate of the chiral ones:

$$\mathbf{\Phi}_A^\dagger(p)=\overline{\mathbf{\Phi}}_A(-p). \qquad (5.8)$$

Chiral and anti-chiral fields are both symmetric or both anti-symmetric under interchange of $p_\pm\to\frac{2\pi}{a}-p_\pm$ and have the periodicity with the period $\frac{4\pi}{a}$.

When only two momenta are involved in "sine" momentum conservation, the equation, $\sin\frac{ap}{2}+\sin\frac{aq}{2}=0$, generally has two solutions and then it can be reduced to the following standard local form:

$$\delta^2\left(\sin\frac{aq}{2}+\sin\frac{ap}{2}\right)=\left(\frac{2}{a}\right)^2\prod_{i=\pm}\left[\frac{1}{\left|\cos\frac{aq_i}{2}\right|_{q_i=-p_i}}\delta(q_i+p_i)+\frac{1}{\left|\cos\frac{aq_i}{2}\right|_{q_i=p_i+\frac{2\pi}{a}}}\delta\left(q_i-p_i-\frac{2\pi}{a}\right)\right]. \qquad (5.9)$$

For the kinetic bi-linear terms these two solutions of "sine" conservation lead to the same terms since the chiral/anti-chiral super fields have antisymmetric properties under the interchange of momentum $p\to\frac{2\pi}{a}-p$ and $\frac{4\pi}{a}$ periodicity. The dimensionless fields $\mathbf{\Phi}(p),\mathbf{F}(p)$ and $\mathbf{\Psi}_i(p)$ can be rescaled into dimensional component fields in terms of lattice constant $a$:

$$\mathbf{\Phi}(p)\to a^{-2}\varphi(p), \qquad \mathbf{\Psi}_i(p)\to a^{-\frac{3}{2}}\psi_i(p), \qquad \mathbf{F}(p)\to a^{-1}f(p). \qquad (5.10)$$

Here it should be noted that there is a difference of the treatment for species doublers between the 1-dimensional and the 2-dimensional super doubler approaches. In 1-dimensional approach $\Phi(p)$ is a composite field to include both scalar and auxiliary fields: $\phi(p)$ and $D(p)$. On the other hand in 2-dimensional approach each of fields $\mathbf{\Phi}(p)$ and $\mathbf{F}(p)$ has 4 species doublers which can be truncated by chiral conditions. See the details in [16, 17]. Even though the species doublers are truncated by chiral conditions, we still call this 2-dimensional lattice supersymmetry formulation as SUPER DOUBLER APPROACH[17].

The kinetic term in momentum representation can be written as

$$S_k=\int_{-\frac{\pi}{a}}^{\frac{\pi}{a}}\frac{d^2\hat{p}}{(2\pi)^2}\left[-\overline{\varphi}(-p)\hat{p}_+\hat{p}_-\varphi(p)-\overline{f}(-p)f(p)+\overline{\psi}_1(-p)\hat{p}_-\psi_1(p)+\overline{\psi}_2(-p)\hat{p}_+\psi_2(p)\right], \qquad (5.11)$$

where $\hat{p}_\pm = \frac{2}{a}\sin\frac{ap_\pm}{2}$.

Interaction terms can also be obtained by $Q$-exact form of supercharges on the products of chiral (anti-chiral) fields $\mathbf{\Phi}$ ($\overline{\mathbf{\Phi}}$),

$$S^n_{int} = g^0_n a^{2n} \int_{-\frac{\pi}{a}}^{\frac{\pi}{a}} \prod_{i=1}^{n} \left[\frac{d^2p_i}{(2\pi)^2}\cos\frac{ap_{i+}}{2}\cos\frac{ap_{i-}}{2}\right](2\pi)^2\delta^2\left(\sum_{i=1}^{n}\sin\frac{ap_i}{2}\right)$$
$$\times\frac{1}{n}\left[Q^{(+)}_+Q^{(+)}_-\left(\mathbf{\Phi}(p_1)\mathbf{\Phi}(p_2)\cdots\mathbf{\Phi}(p_n)\right)+h.c.\right], \tag{5.12}$$

where $g^0_n$ is dimensionless coupling and $h.c.$ denotes hermitian conjugate of chiral combinations. Integration range is already reduced using $p \to \frac{2\pi}{a} - p$ symmetry. The interaction term has manifest supersymmetry invariance since it has Q-exact form. Choosing $n = 2$ the interaction term yields a mass term:

$$S_m = g^0_2 a^4 \int \frac{d^2\hat{p}_1}{(2\pi)^2}\frac{d^2\hat{p}_2}{(2\pi)^2}(2\pi)^2\delta^2\left(\sin\frac{ap_1}{2}+\sin\frac{ap_2}{2}\right)$$
$$\times\left[i\mathbf{F}(p_1)\mathbf{\Phi}(p_2)+\mathbf{\Psi}_2(p_1)\mathbf{\Psi}_1(p_2)+h.c.\right]. \tag{5.13}$$

After rescaling they lead:

$$S_m = m\int\frac{d^2\hat{p}}{(2\pi)^2}\left[if(p)\varphi(p)+\psi_2(-p)\psi_1(p)-i\overline{\varphi}(-p)\overline{f}(p)+\overline{\psi}_1(-p)\overline{\psi}_2(p)\right], \tag{5.14}$$

where we have defined $m \equiv \frac{4g^0_2}{a}$. Similarly we can expand $n \geq 3$ interaction terms by rescaled component fields as:

$$S^n_{int} = g_n \int_{-\frac{\pi}{a}}^{\frac{\pi}{a}} \prod_{i=1}^{n} \left[\frac{d^2p_i}{(2\pi)^2}\cos\frac{ap_{i+}}{2}\cos\frac{ap_{i-}}{2}\right](2\pi)^2\delta^2\left(\frac{2}{a}\sum_{i=1}^{n}\sin\frac{ap_i}{2}\right)$$
$$\times\left[if(p_1)\varphi(p_2)\cdots\varphi(p_n)+(n-1)\psi_2(p_1)\psi_1(p_2)\cdots\varphi(p_n)+h.c.\right], \tag{5.15}$$

where $g_n \equiv \frac{2^{2n}g^0_n}{a}$. We focus on the cubic and quartic interactions for perturbative loop calculations in the following sections.

Explicitly we use the following cubic and quartic interactions:

$$S^{n=3}_{int} = g_3 \int \prod_{i=1}^{3}\left[\frac{d^2p_i}{(2\pi)^2}\cos\frac{ap_{i+}}{2}\cos\frac{ap_{i-}}{2}\right](2\pi)^2\delta^2\left(\sum_{i=1}^{3}\hat{p}_i\right)$$
$$\times\left[if(p_1)\varphi(p_2)\varphi(p_3)-i\overline{f}(p_1)\overline{\varphi}(p_2)\overline{\varphi}(p_3)\right.$$
$$\left.+2\psi_2(p_1)\psi_1(p_2)\varphi(p_3)-2\overline{\psi}_2(p_1)\overline{\psi}_1(p_2)\overline{\varphi}(p_3)\right], \tag{5.16}$$

$$S^{n=4}_{int} = g_4 \int \prod_{i=1}^{4}\left[\frac{d^2p_i}{(2\pi)^2}\cos\frac{ap_{i+}}{2}\cos\frac{ap_{i-}}{2}\right](2\pi)^2\delta^2\left(\sum_{i=1}^{4}\hat{p}_i\right)$$
$$\times\left[if(p_1)\varphi(p_2)\varphi(p_3)\varphi(p_4)-i\overline{f}(p_1)\overline{\varphi}(p_2)\overline{\varphi}(p_3)\overline{\varphi}(p_4)\right.$$
$$\left.+3\psi_2(p_1)\psi_1(p_2)\varphi(p_3)\varphi(p_4)-3\overline{\psi}_2(p_1)\overline{\psi}_1(p_2)\overline{\varphi}(p_3)\overline{\varphi}(p_4)\right]. \tag{5.17}$$

Supertransformation of component fields are shown in Table 3 and Table 4. It is also necessary to rescale supercharges to recover correct canonical dimension: $Q^{(j)}_i \to a^{\frac{1}{2}}Q^{(j)}_i$.

| | $Q_+^{(+)}$ | $Q_+^{(-)}$ | $Q_-^{(+)}$ | $Q_-^{(-)}$ |
|---|---|---|---|---|
| $\varphi(p)$ | $i\psi_1(p)$ | $0$ | $i\psi_2(p)$ | $0$ |
| $\psi_1(p)$ | $0$ | $-i\hat{p}_+\varphi(p)$ | $-f(p)$ | $0$ |
| $\psi_2(p)$ | $f(p)$ | $0$ | $0$ | $-i\hat{p}_-\varphi(p)$ |
| $f(p)$ | $0$ | $\hat{p}_+\psi_2(p)$ | $0$ | $-\hat{p}_-\psi_1(p)$ |

Table 3: Supertransformation for chiral fields.

| | $Q_+^{(+)}$ | $Q_+^{(-)}$ | $Q_-^{(+)}$ | $Q_-^{(-)}$ |
|---|---|---|---|---|
| $\overline{\varphi}(p)$ | $0$ | $i\overline{\psi}_1(p)$ | $0$ | $i\overline{\psi}_2(p)$ |
| $\overline{\psi}_1(p)$ | $-i\hat{p}_+\overline{\varphi}(p)$ | $0$ | $0$ | $-\overline{f}(p)$ |
| $\overline{\psi}_2(p)$ | $0$ | $\overline{f}(p)$ | $-i\hat{p}_-\overline{\varphi}(p)$ | $0$ |
| $\overline{f}(p)$ | $\hat{p}_+\overline{\psi}_2$ | $0$ | $-\hat{p}_-\overline{\psi}_1(p)$ | $0$ |

Table 4: Supertransformation for anti-chiral fields.

### 5.2 Propagators

We can rewrite the kinetic term of the action as follows:

$$\begin{aligned} S_{free} = & \int_{-\frac{\pi}{a}}^{\frac{\pi}{a}} \frac{d^2\hat{p}}{(2\pi)^2} \left[ -\overline{\varphi}(-p)\hat{p}_+\hat{p}_-\varphi(p) - \overline{f}(-p)f(p) + imf(p)\varphi(p) - im\overline{\varphi}(-p)\overline{f}(p) \right. \\ & \left. + \overline{\psi}_1(-p)\hat{p}_-\psi_1(p) + \overline{\psi}_2(-p)\hat{p}_+\psi_2(p) + m\psi_2(-p)\psi_1(p) + m\overline{\psi}_1(-p)\overline{\psi}_2(p) \right], \\ = & \frac{1}{2}\int_{-\frac{\pi}{a}}^{\frac{\pi}{a}} \frac{d^2\hat{p}}{(2\pi)^2} \left[ \Phi_B^\dagger(p) M_B \Phi_B(p) + \Psi_F^\dagger(p) M_F \Psi_F(p) \right], \end{aligned} \tag{5.18}$$

where

$$\Phi_B(p) = \begin{pmatrix} \varphi(p) \\ f(p) \\ \overline{\varphi}(-p) \\ \overline{f}(-p) \end{pmatrix}, \qquad \Phi_B^\dagger(p) = \begin{pmatrix} \overline{\varphi}(-p) & \overline{f}(-p) & \varphi(p) & f(p) \end{pmatrix}, \tag{5.19}$$

$$\Psi_F(p) = \begin{pmatrix} \psi_1(p) \\ \psi_2(p) \\ \overline{\psi}_1(-p) \\ \overline{\psi}_2(-p) \end{pmatrix}, \qquad \Psi_F^\dagger(p) = \begin{pmatrix} \overline{\psi}_1(-p) & \overline{\psi}_2(-p) & \psi_1(p) & \psi_2(p) \end{pmatrix}, \tag{5.20}$$

and

$$M_B = \begin{pmatrix} -\hat{p}_+\hat{p}_- & 0 & 0 & -im \\ 0 & -1 & -im & 0 \\ 0 & im & -\hat{p}_+\hat{p}_- & 0 \\ im & 0 & 0 & -1 \end{pmatrix}, \qquad M_F = \begin{pmatrix} \hat{p}_- & 0 & 0 & m \\ 0 & \hat{p}_+ & -m & 0 \\ 0 & -m & \hat{p}_- & 0 \\ m & 0 & 0 & \hat{p}_+ \end{pmatrix}. \tag{5.21}$$

Inverse matrix of $M_B$ and $M_F$ are given as,

$$M_B^{-1} = \frac{1}{\hat{p}_+\hat{p}_- - m^2}\begin{pmatrix} -1 & 0 & 0 & im \\ 0 & -\hat{p}_+\hat{p}_- & im & 0 \\ 0 & -im & -1 & 0 \\ -im & 0 & 0 & -\hat{p}_+\hat{p}_- \end{pmatrix}, \quad M_F^{-1} = \frac{1}{\hat{p}_+\hat{p}_- - m^2}\begin{pmatrix} \hat{p}_+ & 0 & 0 & -m \\ 0 & \hat{p}_- & m & 0 \\ 0 & m & \hat{p}_+ & 0 \\ -m & 0 & 0 & \hat{p}_- \end{pmatrix}. \tag{5.22}$$

We can thus obtain the following propagators:

$\varphi(p)$  $\overline{\varphi}(-p) = \frac{-1}{D(\hat{p})}$, $f(p)$  $\overline{f}(-p) = \frac{-\hat{p}^2}{D(\hat{p})}$,

$\varphi(p)$  $f(-p) = \frac{im}{D(\hat{p})}$, $\overline{\varphi}(p)$  $\overline{f}(-p) = \frac{-im}{D(\hat{p})}$,

$\psi_1(p)$  $\overline{\psi}_1(-p) = \frac{\hat{p}_+}{D(\hat{p})}$, $\psi_2(p)$  $\overline{\psi}_2(-p) = \frac{\hat{p}_-}{D(\hat{p})}$,

$\psi_1(p)$  $\psi_2(-p) = \frac{-m}{D(\hat{p})}$, $\overline{\psi}_1(p)$  $\overline{\psi}_2(-p) = \frac{m}{D(\hat{p})}$, (5.23)

where

$$\frac{1}{D(\hat{p})} = \frac{1}{\hat{p}_+\hat{p}_- - m^2}. \tag{5.24}$$

Here we do not distinguish lines associated with chiral and anti-chiral fields. Then, Feynman rules for cubic and quartic interaction vertices are assigned as,

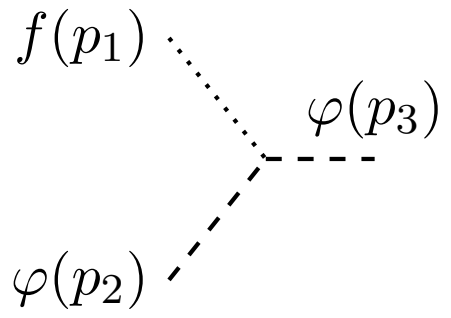

$= -g_3$,

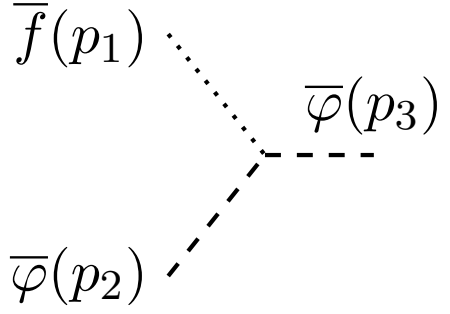

$= g_3$,

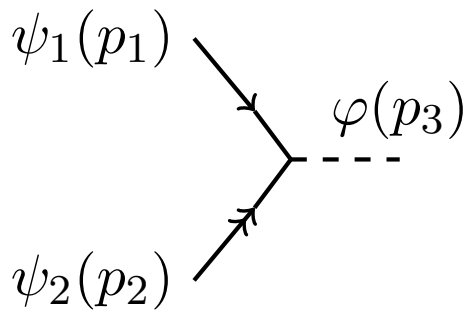

$= 2ig_3$,

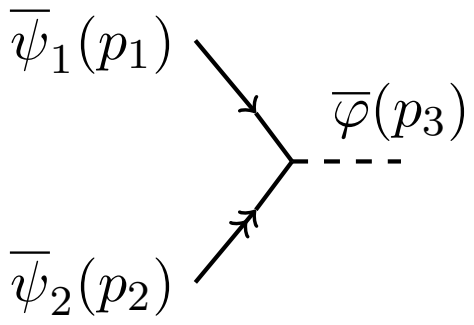

$= -2ig_3$, (5.25)

and

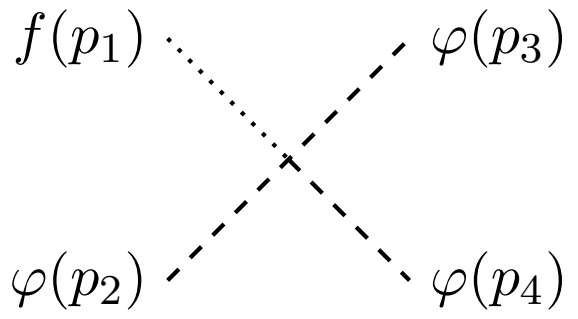

$= -g_4$,

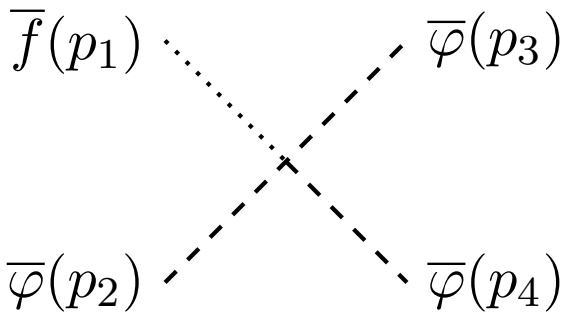

$= g_4$,

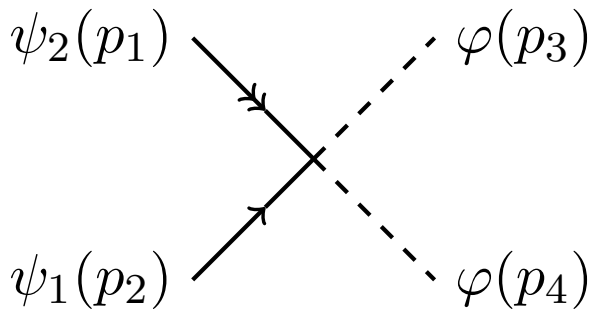

$= 3ig_4$,

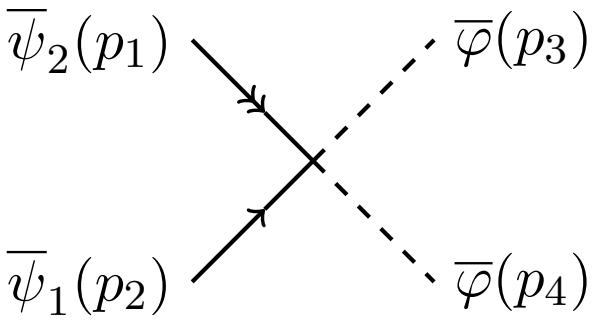

$= -3ig_4$. (5.26)

# 6 Ward-Takahashi identity in 2-dimensions

## 6.1 Tree level Ward-Takahashi identities

To find nontrivial relations between two point functions, we examine all possible fermionic bi-linear operators as $\delta\mathcal{O}$ in (2.4) such as $\left\langle Q_+^{(+)}(\varphi(p)\psi_2(-p))\right\rangle$, $\left\langle Q_-^{(+)}(\overline{\varphi}(p)\overline{\psi}_2(-p))\right\rangle, \cdots$. As for the fermionic bi-linear operators we consider $\mathcal{O} = (\varphi\psi_i), (\overline{\varphi}\overline{\psi}_i), (f,\psi_i), (\overline{f},\overline{\psi}_i), (\varphi,\overline{\psi}_i), (\overline{\varphi},\psi_i), (f,\overline{\psi}_i), (\overline{f},\psi_i)$. Then we find the following 6 independent Ward-Takahasu identities:

$$\left\langle \psi_1(p)\overline{\psi}_1(-p)\right\rangle + \hat{p}_+ \left\langle \varphi(p)\overline{\varphi}(-p)\right\rangle = 0, \tag{6.1a}$$

$$\left\langle \psi_2(p)\overline{\psi}_2(-p)\right\rangle + \hat{p}_- \left\langle \varphi(p)\overline{\varphi}(-p)\right\rangle = 0, \tag{6.1b}$$

$$\hat{p}_- \left\langle \psi_1(p)\overline{\psi}_1(-p)\right\rangle + \left\langle f(p)\overline{f}(-p)\right\rangle = 0, \tag{6.1c}$$

$$\hat{p}_+ \left\langle \psi_2(p)\overline{\psi}_2(-p)\right\rangle + \left\langle f(p)\overline{f}(-p)\right\rangle = 0, \tag{6.1d}$$

$$i\left\langle \psi_2(p)\psi_1(-p)\right\rangle - \left\langle \varphi(p)f(-p)\right\rangle = 0, \tag{6.1e}$$

$$i\left\langle \overline{\psi}_2(p)\overline{\psi}_1(-p)\right\rangle - \left\langle \overline{\varphi}(p)\overline{f}(-p)\right\rangle = 0. \tag{6.1f}$$

Identifying tree level two point function with propagators in (5.23), we find all the Ward-Takahashi identities are satisfied at the tree level which confirms the supersymmetry invariance of the formulation for all super charges.

## 6.2 1-loop corrections

### 6.2.1 1-loop corrections: quartic interaction $\Phi^4$

We now consider one-loop correction with quartic interaction in this subsection. We treat the case of $\Phi^4$ interaction first since the loop corrections of this case is relatively simpler than the $\Phi^3$ case. First of all it is interesting to note that loop corrections for the two point functions of the same chirality vanish since the action possesses properties that kinetic term mixes chiral and anti-chiral fields or different fields with the same chirality while mass term and interaction terms do not change their chirality. For example the following loop corrections vanish:

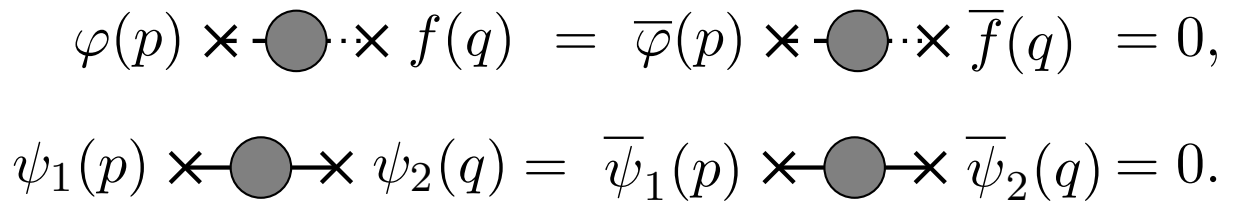

(6.2a)
(6.2b)

The following loop corrections having mixing fields in the loop also vanish because there is a cancellation between loop corrections:

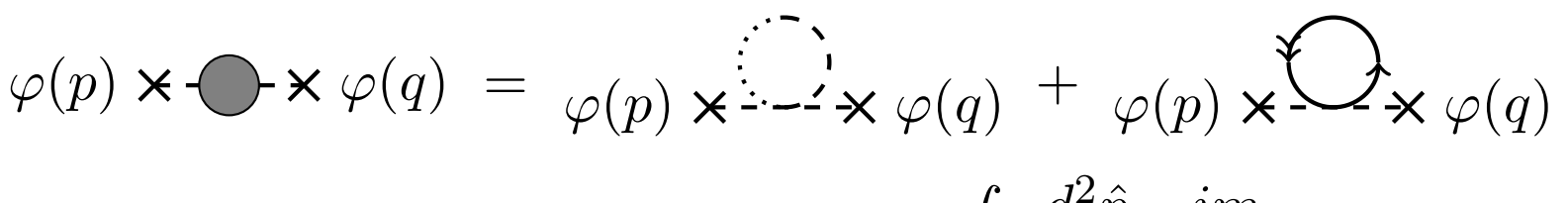

$$= (-g_4)3\varphi(p)\varphi(q)\int \frac{d^2\hat{p}}{(2\pi)^2}\frac{im}{D(\hat{p})} + (3ig_4)\varphi(p)\varphi(q)\int \frac{d^2\hat{p}}{(2\pi)^2}\frac{m}{D(\hat{p})}$$

$$= 0, \tag{6.3a}$$

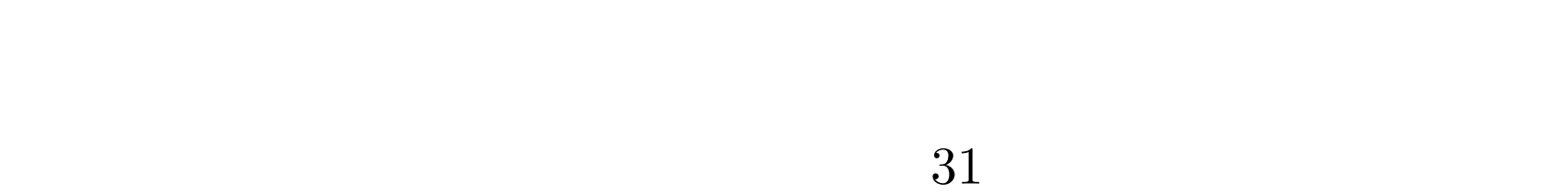

(6.3b)

The result of (6.3b) immediately follows from that of (6.3a), since the anti-chiral fields are hermitian conjugate of corresponding chiral fields. Therefore loop contributions to two point operators must also be hermitian conjugate to each other. Consequently we have no 1-loop correction on any two point functions with quartic interaction.

### 6.2.2 1-loop correction: cubic interaction $\Phi^3$

Next we focus on 1-loop corrections with cubic interaction. Similar to the $\Phi^4$ case in the previous subsection the two point operators of the same chirality vanish with the same reasons: the kinetic term mixes chiral and anti-chiral fields or different fields with the same chirality while mass term and interaction terms do not change their chirality.

$f(p)$ 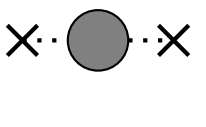 $f(q) = \overline{f}(p)$ 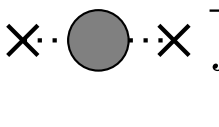 $\overline{f}(q) = 0,$ (6.4a)

$\varphi(p)$ 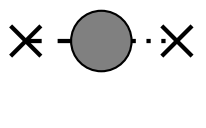 $f(q) = \overline{\varphi}(p)$ 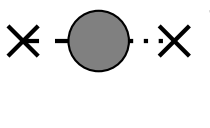 $\overline{f}(q) = 0,$ (6.4b)

$\psi_1(p)$ 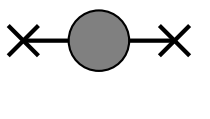 $\psi_1(q) = \overline{\psi}_1(p)$ 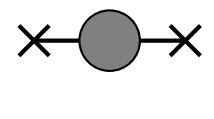 $\overline{\psi}_1(q) = 0,$ (6.4c)

$\psi_2(p)$ 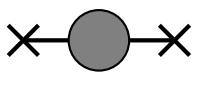 $\psi_2(q) = \overline{\psi}_2(p)$ 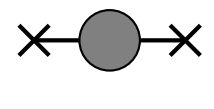 $\overline{\psi}_2(q) = 0,$ (6.4d)

There are the following combination of loop corrections with mixing fields in the loop but they vanish due to the cancellation:

$\varphi(p)$ 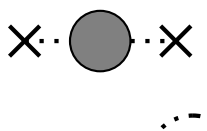 $\varphi(q)$

$= \varphi(p)$ 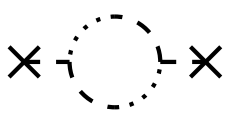 $\varphi(q) + \varphi(p)$ 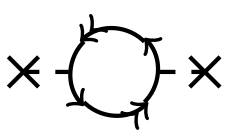 $\varphi(q)$

$$= -2g_3^2 m^2 \varphi(p)\varphi(q) \int \frac{d^2\hat{k}}{(2\pi)^2} \frac{1}{D(\hat{k})D(\hat{q}-\hat{k})} + 2g_3^2 m^2 \varphi(p)\varphi(q) \int \frac{d^2\hat{k}}{(2\pi)^2} \frac{1}{D(\hat{k})D(\hat{q}-\hat{k})}$$

$= 0,$ (6.5a)

$\overline{\varphi}(p)$ 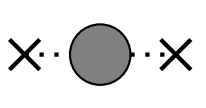 $\overline{\varphi}(q)$

$= \overline{\varphi}(p)$ 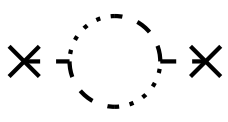 $\overline{\varphi}(q) + \overline{\varphi}(p)$ 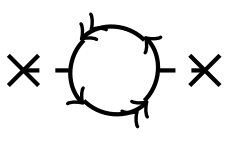 $\overline{\varphi}(q) = 0$ (6.5b)

On the other hand the two point operators of different chirality lead to the following nontrivial contributions:

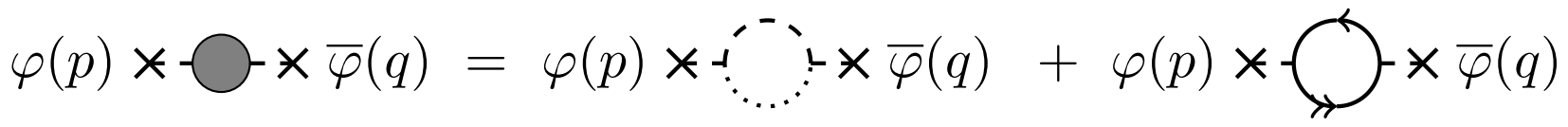

$$= -2g_3^2 \varphi(p)\overline{\varphi}(q) \int \frac{d^2\hat{k}}{(2\pi)^2} \frac{1}{D(\hat{k})D(\hat{q}-\hat{k})} \hat{q}_+ \hat{q}_-, \tag{6.6a}$$

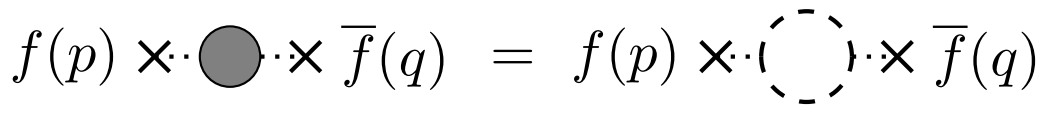

$$= -2g_3^2 f(p)\overline{f}(q) \int \frac{d^2\hat{k}}{(2\pi)^2} \frac{1}{D(\hat{k})D(\hat{q}-\hat{k})}, \tag{6.6b}$$

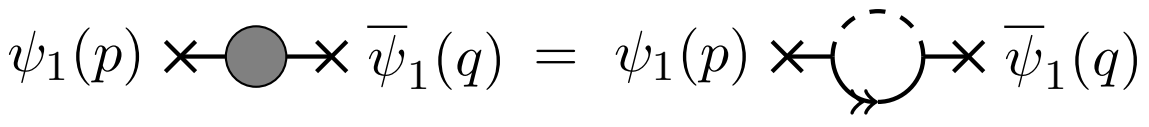

$$= 2g_3^2 \psi_1(p)\overline{\psi}_1(q) \int \frac{d^2\hat{k}}{(2\pi)^2} \frac{1}{D(\hat{k})D(\hat{q}-\hat{k})} \hat{q}_-, \tag{6.6c}$$

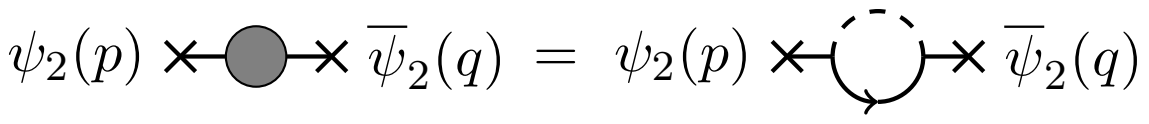

$$= 2g_3^2 \psi_2(p)\overline{\psi}_2(q) \int \frac{d^2\hat{k}}{(2\pi)^2} \frac{1}{D(\hat{k})D(\hat{q}-\hat{k})} \hat{q}_+, \tag{6.6d}$$

where we have renamed loop momenta $\hat{k}_i \to \hat{q}_i - \hat{k}_i$ in the intermediate stage, so that we can extract out the external momentum dependence by using momentum conservation.

Using these relations, we can evaluate two point functions of various pairs of fields. After carrying out Wick contraction for possible combinations of two point operators as (4.14) in 1-dimensional case, we obtain 1-loop corrections to the two point functions as follows:

$$\begin{aligned}
\langle \varphi(p)\overline{\varphi}(-p)\rangle_{\text{1-loop}} &= \langle \varphi(p)\overline{\varphi}(-p)\rangle_{\text{tree}} X(\hat{p}), & \langle \varphi(p) f(-p)\rangle_{\text{1-loop}} &= \langle \varphi(p) f(-p)\rangle_{\text{tree}} Y(\hat{p}),\\
\langle f(p)\overline{f}(-p)\rangle_{\text{1-loop}} &= \langle f(\hat{p})\overline{f}(-p)\rangle_{\text{tree}} X(\hat{p}), & \langle \overline{\varphi}(p)\overline{f}(-p)\rangle_{\text{1-loop}} &= \langle \overline{\varphi}(p)\overline{f}(-p)\rangle_{\text{tree}} Y(\hat{p}),\\
\langle \psi_1(p)\overline{\psi}_1(-p)\rangle_{\text{1-loop}} &= \langle \psi_1(p)\overline{\psi}_1(-p)\rangle_{\text{tree}} X(\hat{p}), & \langle \psi_1(p)\psi_2(-p)\rangle_{\text{1-loop}} &= \langle \psi_1(p)\psi_2(-p)\rangle_{\text{tree}} Y(\hat{p}),\\
\langle \psi_2(p)\overline{\psi}_2(-p)\rangle_{\text{1-loop}} &= \langle \psi_2(p)\overline{\psi}_2(-p)\rangle_{\text{tree}} X(\hat{p}), & \langle \overline{\psi}_1(p)\overline{\psi}_2(-p)\rangle_{\text{1-loop}} &= \langle \overline{\psi}_1(p)\overline{\psi}_2(-p)\rangle_{\text{tree}} Y(\hat{p}),
\end{aligned}$$

where $X(\hat{p})$ and $Y(\hat{p})$ are given in the following table:

| Loop diagram | $X(\hat{p})$ | $Y(\hat{p})$ |
|---|---|---|
| 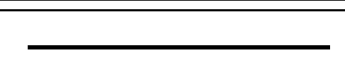 | 1 | 1 |
| 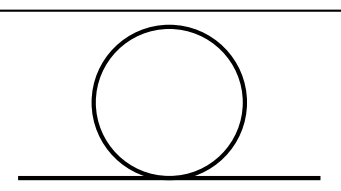 | 0 | 0 |
| 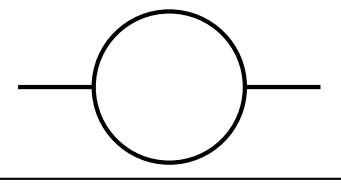 | $-2g_3^2 \frac{\hat{p}_+\hat{p}_- + m^2}{D(\hat{p})} I_1$ | $4g_3^2 \frac{\hat{p}_+\hat{p}_-}{D(\hat{p})} I_1$ |

where

$$I_1 = \int \frac{d^2\hat{k}}{(2\pi)^2} \frac{1}{D(\hat{k})D(\hat{p}-\hat{k})}. \tag{6.7}$$

Here we have summarized the cases for both $\Phi^4$ and $\Phi^3$ with null contribution for $\Phi^4$ case.

The 2-dimensional integral $I_1$ should be evaluated with the care of the integration range since each of the lattice "sine" momenta $\hat{p}$ and $\hat{k}$ has limited integration range in this model. In each direction of the light cone "sine" momentum integration the integration range should be treated as it has been shown in (A.10) of one dimensional model. The current 2-dimensional formulation can be formulated essentially by the direct product of two one-dimensional models.

The proportionality relations between the 1-loop and tree two point functions have the similar structure as in the 1-dimensional case. We can then show that the 1-loop WT identities are proportional to tree level WT identities with multiplicative factors given in the table. For example we obtain

$$\begin{aligned} &\left\langle \psi_1(p)\overline{\psi}_1(-p)\right\rangle_{\text{1-loop}} + \hat{p}_+ \left\langle \varphi(p)\overline{\varphi}(-p)\right\rangle_{\text{1-loop}} \\ &\quad = \left[\left\langle \psi_1(p)\overline{\psi}_1(-p)\right\rangle_{\text{tree}} + \hat{p}_+ \left\langle \varphi(p)\overline{\varphi}(-p)\right\rangle_{\text{tree}}\right] X(\hat{p}) = 0, \end{aligned} \tag{6.8a}$$

$$\begin{aligned} &i\left\langle \psi_2(p)\psi_1(-p)\right\rangle_{\text{1-loop}} - \left\langle \varphi(p)f(-p)\right\rangle_{\text{1-loop}} \\ &\quad = \left[i\left\langle \psi_2(p)\psi_1(-p)\right\rangle_{\text{tree}} - \left\langle \varphi(p)f(-p)\right\rangle_{\text{tree}}\right] Y(\hat{p}) = 0. \end{aligned} \tag{6.8b}$$

The rest of WT identies in (6.1) satisfy the similar proportionality and thus are satisfied in the 1-loop level.

## 6.3 2-loop corrections

Next we investigate 2-loop corrections explicitly with either quartic or cubic interactions.

### 6.3.1 2-loop corrections: quartic interaction $\Phi^4$

We first treat 2-loop diagrams with quartic interaction. There are two types of 2-loop diagrams; tadpole (snow man) diagram in Fig.1-a and overlapping (sunset) diagram in Fig.1-b. The former one includes one-loop tadpole as a subdiagram, therefore it apparently vanishes with the same reasonings as 1-loop arguments of $\Phi^4$ interaction in subsection 6.2.1. We thus focus only on overlapping diagram here.

Similar to 1-loop diagrams, most of two point operators vanish due to vanishing internal propagators. We only have the following non-trivial loop corrections:

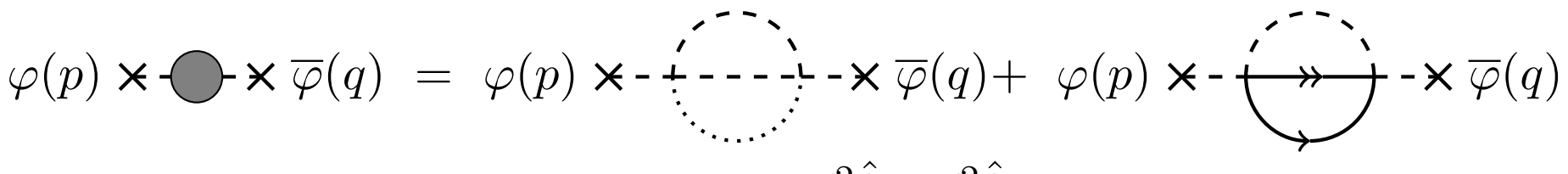

$$= 6g_4^2\varphi(p)\overline{\varphi}(q)\int \frac{d^2\hat{k}_1}{(2\pi)^2}\frac{d^2\hat{k}_2}{(2\pi)^2}\frac{1}{D(\hat{k}_1)}\frac{1}{D(\hat{k}_2)}\frac{1}{D(\hat{q}-\hat{k}_1-\hat{k}_2)}\hat{q}_+\hat{q}_- \tag{6.9a}$$

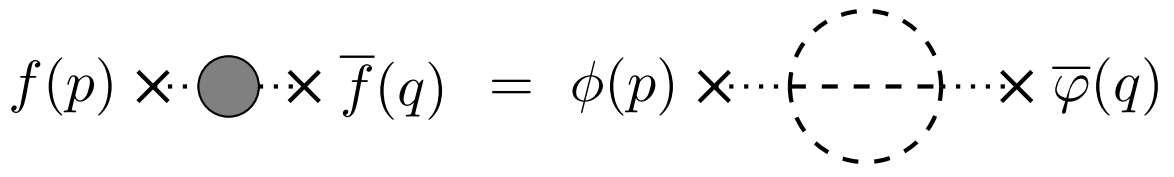

$$= 6g_4^2 f(p)\overline{f}(q)\int \frac{d\hat{k}_1^2}{(2\pi)^2}\frac{d\hat{k}_2^2}{(2\pi)^2}\frac{1}{D(\hat{k}_1)}\frac{1}{D(\hat{k}_2)}\frac{1}{D(\hat{q}-\hat{k}_1-\hat{k}_2)}, \tag{6.9b}$$

$\psi_1(p)$ 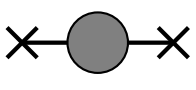 $\overline{\psi}_1(q)$ $=$ $\psi_2(p)$ 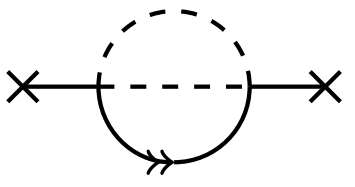 $\overline{\psi}_2(q)$

$$= -6g_4^2 \psi_1(p)\overline{\psi}_1(q) \int \frac{d^2\hat{k}_1}{(2\pi)^2} \frac{d^2\hat{k}_2}{(2\pi)^2} \frac{1}{D(\hat{k}_1)} \frac{1}{D(\hat{k}_2)} \frac{1}{D(\hat{q}-\hat{k}_1-\hat{k}_2)} \hat{q}_-, \tag{6.9c}$$

$\psi_2(p)$ 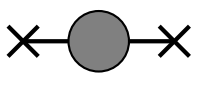 $\overline{\psi}_2(q)$ $=$ $\psi_1(p)$ 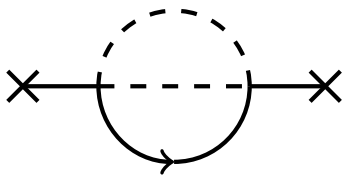 $\overline{\psi}_1(q)$

$$= -6g_4^2 \psi_2(p)\overline{\psi}_2(q) \int \frac{d^2\hat{k}_1}{(2\pi)^2} \frac{d^2\hat{k}_2}{(2\pi)^2} \frac{1}{D(\hat{k}_1)} \frac{1}{D(\hat{k}_2)} \frac{1}{D(\hat{q}-\hat{k}_1-\hat{k}_2)} \hat{q}_+. \tag{6.9d}$$

where we have arranged to rename the integrated loop momenta and the "sine" momentum conservation should be understood: $\hat{q} = \hat{k}_1 + \hat{k}_2 + \hat{k}_3$.

In the evaluation of 2-loop integral we need to take care about the integration range of the "sine" momentum again as we have discussed in the 2-loop correction of 1-dimensional model in (4.39). Here in the current 2-loop expression there is no terms depending on the internal momenta in the numerator so that we simply take care the integration range of two internal loop momenta:

$$\begin{aligned} &\int \frac{d^2\hat{k}_1}{(2\pi)^2} \frac{d^2\hat{k}_2}{(2\pi)^2} \frac{1}{D(\hat{k}_1)} \frac{1}{D(\hat{k}_2)} \frac{1}{D(\hat{q}-\hat{k}_1-\hat{k}_2)} \\ &= \left[ \int_{-\frac{2}{a}}^{\hat{q}} \frac{d\hat{k}_1}{2\pi} \int_{\hat{q}-\hat{k}_1-\frac{2}{a}}^{\frac{2}{a}} \frac{d\hat{k}_2}{2\pi} + \int_{\hat{q}}^{\frac{2}{a}} \frac{d\hat{k}_1}{2\pi} \int_{-\frac{2}{a}}^{\hat{q}-\hat{k}_1+\frac{2}{a}} \frac{d\hat{k}_2}{2\pi} \right] \frac{1}{D(\hat{k}_1)D(\hat{k}_2)D(\hat{q}-\hat{k}_1-\hat{k}_2)}. \end{aligned} \tag{6.10}$$

The details of the treatment of integration range for three internal momenta are given in Appendix A.2.

We eventually have the following two point functions with 2-loop corrections:

$$\langle \varphi(p)\overline{\varphi}(-p)\rangle_{\text{2-loop}} = \langle \varphi(p)\overline{\varphi}(-p)\rangle_{\text{tree}} A(p), \tag{6.11a}$$

$$\langle f(p)\overline{f}(-p)\rangle_{\text{2-loop}} = \langle f(p)\overline{f}(-p)\rangle_{\text{tree}} A(p), \tag{6.11b}$$

$$\langle \psi_1(p)\overline{\psi}_1(-p)\rangle_{\text{2-loop}} = \langle \psi_1(p)\overline{\psi}_1(-p)\rangle_{\text{tree}} A(p), \tag{6.11c}$$

$$\langle \psi_2(p)\overline{\psi}_2(-p)\rangle_{\text{2-loop}} = \langle \psi_2(p)\overline{\psi}_2(-p)\rangle_{\text{tree}} A(p), \tag{6.11d}$$

$$\langle \varphi(p)f(-p)\rangle_{\text{2-loop}} = \langle \varphi(p)f(-p)\rangle_{\text{tree}} B(p), \tag{6.11e}$$

$$\langle \overline{\varphi}(p)\overline{f}(-p)\rangle_{\text{2-loop}} = \langle \overline{\varphi}(p)\overline{f}(-p)\rangle_{\text{tree}} B(p), \tag{6.11f}$$

$$\langle \psi_1(p)\psi_2(-p)\rangle_{\text{2-loop}} = \langle \psi_1(p)\psi_2(-p)\rangle_{\text{tree}} B(p), \tag{6.11g}$$

$$\langle \overline{\psi}_1(p)\overline{\psi}_2(-p)\rangle_{\text{2-loop}} = \langle \overline{\psi}_1(p)\overline{\psi}_2(-p)\rangle_{\text{tree}} B(p). \tag{6.11h}$$

where

$$A(p) = \left[ -6g_4^2 \frac{\hat{p}_+\hat{p}_- + m^2}{D(p)} \int \frac{d\hat{k}_1^2}{(2\pi)^2} \frac{d\hat{k}_2^2}{(2\pi)^2} \frac{1}{D(\hat{k}_1)D(\hat{k}_2)D(\hat{p}-\hat{k}_1-\hat{k}_2)} \right], \tag{6.12a}$$

$$B(p) = \left[ -12g_4^2 \frac{\hat{p}_+\hat{p}_-}{D(\hat{p})} \int \frac{d\hat{k}_1^2}{(2\pi)^2} \frac{d\hat{k}_2^2}{(2\pi)^2} \frac{1}{D(\hat{k}_1)D(\hat{k}_2)D(\hat{p}-\hat{k}_1-\hat{k}_2)} \right]. \tag{6.12b}$$

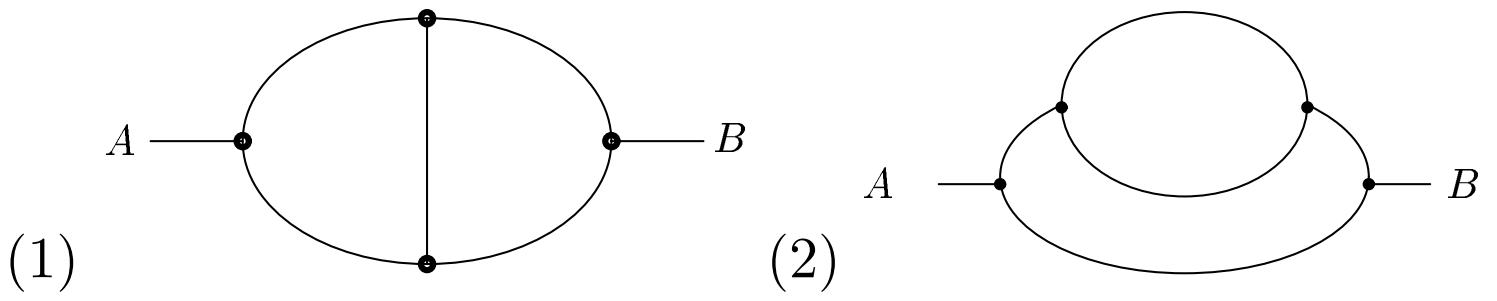

Figure 2: Two points operators with 2-loop corrections

Again the 2-loop corrections of two point functions are proportional to the tree level expressions with multiplicative factors. Since the structure of the proportionality is the same as 1-loop corrections in (6.7), Ward-Takahashi identities are satisfied at the two loop level as well.

### 6.3.2 2-loop correction: cubic interaction $\Phi^3$: (1)

We next consider 2-loop corrections with cubic interaction. Possible non-trivial diagrams are shown in Figure 2. The other diagrams including 1-loop tadpoles as subdiagram trivially vanish since a tadpole contribution itself vanishes. We show explicitly that one point operators of 1-loop and 2-loop tadpole contributions for $\Phi^3$ model all vanish in Appendix B.

We first focus on the diagram (1) in Figure 2. The following diagrams can be considered as possible 2-loop contribution:

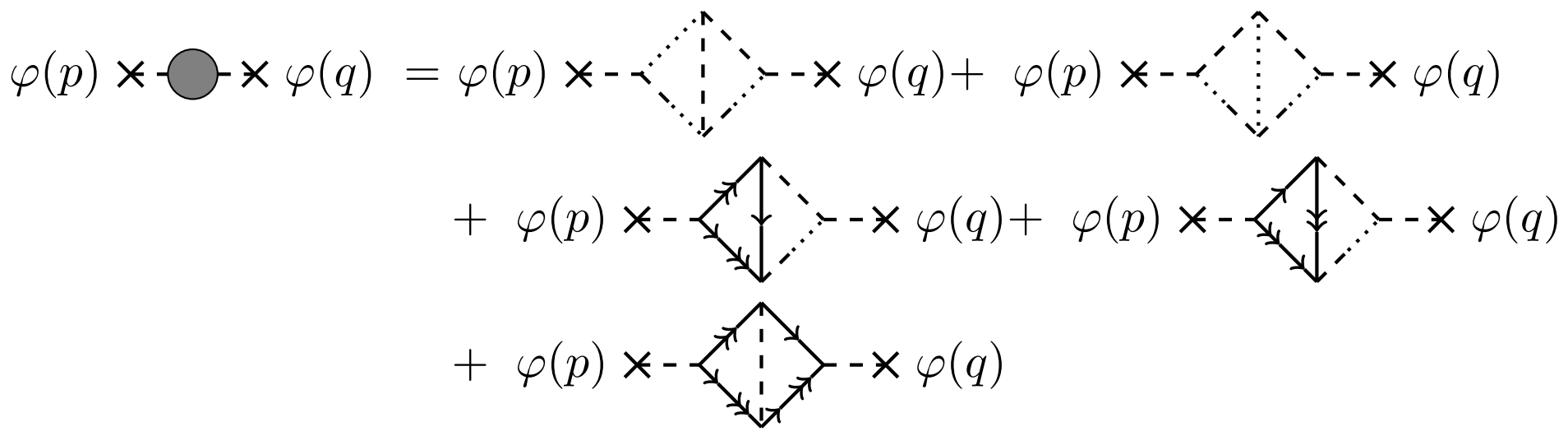

$=0,$ (6.13a)

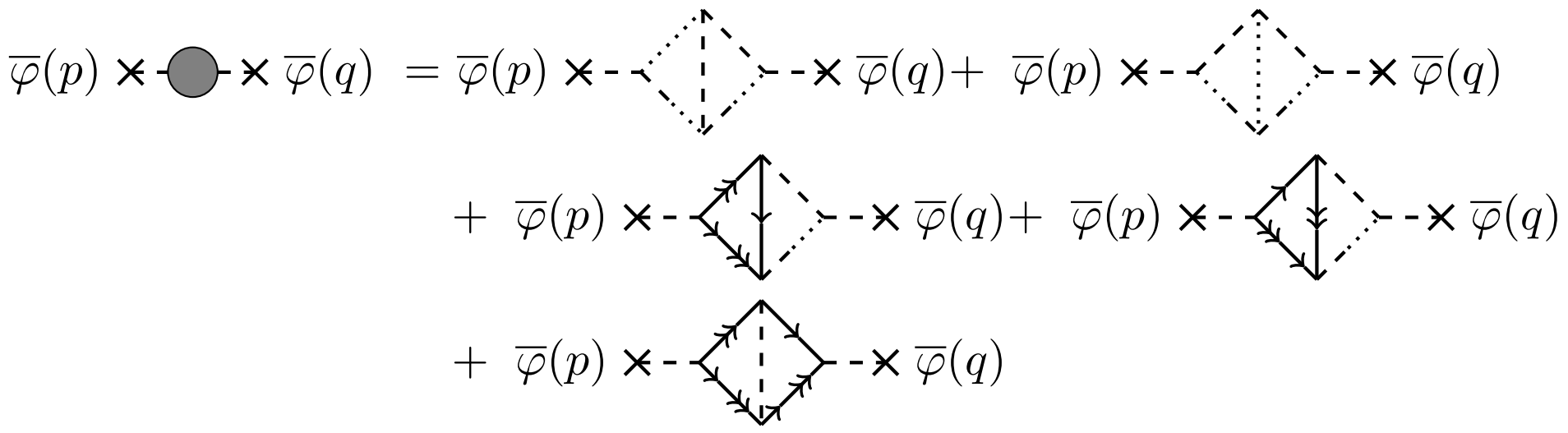

$=0,$ (6.13b)

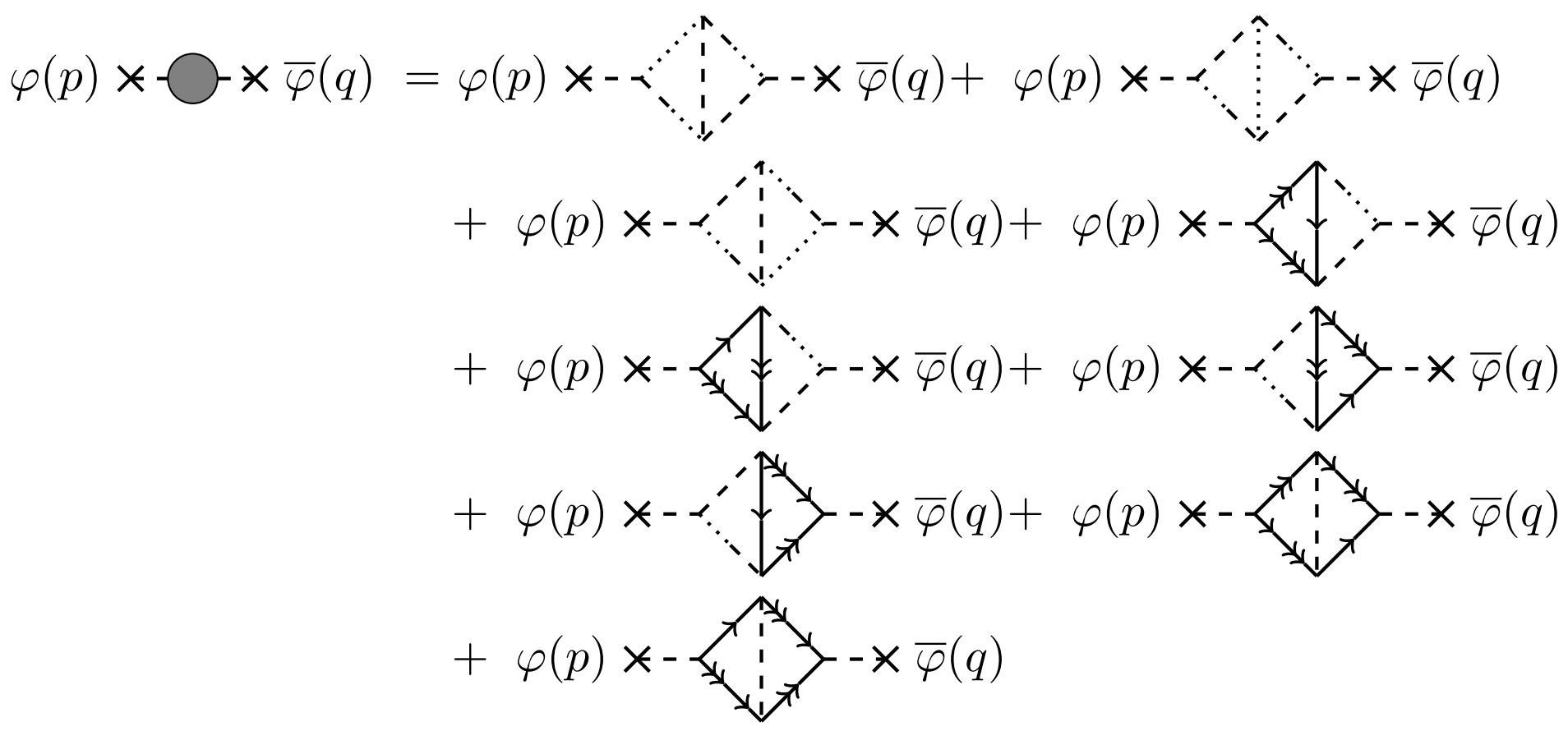

$$= -8m^2 g_3^4 \varphi(p)\overline{\varphi}(q) \int \delta^4 D^{-5} \left[\hat{p}^2 + \hat{q}^2\right], \tag{6.13c}$$

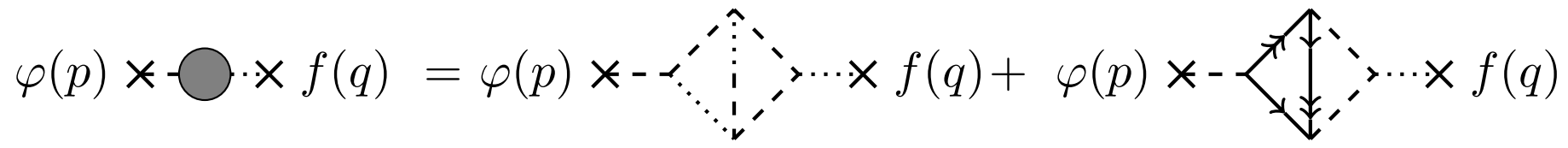

$$= -8img_3^4 \varphi(p) f(q) \int \delta^4 D^{-5} \left[\hat{p}^2\right], \tag{6.13d}$$

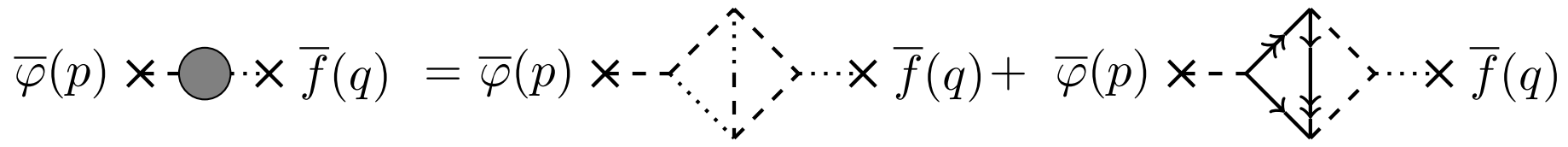

$$= 8img_3^4 \overline{\varphi}(p)\overline{f}(q) \int \delta^4 D^{-5} \left[\hat{p}^2\right], \tag{6.13e}$$

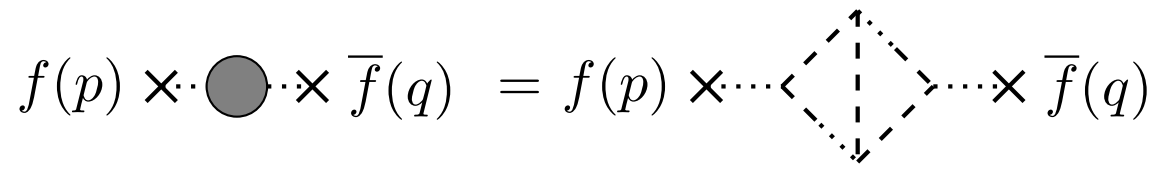

$$= -16m^2 g_3^4 f(p)\overline{f}(q) \int \delta^4 D^{-5}. \tag{6.13f}$$

Next fermionic two point operators receives the following 2-loop contributions:

$\psi_1(p)$ 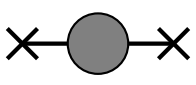 $\overline{\psi}_1(q)$ $= \psi_1(p)$ 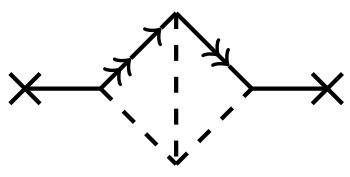 $\overline{\psi}_1(q)$ $+$ $\psi_1(p)$ 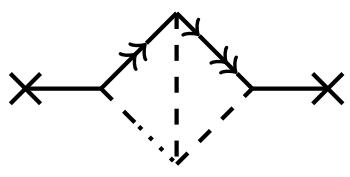 $\overline{\psi}_1(q)$

$+$ $\psi_1(p)$ 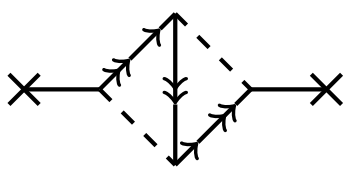 $\overline{\psi}_1(q)$

$$=8m^2 g_3^4 \psi_1(p)\overline{\psi}_1(q) \int \delta^4 D^{-5} \left[-\hat{p}_- + \hat{q}_-\right], \tag{6.13g}$$

$\psi_2(p)$ 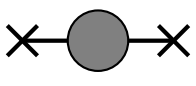 $\overline{\psi}_2(q)$ $= \psi_2(p)$ 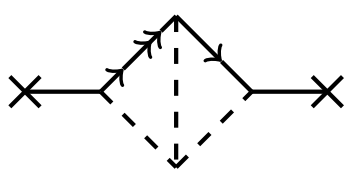 $\overline{\psi}_2(q)$ $+$ $\psi_2(p)$ 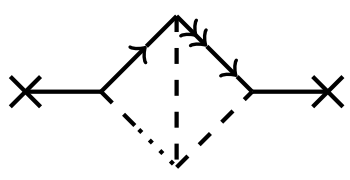 $\overline{\psi}_2(q)$

$+$ $\psi_2(p)$ 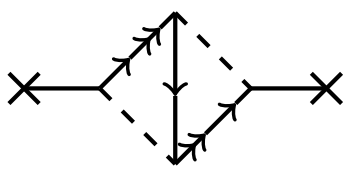 $\overline{\psi}_2(q)$

$$=8m^2 g_3^4 \psi_2(p)\overline{\psi}_2(q) \int \delta^4 D^{-5} \left[-\hat{p}_- + \hat{q}_-\right], \tag{6.13h}$$

$\psi_1(p)$ 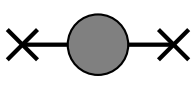 $\psi_2(q)$ $= \psi_1(p)$ 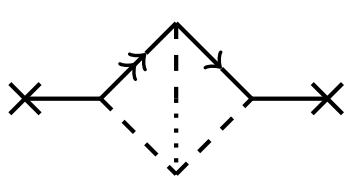 $\psi_2(q)$ $+$ $\psi_1(p)$ 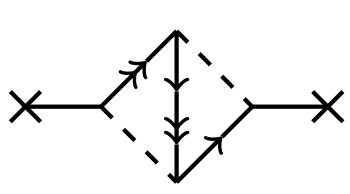 $\psi_2(q)$

$$= -\, 8m g_3^4 \psi_1(p)\psi_2(q) \int \delta^4 D^{-5} \left[\hat{p}_- \hat{q}_+\right], \tag{6.13i}$$

$\overline{\psi}_1(p)$ 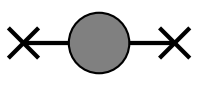 $\overline{\psi}_2(q)$ $= \overline{\psi}_1(p)$ 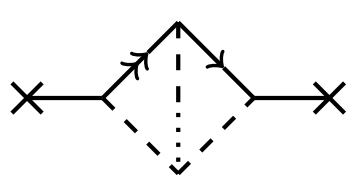 $\overline{\psi}_2(q)$ $+$ $\overline{\psi}_1(p)$ 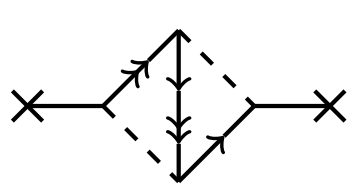 $\overline{\psi}_2(q)$

$$= -\, 8m g_3^4 \psi_1(p)\psi_2(q) \int \delta^4 D^{-5} \left[\hat{p}_- \hat{q}_+\right], \tag{6.13j}$$

where we symbolically denotes the following propagators and delta functions with phase space factors as:

$$\begin{aligned}\int \delta^4 D^{-5} = \int \prod_{i=1}^{5} \left[\frac{d^2\hat{k}_i}{(2\pi)^2}\frac{1}{D(\hat{k}_i)}\right] (2\pi)^8 \delta^2(\hat{p} + \hat{k}_1 + \hat{k}_2) \\ \times\, \delta^2(-\hat{k}_1 + \hat{k}_3 + \hat{k}_4)\delta^2(-\hat{k}_2 - \hat{k}_3 + \hat{k}_5)\delta^2(-\hat{k}_4 - \hat{k}_5 + \hat{q}).\end{aligned} \tag{6.14}$$

The first two operators, (6.13a) and (6.13b), in bosonic operators however vanish after the calculation shown as follows:

$\varphi(p)$ 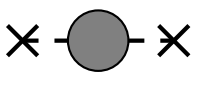 $\varphi(q)$

$$\begin{aligned} &= -16m^2 g_3^4 \varphi(p)\varphi(q) \int \delta^4 D^{-5} \left[\hat{k}_2^2 + \frac{1}{2}\hat{k}_3^2 + \hat{k}_{2-}\hat{k}_{3+} + \hat{k}_{2+}\hat{k}_{3-} - \hat{k}_{2-}\hat{k}_{5+}\right] \\ &= -8m^2 g_3^4 \varphi(p)\varphi(q) \int \delta^4 D^{-5} \left(\hat{k}_{2+} + \hat{k}_{3+} - \hat{k}_{5+}\right)\left(\hat{k}_{2-} + \hat{k}_{3-} - \hat{k}_{5-}\right) = 0. \end{aligned} \tag{6.15}$$

In the first line of (6.15) a factor $\frac{1}{2}$ is inserted because the diagram is symmetric under the exchange of external momentum $\hat{p} \leftrightarrow \hat{q}$. In the second line we use symmetry of the momenta in the delta functions; the product of delta functions is symmetric under the exchange, $(\hat{p}, \hat{k}_1, \hat{k}_2) \leftrightarrow (\hat{q}, -\hat{k}_4, -\hat{k}_5)$. Imposing momentum conservation, we can show that loop contributions vanish. In the similar way the operator with anti-chiral fields vanishes, which is obvious from hermiticity.

The other operators lead non-trivial contributions. One of concrete calculations goes in the following way:

$\varphi(p)$ 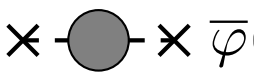 $\overline{\varphi}(q)$

$$
\begin{aligned}
&= -16m^2 g_3^4 \int \delta^4 D^{-5} \Big[\hat{k}_2^2 + \hat{k}_3^2 + \hat{k}_{2-}\hat{k}_{3+} + \hat{k}_{2+}\hat{k}_{3-} + \hat{k}_4^2 \\
&\qquad\qquad + \hat{k}_{3-}\hat{k}_{4+} + \hat{k}_{3+}\hat{k}_{4-} + \hat{k}_{2-}\hat{k}_{4+} + \hat{k}_{2+}\hat{k}_{4-}\Big] \\
&= -8m^2 g_3^4 \int \delta^4 D^{-5} \Big[\hat{k}_1^2 + \hat{k}_2^2 + \hat{k}_4^2 + \hat{k}_5^2 - 2\hat{k}_3^2 + \hat{k}_{2-}\hat{k}_{4+} + \hat{k}_{2+}\hat{k}_{4-} + \hat{k}_{1-}\hat{k}_{5+} + \hat{k}_{1+}\hat{k}_{5-}\Big] \\
&= -8m^2 g_3^4 \int \delta^4 D^{-5} \left[\hat{p}^2 + \hat{q}^2\right]. \qquad (6.16)
\end{aligned}
$$

In most of these calculations we carry out renaming of integrated loop momenta in such a way that the external momentum is extracted out in the numerator. The following "sine" momentum conservations are used repeatedly: $\hat{p} + \hat{k}_1 + \hat{k}_2 = -\hat{k}_1 + \hat{k}_3 + \hat{k}_4 = -\hat{k}_2 - \hat{k}_3 + \hat{k}_5 = -\hat{k}_4 - \hat{k}_5 + \hat{q} = 0$. We can also employ the following symmetry of renaming of momenta in the delta functions: $(\hat{k}_1, \hat{k}_4, \hat{k}_3) \leftrightarrow (\hat{k}_2, \hat{k}_5, -\hat{k}_3)$.

Collecting all these 2-loop contributions for two point functions, we obtain again the proportionality of two point functions of the 2-loop contribution and tree level contribution with multiplicative factors similar as 1-loop case:

$$\langle \varphi(p)\overline{\varphi}(-p)\rangle_{\text{2-loop}} = \langle \varphi(p)\overline{\varphi}(-p)\rangle_{\text{tree}} I(\hat{p}), \qquad (6.17a)$$

$$\left\langle f(p)\overline{f}(-p)\right\rangle_{\text{2-loop}} = \left\langle f(p)\overline{f}(-p)\right\rangle_{\text{tree}} I(\hat{p}), \qquad (6.17b)$$

$$\left\langle \psi_1(p)\overline{\psi}_1(-p)\right\rangle_{\text{2-loop}} = \left\langle \psi_1(p)\overline{\psi}_1(-p)\right\rangle_{\text{tree}} I(\hat{p}), \qquad (6.17c)$$

$$\left\langle \psi_2(p)\overline{\psi}_2(-p)\right\rangle_{\text{2-loop}} = \left\langle \psi_2(p)\overline{\psi}_2(-p)\right\rangle_{\text{tree}} I(\hat{p}), \qquad (6.17d)$$

$$\langle \varphi(p) f(-p)\rangle_{\text{2-loop}} = \langle \varphi(p) f(-p)\rangle_{\text{tree}} J(\hat{p}), \qquad (6.17e)$$

$$\left\langle \overline{\varphi}(p)\overline{f}(-p)\right\rangle_{\text{2-loop}} = \left\langle \overline{\varphi}(p)\overline{f}(-p)\right\rangle_{\text{tree}} J(\hat{p}), \qquad (6.17f)$$

$$\langle \psi_1(p)\psi_2(-p)\rangle_{\text{2-loop}} = \langle \psi_1(p)\psi_2(-p)\rangle_{\text{tree}} J(\hat{p}), \qquad (6.17g)$$

$$\left\langle \overline{\psi}_1(p)\overline{\psi}_2(-p)\right\rangle_{\text{2-loop}} = \left\langle \overline{\psi}_1(p)\overline{\psi}_2(-p)\right\rangle_{\text{tree}} J(\hat{p}), \qquad (6.17h)$$

where

$$I(\hat{p}) = 16m^2 g_3^4 \frac{2\hat{p}_+\hat{p}_- + m^2}{D(\hat{p})} \int \frac{d\hat{k}_1^2}{(2\pi)^2}\frac{d\hat{k}_2^2}{(2\pi)^2} \frac{1}{D(\hat{k}_1)D(\hat{k}_2)D(\hat{k}_1+\hat{p})D(\hat{k}_2+\hat{p})D(\hat{k}_1-\hat{k}_2)}, \qquad (6.18)$$

$$J(\hat{p}) = 8g_3^4 \frac{\hat{p}_+\hat{p}_- + 5m^2}{D(\hat{p})}\hat{p}_+\hat{p}_- \int \frac{d\hat{k}_1^2}{(2\pi)^2}\frac{d\hat{k}_2^2}{(2\pi)^2} \frac{1}{D(\hat{k}_1)D(\hat{k}_2)D(\hat{k}_1+\hat{p})D(\hat{k}_2+\hat{p})D(\hat{k}_1-\hat{k}_2)}. \qquad (6.19)$$

Similar to the previous cases all the two point functions with 2-loop corrections are proportional to tree ones with two multiplicative common factors, we can prove that supersymmetric WT identity is exactly satisfied at 2-loop quantum corrections for $\Phi^3$ model. Integration range of the integrated "sine" momenta should be treated similarly as in the case of (6.10). In the expression of (6.14), all the "sine" momentum conservation with delta functions should be understood with limited integration range as well.

### 6.3.3 2-loop corrections: cubic interaction $\Phi^3$ (2)

We finally focus on the second diagram (2) in Figure 2. This diagram also includes 1-loop as a subdiagram, thus we can utilize the result in subsection 6.2.2. Analogous to previous evaluation we can first extract diagrams with trivial null contribution and then we find the following nontrivial diagrams:

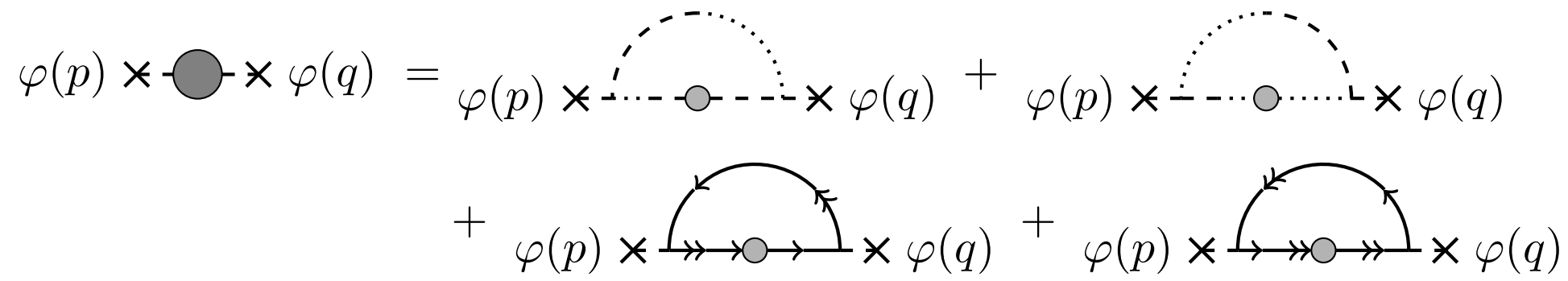

$$=0, \tag{6.20a}$$

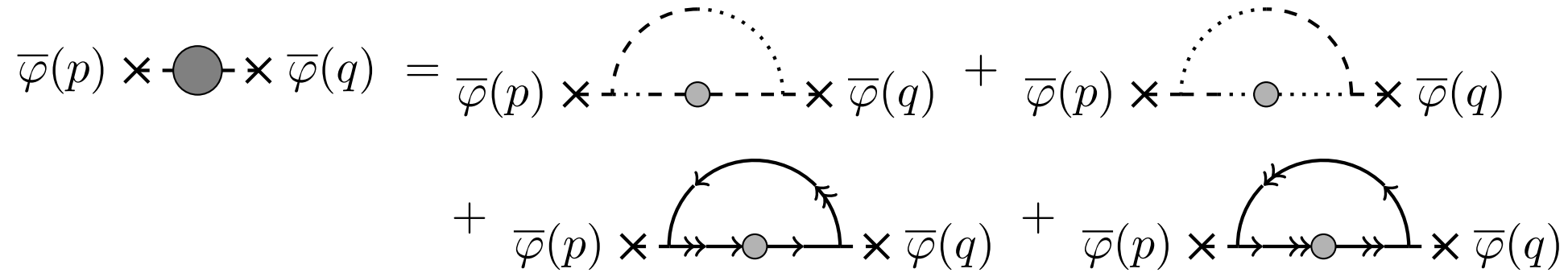

$$=0, \tag{6.20b}$$

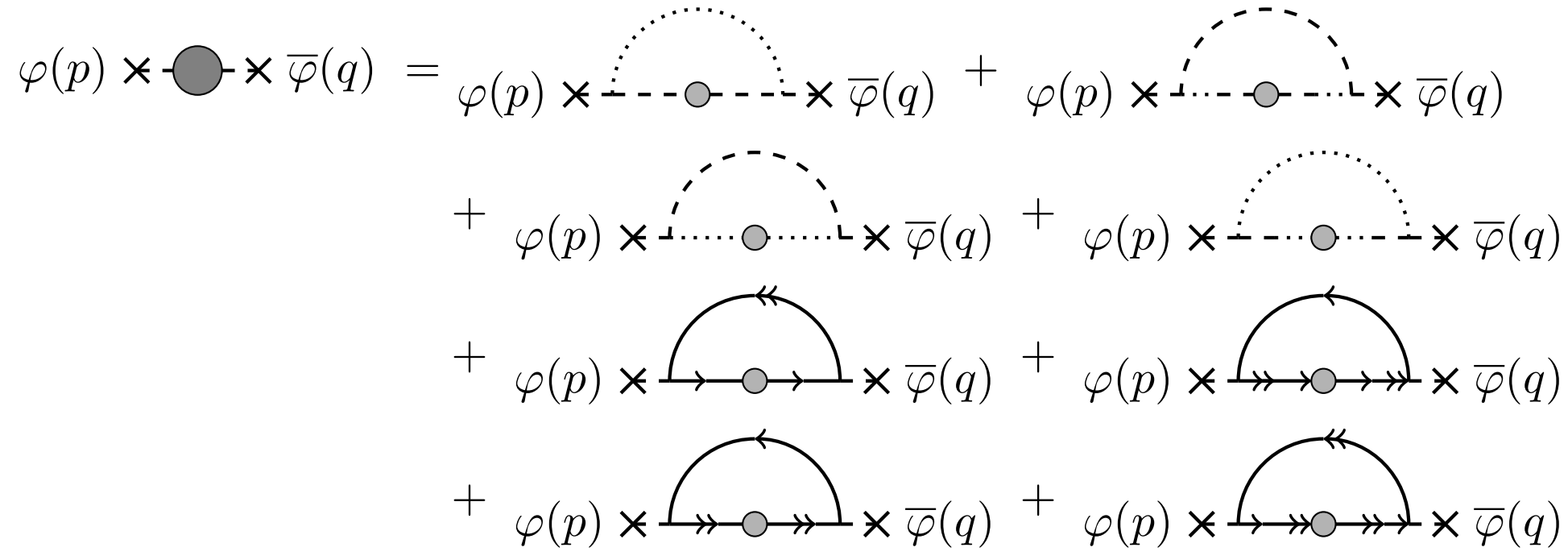

$$=\varphi(p)\overline{\varphi}(q)\left[-4g_3^4(\hat{p}^2+\hat{q}^2)\int\frac{d^2\hat{k}_1 d^2\hat{k}_2}{(2\pi)^2(2\pi)^2}\frac{\hat{k}_1^2+m^2}{D(\hat{k}_1)^2 D(\hat{k}_2)}\int\frac{d^2\hat{k}}{(2\pi)^2}\frac{1}{D(\hat{k})D(\hat{k}_1-\hat{k})}\right], \tag{6.20c}$$

$f(p)$ 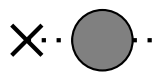 $\overline{f}(q)$ $=$ $f(p)$ 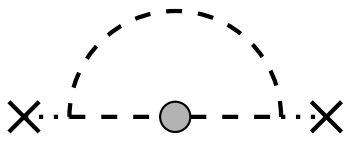 $\overline{f}(q)$ $+$ $f(p)$ 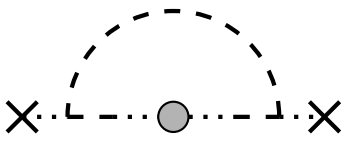 $\overline{f}(q)$

$$= f(p)\overline{f}(q)\left[-8g_3^4\int\frac{d^2\hat{k}_1 d^2\hat{k}_2}{(2\pi)^2(2\pi)^2}\frac{\hat{k}_1^2+m^2}{D(\hat{k}_1)^2D(\hat{k}_2)}\int\frac{d^2\hat{k}}{(2\pi)^2}\frac{1}{D(\hat{k})D(\hat{k}_1-\hat{k})}\right], \tag{6.20d}$$

$\psi_1(p)$ 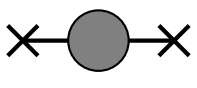 $\overline{\psi}_1(q)$ $=$ $\psi_1(p)$ 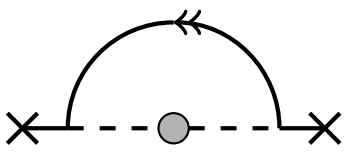 $\overline{\psi}_1(q)$ $+$ $\psi_1(p)$ 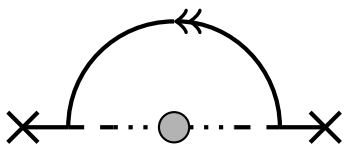 $\overline{\psi}_1(q)$

$+$ $\psi_1(p)$ 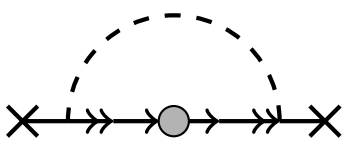 $\overline{\psi}_1(q)$ $+$ $\psi_1(p)$ 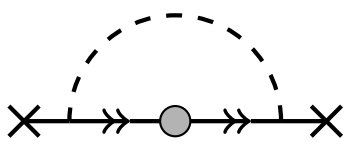 $\overline{\psi}_1(q)$

$$= \psi_1(p)\overline{\psi}_1(q)\left[4g_3^4(\hat{q}_- - \hat{p}_-)\int\frac{d^2\hat{k}_1 d^2\hat{k}_2}{(2\pi)^2(2\pi)^2}\frac{\hat{k}_1^2+m^2}{D(\hat{k}_1)^2D(\hat{k}_2)}\int\frac{d^2\hat{k}}{(2\pi)^2}\frac{1}{D(\hat{k})D(\hat{k}_1-\hat{k})}\right], \tag{6.20e}$$

$\psi_2(p)$ 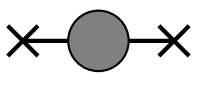 $\overline{\psi}_2(q)$ $=$ $\psi_2(p)$ 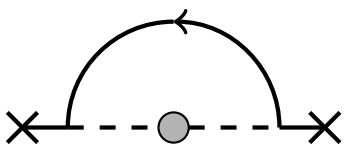 $\overline{\psi}_2(q)$ $+$ $\psi_2(p)$ 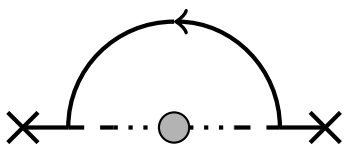 $\overline{\psi}_2(q)$

$+$ $\psi_2(p)$ 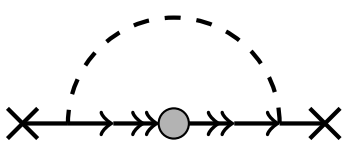 $\overline{\psi}_2(q)$ $+$ $\psi_2(p)$ 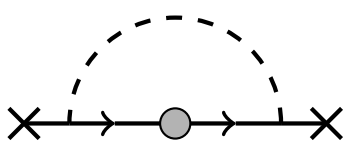 $\overline{\psi}_2(q)$

$$= = \psi_2(p)\overline{\psi}_2(q)\left[4g_3^4(\hat{q}_+ - \hat{p}_+)\int\frac{d^2\hat{k}_1 d^2\hat{k}_2}{(2\pi)^2(2\pi)^2}\frac{\hat{k}_1^2+m^2}{D(\hat{k}_1)^2D(\hat{k}_2)}\int\frac{d^2\hat{k}}{(2\pi)^2}\frac{1}{D(\hat{k})D(\hat{k}_1-\hat{k})}\right]. \tag{6.20f}$$

where ( 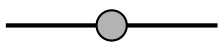 ) denotes one-loop correction calculated in subsection 6.2.2. As in the previous analysis, the first two diagrams vanish and the rest of two point operators give nontrivial results.

Contracting with all possible external combination we eventually obtain the following quantum 2-loop corrections of the type (2) diagrams in Figure2:

$$\langle\varphi(p)\overline{\varphi}(-p)\rangle_{\text{2-loop}} = \langle\varphi(p)\overline{\varphi}(-p)\rangle_{\text{tree}}\,X(\hat{p}), \tag{6.21a}$$
$$\langle f(p)f(-p)\rangle_{\text{2-loop}} = \langle f(p)f(-p)\rangle_{\text{tree}}\,X(\hat{p}), \tag{6.21b}$$
$$\left\langle\psi_1(p)\overline{\psi}_1(-p)\right\rangle_{\text{2-loop}} = \left\langle\psi_1(p)\overline{\psi}_1(-p)\right\rangle_{\text{tree}}\,X(\hat{p}), \tag{6.21c}$$
$$\left\langle\psi_2(p)\overline{\psi}_2(-p)\right\rangle_{\text{2-loop}} = \left\langle\psi_2(p)\overline{\psi}_2(-p)\right\rangle_{\text{tree}}\,X(\hat{p}), \tag{6.21d}$$
$$\langle\varphi(p)f(-p)\rangle_{\text{2-loop}} = \langle\varphi(p)f(-p)\rangle_{\text{tree}}\,Y(\hat{p}), \tag{6.21e}$$
$$\left\langle\overline{\varphi}(p)\overline{f}(-p)\right\rangle_{\text{2-loop}} = \left\langle\overline{\varphi}(p)\overline{f}(-p)\right\rangle_{\text{tree}}\,Y(\hat{p}), \tag{6.21f}$$
$$\langle\psi_1(p)\psi_2(-p)\rangle_{\text{2-loop}} = \langle\psi_1(p)\psi_2(-p)\rangle_{\text{tree}}\,Y(\hat{p}), \tag{6.21g}$$
$$\left\langle\overline{\psi}_1(p)\overline{\psi}_2(-p)\right\rangle_{\text{2-loop}} = \left\langle\overline{\psi}_1(p)\overline{\psi}_2(-p)\right\rangle_{\text{tree}}\,Y(\hat{p}), \tag{6.21h}$$

where

$$X(\hat{p}) = 8g_3^4 \frac{\hat{p}_+\hat{p}_- + m^2}{D(\hat{p})} \int \frac{d^2\hat{k}_1 d^2\hat{k}_2}{(2\pi)^2(2\pi)^2} \frac{\hat{k}_1^2 + m^2}{D(\hat{k}_1)^2 D(\hat{k}_2)} \int \frac{d^2\hat{k}}{(2\pi)^2} \frac{1}{D(\hat{k})D(\hat{k}_1 - \hat{k})}, \tag{6.22}$$

$$Y(\hat{p}) = 16g_3^4 \frac{\hat{p}_+\hat{p}_-}{D(\hat{p})} \int \frac{d^2\hat{k}_1 d^2\hat{k}_2}{(2\pi)^2(2\pi)^2} \frac{\hat{k}_1^2 + m^2}{D(\hat{k}_1)^2 D(\hat{k}_2)} \int \frac{d^2\hat{k}}{(2\pi)^2} \frac{1}{D(\hat{k})D(\hat{k}_1 - \hat{k})}. \tag{6.23}$$

The integration range is again constrained to satisfy all "sine" momentum conservations of given diagrams. Since the 2-loop corrections of the two point functions are proportional to tree ones, supersymmetric WT identity is satisfied in a similar way.

In conclusion, we have shown that supersymmetric WT identities are satisfied up to 2-loop orders of quantum corrections. The two point functions with loop corrections are always proportional to tree counter parts with a multiplicative factor. Two-loop contributions are summarized in the followings:

$$\langle\varphi(p)\overline{\varphi}(-p)\rangle_{\text{2-loop}} = \langle\varphi(p)\overline{\varphi}(-p)\rangle_{\text{tree}} X(\hat{p}), \qquad \langle\varphi(p)f(-p)\rangle_{\text{2-loop}} = \langle\varphi(p)f(-p)\rangle_{\text{tree}} Y(\hat{p}),$$
$$\langle f(p)\overline{f}(-p)\rangle_{\text{2-loop}} = \langle f(\hat{p})\overline{f}(-p)\rangle_{\text{tree}} X(\hat{p}), \qquad \langle\overline{\varphi}(p)\overline{f}(-p)\rangle_{\text{2-loop}} = \langle\overline{\varphi}(p)\overline{f}(-p)\rangle_{\text{tree}} Y(\hat{p}),$$
$$\langle\psi_1(p)\overline{\psi}_1(-p)\rangle_{\text{2-loop}} = \langle\psi_1(p)\overline{\psi}_1(-p)\rangle_{\text{tree}} X(\hat{p}), \quad \langle\psi_1(p)\psi_2(-p)\rangle_{\text{2-loop}} = \langle\psi_1(p)\psi_2(-p)\rangle_{\text{tree}} Y(\hat{p}),$$
$$\langle\psi_2(p)\overline{\psi}_2(-p)\rangle_{\text{2-loop}} = \langle\psi_2(p)\overline{\psi}_2(-p)\rangle_{\text{tree}} X(\hat{p}), \quad \langle\overline{\psi}_1(p)\overline{\psi}_2(-p)\rangle_{\text{2-loop}} = \langle\overline{\psi}_1(p)\overline{\psi}_2(-p)\rangle_{\text{tree}} Y(\hat{p}),$$

| Loop diagram | $X(\hat{p})$ | $Y(\hat{p})$ |
|---|---|---|
|  | 1 | 1 |
| 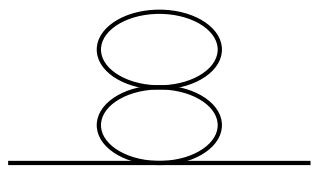 | 0 | 0 |
| 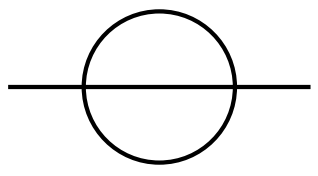 | $-6g_4^2 \frac{\hat{p}_+\hat{p}_- + m^2}{D(p)} I_2$ | $-12g_4^2 \frac{\hat{p}_+\hat{p}_-}{D(\hat{p})} I_2$ |
| 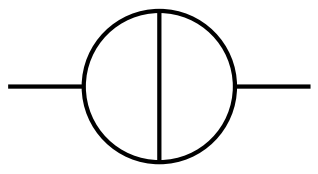 | $16m^2 g_3^4 \frac{2\hat{p}_+\hat{p}_- + m^2}{D(\hat{p})} I_3$ | $8g_3^4 \hat{p}_+\hat{p}_- \frac{\hat{p}_+\hat{p}_- + 5m^2}{D(\hat{p})} I_3$ |
| 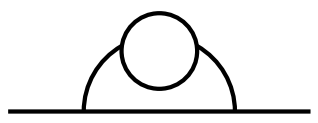 | $8g_3^4 \frac{\hat{p}_+\hat{p}_- + m^2}{D(\hat{p})} I_4$ | $16g_3^4 \frac{\hat{p}_+\hat{p}_-}{D(\hat{p})} I_4$ |

where

$$I_2 = \int \frac{d\hat{k}_1^2}{(2\pi)^2} \frac{d\hat{k}_2^2}{(2\pi)^2} \frac{1}{D(\hat{k}_1)D(\hat{k}_2)D(\hat{p} - \hat{k}_1 - \hat{k}_2)}, \tag{6.24}$$

$$I_3 = \int \frac{d\hat{k}_1^2}{(2\pi)^2} \frac{d\hat{k}_2^2}{(2\pi)^2} \frac{1}{D(\hat{k}_1)D(\hat{k}_2)D(\hat{k}_1 + \hat{p})D(\hat{k}_2 + \hat{p})D(\hat{k}_1 - \hat{k}_2)}, \tag{6.25}$$

$$I_4 = \int \frac{d^2\hat{k}_1 d^2\hat{k}_2}{(2\pi)^2(2\pi)^2} \frac{\hat{k}_1^2 + m^2}{D(\hat{k}_1)^2 D(\hat{k}_2)} \int \frac{d^2\hat{k}}{(2\pi)^2} \frac{1}{D(\hat{k})D(\hat{k}_1 - \hat{k})}. \tag{6.26}$$

## 7 Breakdown of associativity for the super doubler approach

As we have shown in the previous sections that Ward-Takahashi identities of Wess-Zumino models of D=1,2 and N=2 are exactly satisfied even at the quantum level. In the super doubler approach we impose an ansatz that the lattice "sine" momenta are the conserved quantity. Due to the change of this interpretation there appear unusual properties for the formulation.

Let us now reconsider the product of fields $\Phi_1$ and $\Phi_2$ in the super doubler approach:

$$(\Phi_1 \star \Phi_2)(p_{12}) = \int_0^{4\frac{\pi}{a}} dp_1 dp_2 \Phi_1(p_1)\Phi_2(p_2)\delta(\hat{p}_{12} - \hat{p}_1 - \hat{p}_2). \tag{7.1}$$

When $\hat{p}_i = p_i$ the above definition of the product of $\Phi_1$ and $\Phi_2$ leads to the standard convolution and thus normal product $(\Phi_1 \cdot \Phi_2)$ in the momentum space. When $\hat{p}_i = \frac{2}{a}\sin\frac{ap}{2}$ is chosen as the lattice momentum as in (3.12) the product is changed into $\star$-product. The coordinate lattice representation of the product has the following non-local form:

$$(\Phi_1 \star \Phi_2)(\frac{na}{2}) = \sum_{n_1,n_2} K(n,n_1,n_2)\Phi_1(\frac{n_1 a}{2})\Phi_2(\frac{n_2 a}{2}), \tag{7.2}$$

where

$$K(n,n_1,n_2) = \int_{-\infty}^{\infty} J(\lambda,\frac{na}{2})J(\lambda,\frac{n_1 a}{2})J(\lambda,\frac{n_2 a}{2}) \tag{7.3}$$

and $J$ is the Bessel function:

$$J(\lambda,\frac{na}{2}) = \frac{a}{2}\int_0^{\frac{4\pi}{a}} \frac{dp}{2\pi} e^{-i(n\frac{ap}{2} - \lambda\frac{n\hat{p}}{2})}. \tag{7.4}$$

Obviously the product is commutative:

$$(\Phi_1 \star \Phi_2)(p_{12}) = (\Phi_2 \star \Phi_1)(p_{21}) \tag{7.5}$$

where

$$\hat{p}_{12} = \hat{p}_{21} = \hat{p}_1 + \hat{p}_2. \tag{7.6}$$

We can show that Leibniz rule is satisfied for the difference operator on the $\star$-product:

$$\hat{\partial}(\Phi_1 \star \Phi_2) = (\hat{\partial}\Phi_1) \star \Phi_2 + \Phi_1 \star (\hat{\partial}\Phi_2), \tag{7.7}$$

where

$$\hat{\partial}\Phi\left(\frac{na}{2}\right) = \Phi\left(\frac{(n+1)a}{2}\right) - \Phi\left(\frac{(n-1)a}{2}\right). \tag{7.8}$$

The $\star$-product is, however, non-associative:

$$\begin{aligned}
&(\Phi_1 \star (\Phi_2 \star \Phi_3))(p_{123}) \\
&= \int dp_{23} \int dp_1 \int dp_2 \int dp_3 \tilde{\Phi}_1(p_1)\tilde{\Phi}_2(p_2)\tilde{\Phi}_3(p_3)\delta(\hat{p}_{23} - \hat{p}_2 - \hat{p}_3)\delta(\hat{p}_{123} - \hat{p}_1 - \hat{p}_{23}) \\
&\neq \int dp_{12} \int dp_1 \int dp_2 \int dp_3 \tilde{\Phi}_1(p_1)\tilde{\Phi}_2(p_2)\tilde{\Phi}_3(p_3)\delta(\hat{p}_{12} - \hat{p}_1 - \hat{p}_2)\delta(\hat{p}_{123} - \hat{p}_3 - \hat{p}_{12}) \\
&= ((\Phi_1 \star \Phi_2) \star \Phi_3)(p_{123}).
\end{aligned} \tag{7.9}$$

This is because the phase space of covering region on the left-hand side and right-hand side are different:

$$dp_1 dp_2 dp_3 \delta(\hat{p}_{23} - \hat{p}_2 - \hat{p}_3)\delta(\hat{p}_{123} - \hat{p}_1 - \hat{p}_{23}) \neq dp_1 dp_2 dp_3 \delta(\hat{p}_{12} - \hat{p}_1 - \hat{p}_2)\delta(\hat{p}_{123} - \hat{p}_3 - \hat{p}_{12}), \tag{7.10}$$

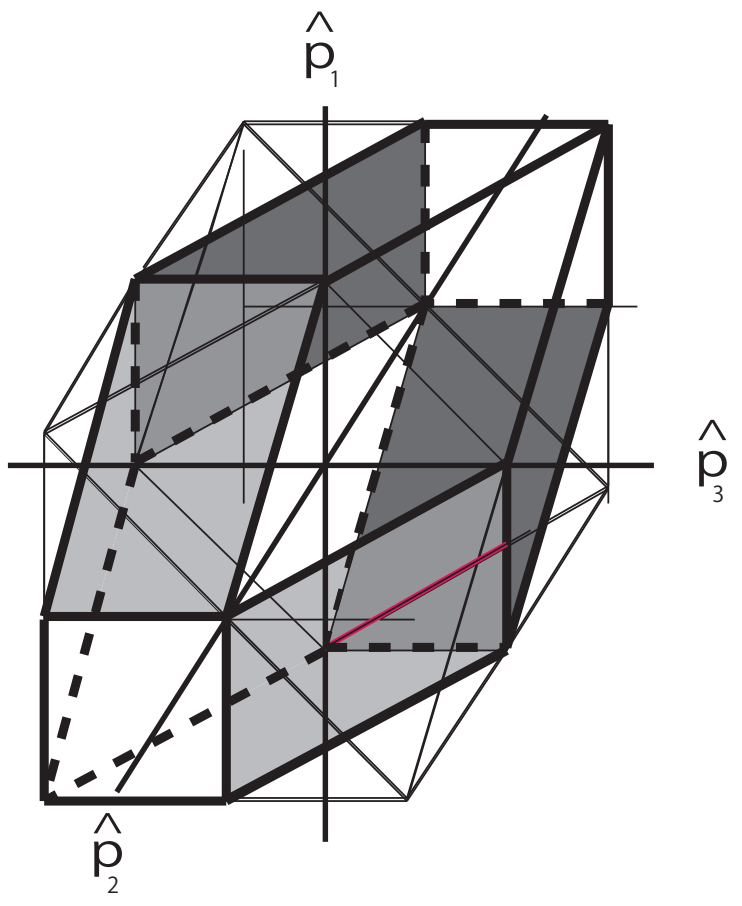

Figure 3: Non-associative phase space region: four solid pentagons framed by thin solid lines with shaded square bottom located outside of the polygon framed by fat solid lines.

which originates from the limited momentum region of lattice momentum:

$$-\frac{2}{a} \leq \hat{p}_i \leq \frac{2}{a}, \quad |\hat{p}_i + \hat{p}_j| \leq \frac{2}{a}. \tag{7.11}$$

We show the phase space region where the associativity of (7.9) is not satisfied, in fig. 3.

The breakdown of the associativity for the $\star$-product makes it difficult to extend this formulation into gauge theory since associativity is crucial for the gauge invariance proof as we can see in the following :

$$\begin{aligned}\Phi^\dagger(x) \star \Phi(x) &\to (\Phi^\dagger(x) \star e^{-i\alpha(x)}) \star (e^{i\alpha(x)} \star \Phi(x)) \\ &\neq \Phi^\dagger(x) \star (e^{-i\alpha(x)} \star e^{i\alpha(x)}) \star \Phi(x) = \Phi^\dagger(x) \star \Phi(x),\end{aligned} \tag{7.12}$$

where all the product are understood as $\star$-product.

One may wonder how this breakdown of the associativity influences to the exact lattice supersymmetry. There is no ambiguity of defining a product of several fields by $\star$-product.

Once product is well defined for each term of action, supersymmetry transformation for the action can be examined without any problem. The exact supersymmetry invariance of the action on the lattice with $\star$-product is assured since supersymmetry transformation is linear for each field for Wess-Zumino models. Algebraic consistency of supersymmetry algebra is assured since Leibniz rule is satisfied for difference operator on the $\star$-product. As far as gauge symmetry is not introduced in the formulation, lattice SUSY can be exactly kept in the formulation.

In order to generalize this super doubler approach of lattice supersymmetry into super Yang-Mills we have to find out a formulation where associativity is recovered. As we saw in the above the origin of the breakdown of the supersymmetry is due to the finite range of the lattice momentum $\hat{p}$. In order to find out an alternative of lattice momentum $\hat{p}$, we need to find a lattice momentum expression $\hat{\hat{p}}(p)$ satisfying;

1. $\hat{\hat{p}}(-p) = -\hat{\hat{p}}(p) \quad -\infty < \hat{\hat{p}}(p) < \infty$,
2. $\hat{\hat{p}}(p) \to p \ \ (a \to 0)$,
3. $\hat{\hat{p}}(\frac{2\pi}{a} - p) = \hat{\hat{p}}(p)$.

The last item 3 is also important to take into account the species doubler nature of super doubler approach. We found interesting solution which satisfies all these criteria:

$$\hat{\hat{p}}(p) = \frac{1}{a} \log \frac{1 + \sin\frac{ap}{2}}{1 - \sin\frac{ap}{2}}. \tag{7.13}$$

With the use of this expression as conserved momentum we can yet define another $\star$-product which satisfies associativity. A new type of difference operator can be defined and satisfies Leibniz rule on the $\star$-product. The details of this formulation will be soon given elsewhere[41].

# 8 Conclusion and Discussions

In this paper we have explicitly shown in details that the lattice version of supersymmetric Ward-Takahashi identities are exactly satisfied at the quantum level up to two loops of quantum corrections for super doubler approach of D=1,2 and N=2 Wess-Zumino models with interactions of $\Phi^3$ and $\Phi^4$. This means that the supersymmetry on the lattice for these models is kept exactly even at the quantum level. In the proof of the Ward-Takahashi identities at the 1- and 2-loop level of two point functions, we found universal structure that loop level WT identities are proportional to the corresponding tree level WT identities with common multiplicative factors. A brief summary of the results of this paper were reported in [45]. This result remind us of the statement of non-renormalization theorem on supersymmetric field theory models: loop level amplitudes are not renormalized and thus proportional to tree level amplitudes[46, 47].

It should be pointed out at this stage that a proper choice of overall function $G(p_1, \cdots, p_n)$ as in (3.2) makes the formulation similar to the cut off theory where the lattice "sine" momentum $\hat{p} = \frac{2}{a}\sin\frac{ap}{2}$ can be identified as momentum having cut-off of $|\hat{p}| \leq \frac{2}{a}$. It is also possible to see that identifying the lattice "sine" momentum as periodic lattice momentum the formulation is similar to SLAC derivative formulation of lattice theory. WT identities of Wess-Zumino models by cut-off theory have been intensively investigated in [42, 48] and gave similar relations between loop level and tree level WT identities.

The super doubler approach imposes an ansatz that lattice "sine" momentum is the conserved quantity. Due to this change of the formulation there appears unusual characteristics to the formulation, especially in the coordinate terminology:

1. Difference operator satisfies Leibniz rule on the $\star$-product,
2. Product of fields is non-local $\star$-product,
3. Discrete lattice translational invariance is broken,
4. Associativity for the $\star$-product is broken.

In particular the item 4 breaks gauge invariance and thus the super doubler approach cannot be used for lattice super Yang-Mills formulation if we want to keep exact super symmetry and gauge invariance at the same time. In the previous section we have proposed a new type of lattice momentum function (7.13) which cures the breakdown of the associativity of super doubler approach. Details of this formulation will be given elsewhere[41].

# Acknowledgment

We would like to thank I. Kanamori, E. Giguere, F. Sugino and Hiroshi Suzuki for fruitful discussions and useful comments. This work was supported in part by Japanese Ministry of Education, Science, Sports and Culture under the grant number 22540261 and also by the research funds of Instituto Nazionale di Fisica Nucleare (INFN) of Italy.

# Appendix

# A Integration domain

In this Appendix we discuss subtleties of integration range originated from lattice "sine" momentum conservation $\delta\left(\sum_i \hat{p}_i\right)$. In our formulation $\hat{p}$ is not linear with respect to the periodic momentum $p$, thus we cannot re-parametrize it freely by shifting the momentum unless we take care of the integration range carefully. In particular we need to treat the integration range carefully for the type of diagrams as given in subsections 4.2.1 and 4.3.2 for 1-dimension and 6.2.2 and 6.3.1 for 2-dimensions. In deriving multiplicative factors of WT identities we have used several nontrivial equalities which look not obvious but turn out to be equivalent if we take care integration range carefully. In the following we consider some examples to show the structure by investigating 1-dimensional cases.

## A.1 Cubic interaction $\Phi^3$

We begin with cubic interaction $\Phi^3$. A possible problem arises when we encounter the following integration in loop calculation:

$$\int_{-\frac{2}{a}}^{\frac{2}{a}} \frac{d\hat{k}_1}{2\pi}\frac{d\hat{k}_2}{2\pi}\frac{F(\hat{k}_1,\hat{k}_2)}{D(\hat{k}_1)D(\hat{k}_2)}(2\pi)^2\delta(\hat{k}_1+\hat{k}_2-\hat{p})\delta(\hat{q}-\hat{k}_1-\hat{k}_2), \tag{A.1}$$

where $F(\hat{k}_1;\hat{k}_2)$ is any arbitrary function of $\hat{k}_1$ and $\hat{k}_2$. If $F(\hat{k}_1,\hat{k}_2)$ has the following form as an example:

$$F(\hat{k}_1,\hat{k}_2) \equiv C^2(\hat{k}_2), \tag{A.2}$$

it is natural to expect

$$\begin{aligned} &\int_{-\frac{2}{a}}^{\frac{2}{a}} \frac{d\hat{k}_1}{2\pi}\frac{d\hat{k}_2}{2\pi}\frac{C^2(\hat{k}_1)}{D(\hat{k}_1)D(\hat{k}_2)}(2\pi)^2\delta(\hat{k}_1+\hat{k}_2-\hat{p})\delta(\hat{q}-\hat{k}_1-\hat{k}_2) \\ &= \int_{-\frac{2}{a}}^{\frac{2}{a}} \frac{d\hat{k}_1}{2\pi}\frac{d\hat{k}_2}{2\pi}\frac{C^2(\hat{k}_2)}{D(\hat{k}_1)D(\hat{k}_2)}(2\pi)^2\delta(\hat{k}_1+\hat{k}_2-\hat{p})\delta(\hat{q}-\hat{k}_1-\hat{k}_2), \end{aligned} \tag{A.3}$$

since the second expression is simply the label change of $\hat{k}_1 \leftrightarrow \hat{k}_2$ of the first expression. When we naively perform integration over either $\hat{k}_1$ or $\hat{k}_2$ we obtain,

$$\int_{-\frac{2}{a}}^{\frac{2}{a}} \frac{d\hat{k}}{2\pi}\frac{C^2(\hat{k})}{D(\hat{k})D(\hat{p}-\hat{k})}(2\pi)\delta(\hat{p}-\hat{q}), \tag{A.4}$$

$$= \int_{-\frac{2}{a}}^{\frac{2}{a}} \frac{d\hat{k}}{2\pi}\frac{C^2(\hat{p}-\hat{k})}{D(\hat{k})D(\hat{p}-\hat{k})}(2\pi)\delta(\hat{p}-\hat{q}). \tag{A.5}$$

In the continuum formulation these two results are shown to be equivalent by shifting momentum variables while now these two expression don't look to give the same result. This is because the momentum $\hat{p}$ is bounded in the range $[-\frac{2}{a},\frac{2}{a}]$.

The resolution to this puzzle is obtained by correctly taking into account the finite range of the integrated "sine" momenta. First of all each momentum is bounded by $|\hat{k}_i|, |\hat{p}| \le \frac{2}{a}$. Then the following inequality should be satisfied:

$$-\frac{2}{a} \le \hat{k}_2 = \hat{p} - \hat{k}_1 \le \frac{2}{a}. \tag{A.6}$$

The allowed integration range satisfying these constraints is shown in Figure 4(a). The above integration range of $\hat{k}_2$ in (A.6) in turn gives another range constraint of $\hat{k}_1$ as

$$\hat{p} - \frac{2}{a} \le \hat{k}_1 \le \hat{p} + \frac{2}{a}. \tag{A.7}$$

Since the integration range of $\hat{k}_1$ is $-\frac{2}{a} \le \hat{k}_1 \le \frac{2}{a}$ either of the expressions of limits in (A.7) is exceeding the range of $\hat{k}_1$. For example if $0 \le \hat{p}$ then $\frac{2}{a} \le \hat{p} + \frac{2}{a}$, the $\hat{k}_1$ integration range is limited as

$$\hat{p} - \frac{2}{a} \le \hat{k}_1 \le \frac{2}{a}. \tag{A.8}$$

On the other hand if $\hat{p} \le 0$ then $\hat{p} - \frac{2}{a} \le -\frac{2}{a}$ the $\hat{k}_1$ integration range is changed into

$$-\frac{2}{a} \le \hat{k}_1 \le \hat{p} + \frac{2}{a}. \tag{A.9}$$

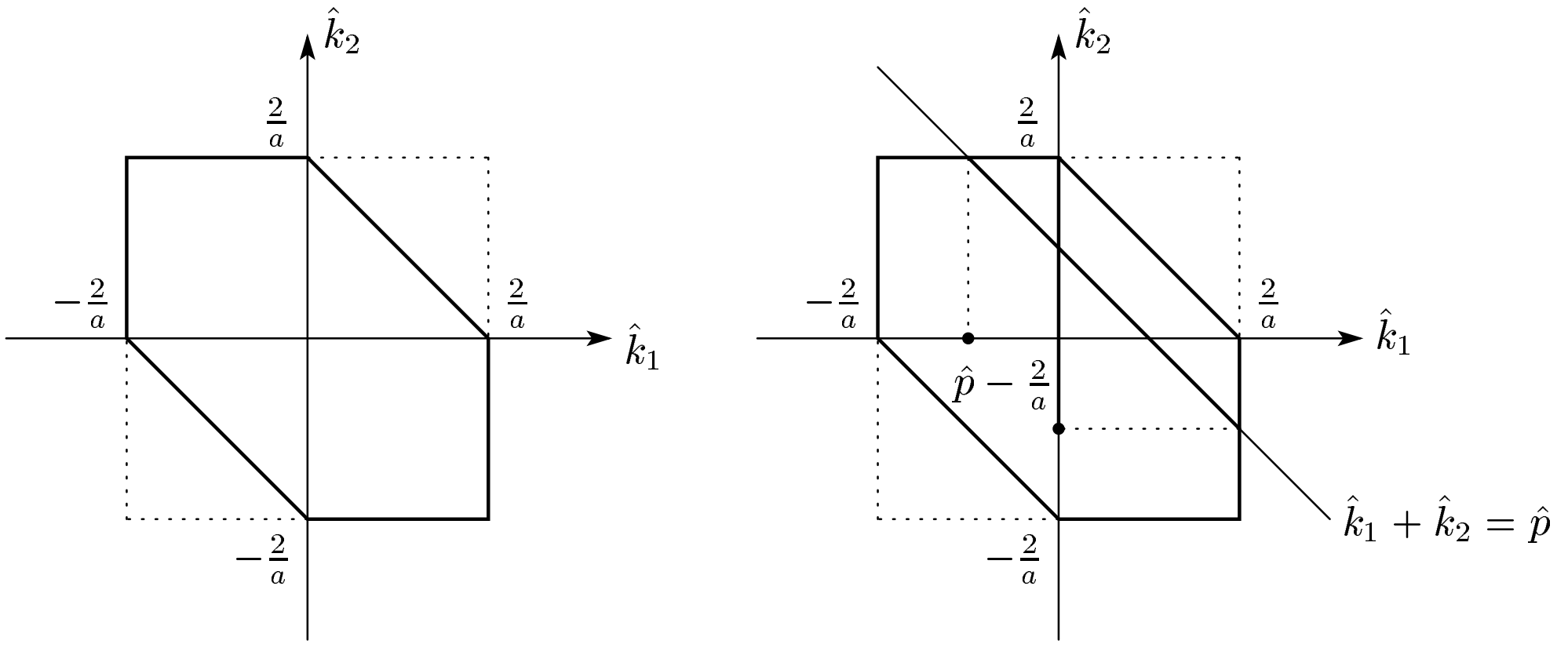

Figure 4: (a) The region bounded by the conditions: $|\hat{k}_i| < \frac{2}{a}$ and $|\hat{k}_1 + \hat{k}_2| < \frac{2}{a}$. (b) The line represents the condition $\hat{k}_1 + \hat{k}_2 = \hat{p}$ (for $\hat{p} > 0$).

See Figure 4(b). We can now show explicitly the equality of the following relation:

$$\Big[\Theta(\hat{p})\int_{\hat{p}-\frac{2}{a}}^{\frac{2}{a}} \frac{d\hat{k}}{2\pi} + \Theta(-\hat{p})\int_{-\frac{2}{a}}^{\hat{p}+\frac{2}{a}} \frac{d\hat{k}}{2\pi}\Big] \frac{C^2(\hat{k})}{D(\hat{k})D(\hat{p}-\hat{k})}(2\pi)\delta(\hat{p}-\hat{q})$$
$$= \Big[\Theta(\hat{p})\int_{\hat{p}-\frac{2}{a}}^{\frac{2}{a}} \frac{d\hat{k}}{2\pi} + \Theta(-\hat{p})\int_{-\frac{2}{a}}^{\hat{p}+\frac{2}{a}} \frac{d\hat{k}}{2\pi}\Big] \frac{C^2(\hat{p}-\hat{k})}{D(\hat{k})D(\hat{p}-\hat{k})}(2\pi)\delta(\hat{p}-\hat{q}), \tag{A.10}$$

where $\Theta(\hat{p})$ is a step function. Therefore the integrations (A.4) and (A.5) should be understood as the integration of the limited range in (A.10).

## A.2 Quartic interaction $\Phi^4$

Similar to the 1-loop diagram of $\Phi^3$ model in the previous Appendix A.1, we encounter a similar type of integration in the evaluation of 2-loop diagrams of $\Phi^4$ model. We again need to consider the integration range carefully for the following expression:

$$\int \prod_{i=1}^{3} \frac{d\hat{k}_i}{2\pi} \frac{F(\hat{k}_1,\hat{k}_2,\hat{k}_3)}{D(\hat{k}_1)D(\hat{k}_2)D(\hat{k}_3)}(2\pi)^2\delta(\hat{k}_1+\hat{k}_2+\hat{k}_3-\hat{p})\delta(\hat{q}-\hat{k}_1-\hat{k}_2-\hat{k}_3), \tag{A.11}$$

where $F(\hat{k}_1,\hat{k}_2,\hat{k}_3)$ can in principle be arbitrary function of $\hat{k}_1$ and $\hat{k}_2$.

We extend the previous treatment in Appendix A.1 to three dimensional lattice momentum space. The Figure 5(a) shows the allowed momentum region bounded by $|\hat{k}_i| \le \frac{2}{a}$ and the condition $|\hat{k}_1+\hat{k}_2+\hat{k}_3| \le \frac{2}{a}$. The shaded region represents a surface satisfying the condition $\hat{k}_1+\hat{k}_2+\hat{k}_3 = \hat{p}$ for arbitrary positive $\hat{p}$.

To see relations between the constraints and the integration range we consider the allowed range of momentum $\hat{k}_3$ which is bounded for a given external momentum $\hat{p}$ as:

$$-\frac{2}{a} \le \hat{k}_3 = \hat{p}-\hat{k}_1-\hat{k}_2 \le \frac{2}{a}, \tag{A.12}$$

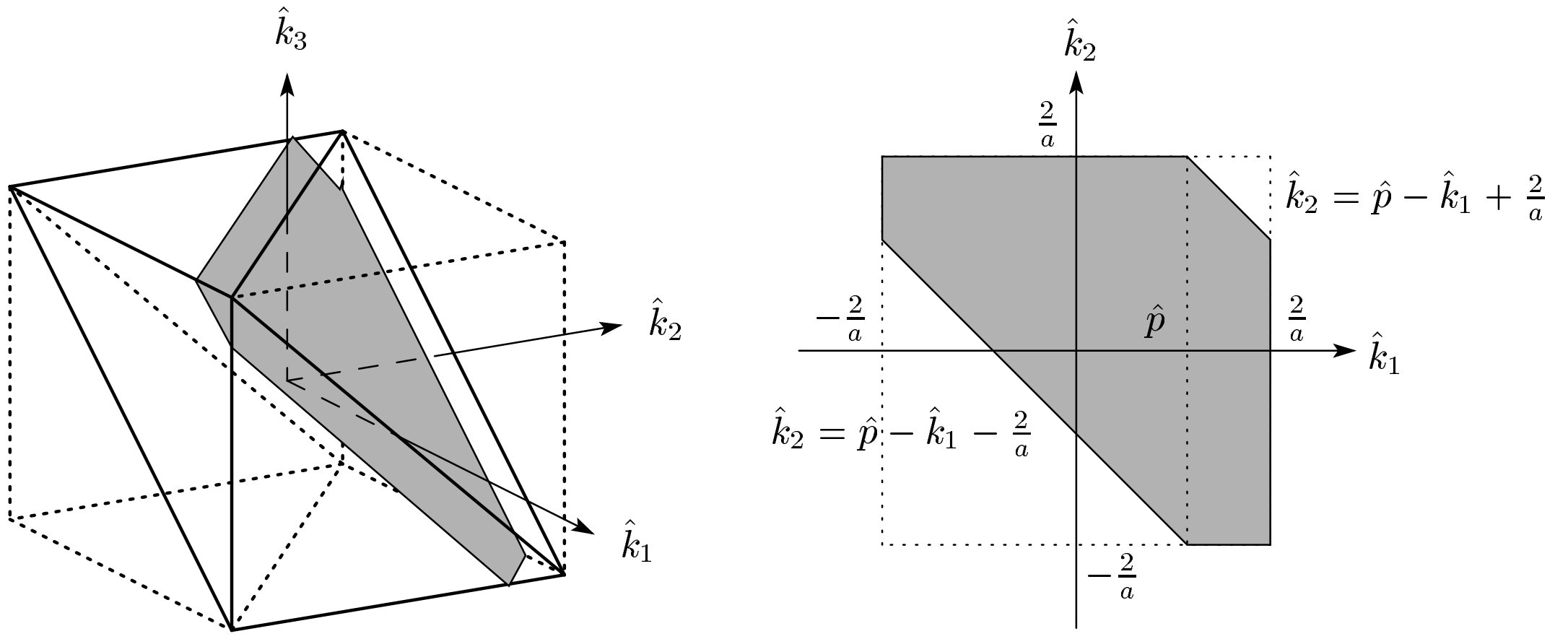

Figure 5: [a] The shaded plane represents $\hat{k}_1 + \hat{k}_2 + \hat{k}_3 = \hat{p}$ for some $\hat{p} > 0$. [b] Projected region after the integration over $\hat{k}_3$. Upper-right and lower-right lines satisfy the condition $\hat{k}_2 = \hat{p} - \hat{k}_1 + \frac{2}{a}$ and $\hat{k}_2 = \hat{p} - \hat{k}_1 - \frac{2}{a}$ respectively.

and thus

$$\hat{p} - \frac{2}{a} - \hat{k}_1 \leq \hat{k}_2 \leq \hat{p} + \frac{2}{a} - \hat{k}_1. \tag{A.13}$$

After integrating out one of the momenta, for example $\hat{k}_3$, we obtain an allowed region projected on $(\hat{k}_1, \hat{k}_2)$ plane which is shown in Figure 5(b). We thus have to consider the following integration regions:

$$\left[\int_{-\frac{2}{a}}^{\hat{p}} \frac{d\hat{k}_1}{2\pi} \int_{\hat{p}-\hat{k}_1-\frac{2}{a}}^{\frac{2}{a}} \frac{d\hat{k}_2}{2\pi} + \int_{\hat{p}}^{\frac{2}{a}} \frac{d\hat{k}_1}{2\pi} \int_{-\frac{2}{a}}^{\hat{p}-\hat{k}_1+\frac{2}{a}} \frac{d\hat{k}_2}{2\pi}\right] \frac{F(\hat{k}_1, \hat{k}_2, \hat{p} - \hat{k}_1 - \hat{k}_2)}{D(\hat{k}_1)D(\hat{k}_2)D(\hat{p} - \hat{k}_1 - \hat{k}_2)}(2\pi)\delta(\hat{p} - \hat{q}). \tag{A.14}$$

For $\hat{p} < 0$, we have the same representation with replacement $\hat{p} = -|\hat{p}|$. Since the resulting region is symmetric under $\hat{k}_1 \leftrightarrow \hat{k}_2$ as shown in Figure 5, thus we can change the above integration as.

$$\left[\int_{-\frac{2}{a}}^{\hat{p}} \frac{d\hat{k}_2}{2\pi} \int_{\hat{p}-\hat{k}_2-\frac{2}{a}}^{\frac{2}{a}} \frac{d\hat{k}_1}{2\pi} + \int_{\hat{p}}^{\frac{2}{a}} \frac{d\hat{k}_2}{2\pi} \int_{-\frac{2}{a}}^{\hat{p}-\hat{k}_2+\frac{2}{a}} \frac{d\hat{k}_1}{2\pi}\right] \frac{F(\hat{k}_1, \hat{k}_2, \hat{p} - \hat{k}_1 - \hat{k}_2)}{D(\hat{k}_1)D(\hat{k}_2)D(\hat{p} - \hat{k}_1 - \hat{k}_2)}(2\pi)\delta(\hat{p} - \hat{q}). \tag{A.15}$$

In realistic examples the arbitrary function $F(\hat{k}_1, \hat{k}_2, \hat{k}_3)$ in (A.11) falles into one of cases of the following:

(1) $F$ is totally symmetric function of $\hat{k}_1$ and $\hat{k}_2$.

(2) $F$ depends only on single momentum.

The second case needs some care to show the equivalence of different looking expressions. For example when we take $F = C^2(\hat{k}_2)$ we can show the following equivalence if we take care about the proper integrationn range:

$$\left[\int_{-\frac{2}{a}}^{\hat{p}} \frac{d\hat{k}_1}{2\pi} \int_{\hat{p}-\hat{k}_1-\frac{2}{a}}^{\frac{2}{a}} \frac{d\hat{k}_2}{2\pi} + \int_{\hat{p}}^{\frac{2}{a}} \frac{d\hat{k}_1}{2\pi} \int_{-\frac{2}{a}}^{\hat{p}-\hat{k}_1+\frac{2}{a}} \frac{d\hat{k}_2}{2\pi}\right] \frac{C^2(\hat{k}_2)}{D(\hat{k}_1)D(\hat{k}_2)D(\hat{p}-\hat{k}_1-\hat{k}_2)}, \tag{A.16}$$

$$= \left[\int_{-\frac{2}{a}}^{\hat{p}} \frac{d\hat{k}_2}{2\pi} \int_{\hat{p}-\hat{k}_2-\frac{2}{a}}^{\frac{2}{a}} \frac{d\hat{k}_1}{2\pi} + \int_{\hat{p}}^{\frac{2}{a}} \frac{d\hat{k}_2}{2\pi} \int_{-\frac{2}{a}}^{\hat{p}-\hat{k}_2+\frac{2}{a}} \frac{d\hat{k}_1}{2\pi}\right] \frac{C^2(\hat{k}_1)}{D(\hat{k}_1)D(\hat{k}_2)D(\hat{p}-\hat{k}_1-\hat{k}_2)}, \tag{A.17}$$

$$= \left[\int_{-\frac{2}{a}}^{\hat{p}} \frac{d\hat{k}_1}{2\pi} \int_{\hat{p}-\hat{k}_1-\frac{2}{a}}^{\frac{2}{a}} \frac{d\hat{k}_2}{2\pi} + \int_{\hat{p}}^{\frac{2}{a}} \frac{d\hat{k}_1}{2\pi} \int_{-\frac{2}{a}}^{\hat{p}-\hat{k}_1+\frac{2}{a}} \frac{d\hat{k}_2}{2\pi}\right] \frac{C^2(\hat{p}-\hat{k}_1-\hat{k}_2)}{D(\hat{k}_1)D(\hat{k}_2)D(\hat{p}-\hat{k}_1-\hat{k}_2)}. \tag{A.18}$$

The equivalence between the first and the second integrants can be shown by interchanging the integrated momenta $\hat{k}_1 \leftrightarrow \hat{k}_2$. And the equivalence of the first and the third can be shown by a redefinition of the momentum $\hat{k}_2' \equiv \hat{p} - \hat{k}_1 - \hat{k}_2$.

# B Tadpole corrections

## B.1 Cubic interaction $\Phi^3$ in 2-dimensions: tadpole diagram with 1 loop correction

Here we consider 1-loop tadpole corrections. The following 1-loop tadpole corrections vanish trivially since there are no surviving vertex and propagator:

$f(p)$ 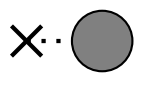 $=$ $\overline{f}(p)$ 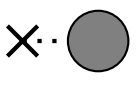 $= 0,$ (B.1a)

The remaining diagrams also vanish by the cancellation between loop contributions:

$\varphi(p)$ 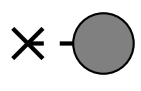 $=$ $\varphi(p)$ 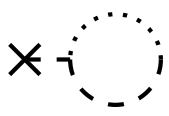 $+$ $\varphi(p)$ 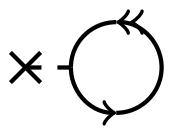

$$= 2(-g_3)\varphi(p)\int \frac{d^2\hat{k}}{(2\pi)^2}\frac{im}{D(\hat{k})} + (2ig_3)\varphi(p)\int \frac{d^2\hat{k}}{(2\pi)^2}\frac{m}{D(\hat{k})} = 0, \tag{B.2a}$$

$\overline{\varphi}(p)$ 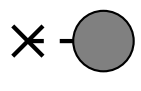 $=$ $\overline{\varphi}(p)$ 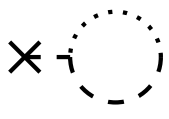 $+$ $\overline{\varphi}(p)$ 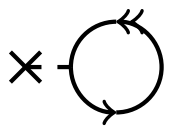

$$= (g_3)2\overline{\varphi}(p)\int \frac{d^2\hat{k}}{(2\pi)^2}\frac{-im}{D(\hat{k})} + (-2ig_3)\overline{\varphi}(p)\int \frac{d^2\hat{k}}{(2\pi)^2}\frac{-m}{D(\hat{k})} = 0. \tag{B.2b}$$

Fermionic tadpole contributions vanish with an obvious reason. Thus all the 1-loop tadpole contribution to the propagators vanish. Similar to 1-loop corrections for quartic interaction, the result is a consequence of the supersymmetry.

## B.2 Cubic interaction $\Phi^3$: 2 point operators with two loop tadpole

Next we consider 2-loop contributions on propagators. Since 2-loop tadpole diagram includes 1-loop diagram as a subdiagram, we can utilize the results in subsection 6.2.2. We thus consider the following diagrams:

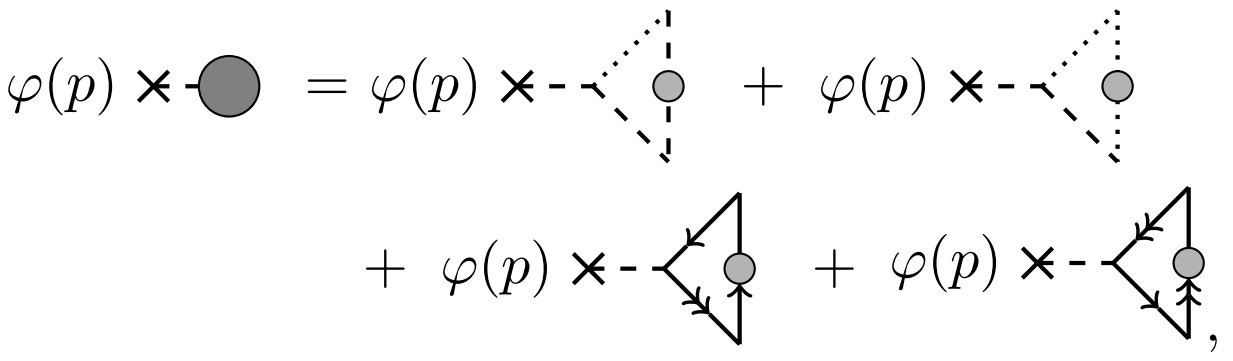

(B.3a)

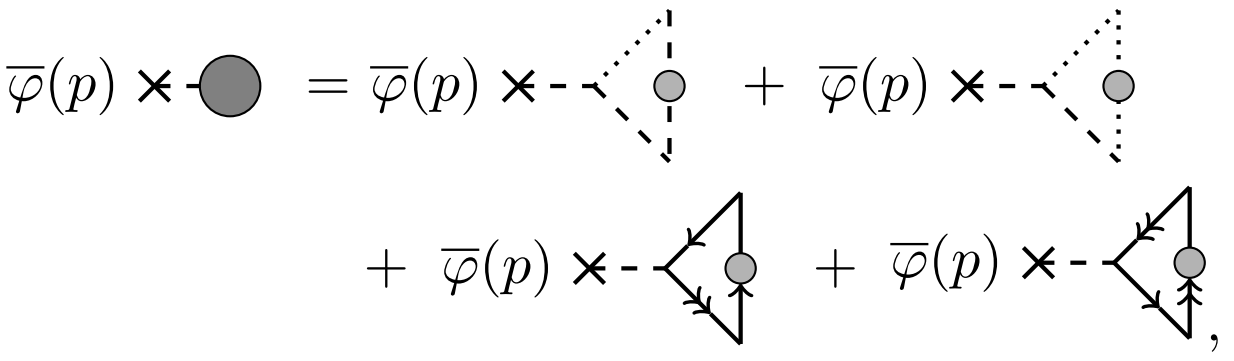

(B.3b)

where (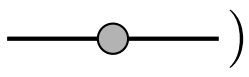) denotes one-loop correction calculated in subsection 6.2.2. Since $\varphi$ and $\overline{\varphi}$ are Hermitian conjugate each other, the above two corrections are also conjugate. It is thus sufficient to calculate either of the above contributions. Explicit calculation leads

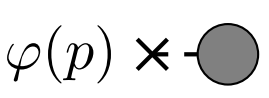

$$= \varphi(p)(1+1-1-1) \times \left[ 4img_3^3 \int \frac{d^2\hat{q}}{(2\pi)^2} \frac{\hat{q}^2}{D(\hat{q})^2} \int \frac{d^2\hat{k}}{(2\pi)^2} \frac{1}{D(\hat{k})D(\hat{q}-\hat{k})} \right] = 0. \tag{B.4}$$

Therefore all the tadpole contributions vanish.

# References

[1] J. Wess, and B. Zumino, A Lagrangian Model Invariant Under Supergauge Transformation, Phys. Lett. **B 49** (1974) 52.

[2] E. Witten, Topological Quantum Field Theory, Commun. Math. Phys. **117** (1988) 353.

[3] N. Kawamoto and T. Tsukioka, N=2 supersymmetric model with Dirac-Kaehler fermions from generalized gauge theory in two-dimensions, Phys. Rev.**D 61** (2000) 105009, [hep-th/0502119],
J. Kato, N. Kawamoto and Y. Uchida, Twisted superspace for N=D=2 super BF and Young-Mills with Dirac-Kahler fermion mechanism, Int.J.Mod.Phys.**A 19** (2004) 2149-2182,
J. Kato, N. Kawamoto and A. Miyake, N=4 twisted superspace from Dirac-Kahler twist and off-shell SUSY invariant actions in four dimensions, Nucl. Phys. **B 721** (2005) 229-286.

[4] P. Dondi and H. Nicolai, Lattice Supersymmetry, Nuovo Cimento A 41 (1977) 1.

[5] P. Hasenfratz, Nucl. Phys. **B525** (1988) 401,
H. Neuberger, Phys. Lett. **417B** (1988) 141; **427B** (1988) 353,
M. Luscher, Phys. Lett. **428B** (1988) 342.

[6] S. Nojiri, Continuous 'Translation' and Supersymmetry on the lattice, Prog. Theor. Phys. 74 (1985) 819; The Spontaneous Breaking of Supersymmetry on the Finite Lattice, Prog. Theor. Phys. 74 (1985) 1124.

[7] H. Nicolai, Phys. Lett. **89B** (1980) 341; Nucl. Phys. **B176** (1980) 419.

[8] N. Sakai and M. Sakamoto, Lattice Supersymmetry and the Nicolai mapping, Nucl.Phys. B229 (1983) 173.

[9] G. Curci and G. Veneziano, Supersymmetry and the lattice: A reconciliation ?, Nucl. Phys. **B292** (1987) 555.

[10] I. Montvay, An algorithm for gluinos on the lattice, Nucl. Phys. **B466** (1996) 259; Supersymmetric Yang-Mills theory on the lattice, Int. J. Mod. Phys. **A17** (2002)2377.

[11] Y. Taniguchi, One loop calculation of SUSY Ward-Takahashi identity on lattice with Wilson fermion, Phys. Rev. **D63** (2001) 014502 [hep-lat/0305020].

[12] A. Feo, The supersymmetric Ward-Takahashi identity in 1-loop lattice perturbation theory, I: General procedure, Phys. Rev. **D70** (2004) 054504 [hep-lat/0305020], Supersymmetry on the lattice, Nucl. Phys. Proc. Suppl. **119** (2003) 198, [arXiv:hep-lat/0210015]; Predictions and recent results in SUSY on the lattice, Mod. Phys. Lett. **A 19** (2004) 2387 [arXiv:hep-lat/0410012].

[13] J. Bartels and G. Kramer, A lattice version of the wess-zumino model, Z. Phys. C 20 (1983) 159.

[14] J. Giedt, R. Koniuk, E. Poppitz and T. Yavin, Less naive about supersymmetric lattice quantum mechanics, JHEP 0412 (2004) 033, [arXiv:hep-th/0410041].

[15] M. Caselle, A. D'Adda and L. Magnea, Doubling of All Matter Fields Coupled with Gravity on a Lattice, Phys. Lett. **B192** (1987) 411.

[16] A. D'Adda, A. Feo, I. Kanamori, N. Kawamoto and J. Saito, Species Doublers as Super Multiplets in Lattice Supersymmetry: Exact Supersymmetry with interactions for $D = 1, N = 2$, JHEP 1009 (2010) 059, [arXiv:1006.2046].

[17] A. D'Adda, I. Kanamori, N. Kawamoto and J. Saito, Species Doublers as Super Multiplets in Lattice Supersymmetry: Chiral Conditions of Wess-Zumino Model for $D = N = 2$, [arXiv:1107.1629].

[18] K. Fujikawa, Supersymmetry on the lattice and the Leibniz rule, Nucl Phys. **B 636** (2002) 80 [arXiv:hep-th/0205095].

[19] K. Fujikawa, M. Ishibashi, Lattice chiral symmetry, Yukawa couplings and the Majorana condition, Phys. Lett. B 528 (2002) 295 [arXiv:hep-lat/0112050]; K. Fujikawa and M. Ishibashi, Lattice chiral symmetry and the Wess-Zumino model, Nucl Phys. **B 622** (2002) 115 [arXiv:hep-th/0109156].

[20] Y. Kikukawa and H. Suzuki, A local formulation of lattice Wess-Zumino model with exact $U(1)_R$ symmetry, JHEP 0502 (2005) 012, [arXiv:hep-lat/0412042].

[21] A. Feo, Exact Ward-Takahashi identity for the lattice $N = 1$ Wess-Zumino model, J. Phys. Conf. Ser. 33 (2006) 405, [arXiv;hep-lat/0512028], M. Bonini and A. Feo, Exact lattice Ward-Takahashi identity for the $N = 1$ Wess-Zumino model, Phys. Rev. **D 71**

(2005) 114512 [arXiv:hep-lat/0504010], Wess-Zumino model with exact supersymmetry on the lattice, JHEP 09 (2004) 011 [arXiv:hep-lat/0402034].

[22] T. Aoyama and Y. Kikukawa, Overlap formula for the chiral multiplet, Phys.Rev.**D59**(1999)054507, H. So and N. Ukita, Gansparg-Wilson relation and lattice supersymmetry, Phys. Lett. **B457** (1999) 314.

[23] G. Bergner, F. Bruckmann and J. M. Pawlowski, Generalising the Ginsparg-Wilson relation: Lattice Supersymmetry from Blocking Transformations, Phys. Rev. **D79** (2009) 115007

[24] D.B. Kaplan, E. Katz and M. Unsal, Supersymmetry on a spatial lattice, JHEP 0305 (2003) 037, [arXiv:hep-lat/0206019]; A.G. Cohen, D.B. Kaplan, E. Katz and M. Unsal, Supersymmetry on a Euclidean spacetime lattice. I: A target theory with four supercharges, JHEP 0308 (2003) 024, [arXiv:hep-lat/0302017]; A.G. Cohen, D.B. Kaplan, E. Katz and M. Unsal, Supersymmetry on a Euclidean spacetime lattice. II: Target theories with eight supercharges, JHEP 0312 (2003) 031, [arXiv:hep-lat/03070120].

[25] D.B. Kaplan, Recent developments in lattice supersymmetry, Nucl. Phys. Proc. Suppl. **129** (2004) 109, [arXiv:hep-lat/0309099].

[26] S. Catterall, Lattice supersymmetry and topological field theory, JHEP 0305 (2003) 038, [arXiv:hep-lat/0301028]; S. Catterall, Lattice supersymmetry and topological field theory, Nucl. Phys. Proc. Suppl. **129** (2004) 871, [arXiv:hep-lat/0309040].

[27] S. Catterall, PoS LAT2005 (2005) 006; S. Catteral, A geometrical approach to $N = 2$ super Yang-Mills theory on the 2-dimensional lattice, JHEP 0411 (2004) 006 [arXiv:hep-lat/0410052]; S. Catteral, Lattice formulation of $N = 4$ super Yang-Mills theory, JHEP 0506 (2005) 027 [arXiv:hep-lat/0503036].

[28] S. Catterall, D.B. Kaplan and M. Ünsal, Exact lattice supersymmetry, Phys. Rept. **484** (2009) 71, [arXiv:0903.4881[hep-lat]].

[29] J. Giedt, Advances and applications of lattice supersymmetry, PoS LATTICE2006 (2006) 008, [arXiv:hep-lat/0701006].

[30] F. Sugino, A lattice formulation of super Yang-Mills theories with exact supersymmetry, JHEP 0401 (2004) 015 [arXiv:hep-lat/0311021]; F. Sugino, Super Yang-Mills theories on the two-dimensional lattice with exact supersymmetry, JHEP 0403 (2004) 067, [arXiv:hep-lat/0401017]; F. Sugino, Various super Yang-Mills theories with exact supersymmetry on the lattice , JHEP 0501 (2005) 016, [arXiv:hep-lat/0410035].

[31] M. Kato, M. Sakamoto and H. So, Taming the Leibniz Rule on the Lattice, JHEP 05 (2008) 057, [arXiv:0803.3121 [hep-lat]] M. Kato, M. Sakamoto and H. So, No-Go Theorem of Leibniz Rule and Supersymmetry on the Lattice, PoS LATTICE 2008 233, [arXiv:0810.2360 [hep-lat]].

[32] M. Kato, M. Sakamoto and H. So, Leibniz rule and exact supersymmetry on lattice: a case of supersymmetrical quantum mechanics, PoS LAT2005 (2006) 274, [arXiv:hep-lat/0509149].

[33] S. Catterall and E. Gregory, A lattice path integral for supersymmetry quantum mechanics, Phys.Lett. B487 (2000) 349, [arXiv:hep-lat/0006013],
S. Catterall and S. Karamaov, Exact Lattice Supersymmetry: the 2-dimensional $N = 2$ Wess-Zumino Model, [arXiv:hep-lat/0108024]; A Two-Dimensional Lattice Model with Exact Supersymmetry, Nucl. Phys. Proc. Suppl. 106 (2002) 935, [arXiv:hep-lat/0110071].

[34] A. D'Adda, I. Kanamori, N. Kawamoto and K. Nagata, Twisted superspace on a lattice, Nucl. Phys. **B 707** (2005) 100, [arXiv:hep-lat/0406029]; A. D'Adda, I. Kanamori, N. Kawamoto and K. Nagata, Twisted $N = 2$ Exact SUSY on the Lattice for BF and Wess-Zumino, Nucl. Phys. Proc. Suppl. **140** (2005) 754, [arXiv:hep-lat/0409092]; A. D'Adda, I. Kanamori, N. Kawamoto and K. Nagata, Exact extended supersymmetry on a lattice: Twisted $N = 2$ super Yang-Mills in 2-dimensions, Phys. Lett. B633 (2006) 645 [arXiv:hep-lat/0507029].

[35] F. Bruckmann and M. de Kok, Noncommutativity approach to supersymmetry on the lattice: SUSY quantum mechanics and an inconsistency, Phys. Rev. **D73** (2006) 074511, [arXiv:hep-lat/0603003].

[36] F. Bruckmann, S. Catterall and M. de Kok, Critique of the link approach to exact lattice supersymmetry, Phys. Rev. **D 75** (2007) 045016, [arXiv:hep-lat/0611001].

[37] A. D'Adda, N. Kawamoto and J. Saito, Formulation of Supersymmetry on a Lattice as a Representation of a Deformed Superalgebra, Phys. Rev. **D 81** (2010) 065001, [arXiv:0907.4137].

[38] P. H. Damgaard and S. Matsuura, Classification of Supersymmetric Lattice Gauge Theories by Orbifolding, JHEP **0707** (2007) 051 [arXiv:0704.2696 [hep-lat]]; Relations among Supersymmetric Lattice Gauge Theories via Orbifolding, JHEP **0708** (2007) 087 [arXiv:0706.3007 [hep-lat]]; Lattice Supersymmetry: Equivalence between the Link Approach and Orbifolding, JHEP **0709** (2007) 097 [arXiv:0708.4129 [hep-lat]]; Geometry of Orbifolded Supersymmetric Lattice Gauge Theories, Phys. Lett. B **661** (2008) 52 [arXiv:0801.2936 [hep-th]].

[39] H.B. Nielsen and M. Ninomiya, Absence Of Neutrinos On A Lattice. 1. Proof By Homotopy Theory, Nucl. Phys. **B 185** (1981) 20 [Erratum-ibid. **B 195** (1982) 541]; H.B. Nielsen and M. Ninomiya, Absence Of Neutrinos On A Lattice. 2. Intuitive Topological Proof, Nucl. Phys. **B 193** (1981) 173.

[40] L.H. Karsten and J. Smit, Lattice Fermions :Species Doubling, Chiral Invariance, And The Triangle Anomaly, Nucl. Phys. **B 183** (1981) 103.

[41] A. D'Adda, N. Kawamoto and J. Saito, to appear.

[42] G. Bergner, Complete supersymmetry on the lattice and a No-Go theorem, JHEP 1001 (2010) 024, [arXiv:0909.4791].

[43] G. Bergner, T. Kaestner, S. Uhlman, A. Wipf and C. Wozar, Supersymmetric lattice models in 1- and 2-dimensions, PoS LAT 2007 (2007) 265, [arXiv:0709.0822].

[44] G. Bergner, T. Kaestner, S Uhlmann and A. Wipf, Low-dimensional Supersymmetric Lattice Models, Annals. Phys. **323** (2008) 946, [arXiv:0705.2212].

[45] K. Asaka, A. D'Adda, N. Kawamoto and Y. Kondo, Nat. Acad.of Sci. Kazakhstan report Phys. and Math., No2, 288, March-April 2013,P.18 [arXiv:1302.1268[hep-lat]]

[46] J. Iliopoulos and B. Zumino, Nucl. Phys. **B76** (1974) 310.

[47] K. Fujikawa and W. Lang, Nucl. Phys. **B88** (1975) 61,
P. West, Nucl. Phys. **B106** (1976) 219,
M. T. Grisaru, N. Rocek and W. Siegel, Nucl. Phys. **B519** (1979) 429.

[48] D. Kadoh and H. Suzuki, Supersymmetric nonperturbative formulation of the WZ model in lower dimensions, Phys.Lett. B684 (2010) 167 [arXiv:0909.3686[hep-th]]; Supersymmetry restoration in lattice formulations of 2D $N = (2, 2)$ WZ model based on the Nicolai map, Phys.Lett. B696 (2011) 163 [arXiv:1011.0788].

[49] I. Kanamori and H. Suzuki, Restoration of supersymmetry on the lattice:Two-dimensional $N = (2, 2)$ supersymmetric Yang-Mills theory, Nucl. Phys. **B811** (2009) 420, [arXiv:0809.2856[hep-lat]]; Some physics of the two-dimensional $N = (2, 2)$ supersymmetric Yang-Mills theory: Lattice Monte Carlo study, Phys. Lett. **B672** (2009) 307, [arXiv:0811.2851 [hep-lat]].